\documentclass[12pt]{article}
\usepackage{amssymb}
\usepackage{graphicx}
\begin{document}
\newcommand{\beq}{\begin{equation}}
\newcommand{\eeq}{\end{equation}}
\newcommand{\beqa}{\begin{eqnarray}}
\newcommand{\eeqa}{\end{eqnarray}}
\newcommand{\beqar}{\begin{eqnarray*}}
\newcommand{\eeqar}{\end{eqnarray*}}
\newcommand{\al}{\alpha}
\newcommand{\be}{\beta}
\newcommand{\del}{\delta}
\newcommand{\D}{\Delta}
\newcommand{\eps}{\epsilon}
\newcommand{\ga}{\gamma}
\newcommand{\Ga}{\Gamma}
\newcommand{\ka}{\kappa}
\newcommand{\nn}{\nonumber}
\newcommand{\inn}{\!\cdot\!}
\newcommand{\h}{\eta}
\newcommand{\ii}{\iota}
\newcommand{\kk}{\varphi}
\newcommand\F{{}_3F_2}
\newcommand{\la}{\lambda}
\newcommand{\La}{\Lambda}
\newcommand{\na}{\prt}
\newcommand{\Om}{\Omega}
\newcommand{\om}{\omega}
\newcommand{\p}{\phi}
\newcommand{\sig}{\sigma}
\renewcommand{\t}{\theta}
\newcommand{\z}{\zeta}
\newcommand{\ssc}{\scriptscriptstyle}
\newcommand{\eg}{{\it e.g.,}\ }
\newcommand{\ie}{{\it i.e.,}\ }
\newcommand{\labell}[1]{\label{#1}} 
\newcommand{\reef}[1]{(\ref{#1})}
\newcommand\prt{\partial}
\newcommand\veps{\varepsilon}
\newcommand{\pol}{\varepsilon}
\newcommand\vp{\varphi}
\newcommand\ls{\ell_s}
\newcommand\cF{{\cal F}}
\newcommand\cA{{\cal A}}
\newcommand\cS{{\cal S}}
\newcommand\cT{{\cal T}}
\newcommand\cV{{\cal V}}
\newcommand\cL{{\cal L}}
\newcommand\cM{{\cal M}}
\newcommand\cN{{\cal N}}
\newcommand\cG{{\cal G}}
\newcommand\cH{{\cal H}}
\newcommand\cI{{\cal I}}
\newcommand\cJ{{\cal J}}
\newcommand\cl{{\iota}}
\newcommand\cP{{\cal P}}
\newcommand\cQ{{\cal Q}}
\newcommand\cg{{\it g}}
\newcommand\cR{{\cal R}}
\newcommand\cB{{\cal B}}
\newcommand\cO{{\cal O}}
\newcommand\tcO{{\tilde {{\cal O}}}}
\newcommand\bg{\bar{g}}
\newcommand\bb{\bar{b}}
\newcommand\bH{\bar{H}}
\newcommand\bX{\bar{X}}
\newcommand\bK{\bar{K}}
\newcommand\bA{\bar{A}}
\newcommand\bZ{\bar{Z}}
\newcommand\bxi{\bar{\xi}}
\newcommand\bphi{\bar{\phi}}
\newcommand\bpsi{\bar{\psi}}
\newcommand\bprt{\bar{\prt}}
\newcommand\bet{\bar{\eta}}
\newcommand\btau{\bar{\tau}}
\newcommand\hF{\hat{F}}
\newcommand\hA{\hat{A}}
\newcommand\hT{\hat{T}}
\newcommand\htau{\hat{\tau}}
\newcommand\hD{\hat{D}}
\newcommand\hf{\hat{f}}
\newcommand\hg{\hat{g}}
\newcommand\hp{\hat{\phi}}
\newcommand\hi{\hat{i}}
\newcommand\ha{\hat{a}}
\newcommand\hb{\hat{b}}
\newcommand\hQ{\hat{Q}}
\newcommand\hP{\hat{\Phi}}
\newcommand\hS{\hat{S}}
\newcommand\hX{\hat{X}}
\newcommand\tL{\tilde{\cal L}}
\newcommand\hL{\hat{\cal L}}
\newcommand\tG{{\widetilde G}}
\newcommand\tg{{\widetilde g}}
\newcommand\tphi{{\widetilde \phi}}
\newcommand\tPhi{{\widetilde \Phi}}
\newcommand\te{{\tilde e}}
\newcommand\tk{{\tilde k}}
\newcommand\tf{{\tilde f}}
\newcommand\ta{{\tilde a}}
\newcommand\tb{{\tilde b}}
\newcommand\tR{{\tilde R}}
\newcommand\teta{{\tilde \eta}}
\newcommand\tF{{\widetilde F}}
\newcommand\tK{{\widetilde K}}
\newcommand\tE{{\widetilde E}}
\newcommand\tpsi{{\tilde \psi}}
\newcommand\tX{{\widetilde X}}
\newcommand\tD{{\widetilde D}}
\newcommand\tO{{\widetilde O}}
\newcommand\tS{{\tilde S}}
\newcommand\tB{{\widetilde B}}
\newcommand\tA{{\widetilde A}}
\newcommand\tT{{\widetilde T}}
\newcommand\tC{{\widetilde C}}
\newcommand\tV{{\widetilde V}}
\newcommand\thF{{\widetilde {\hat {F}}}}
\newcommand\Tr{{\rm Tr}}
\newcommand\tr{{\rm tr}}
\newcommand\STr{{\rm STr}}
\newcommand\hR{\hat{R}}
\newcommand\MZ{\mathbb{Z}}
\newcommand\MR{\mathbb{R}}
\newcommand\M[2]{M^{#1}{}_{#2}}

\newcommand\bS{\textbf{ S}}
\newcommand\bI{\textbf{ I}}
\newcommand\bJ{\textbf{ J}}

\begin{titlepage}
\begin{center}

\vskip 2 cm
{\LARGE \bf   An NS-NS basis for odd-parity couplings\\  \vskip 0.25 cm at  order $\alpha'^3$ 
 }\\
\vskip 1.25 cm
 Mohammad R. Garousi\footnote{garousi@um.ac.ir}
 
\vskip 1 cm
{{\it Department of Physics, Faculty of Science, Ferdowsi University of Mashhad\\}{\it P.O. Box 1436, Mashhad, Iran}\\}
\vskip .1 cm
 \end{center}
\begin{abstract}

In this study, we thoroughly investigate the covariant and $B$-field gauge invariant odd-parity NS-NS couplings at order $\alpha'^3$, while considering the removal of field redefinitions, Bianchi identities, and total derivative freedoms. Our comprehensive analysis reveals the existence of 477 independent couplings. To establish a specific basis, we construct it in such a way that none of the couplings contain terms involving structures such as $R$, $R_{\mu\nu}$, $\nabla_\mu H^{\mu\alpha\beta}$, $\nabla_\mu\nabla^\mu\Phi$, or terms with more than two derivatives, except for one term that possesses three derivatives on $H$. Interestingly, the mentioned coupling with the four-derivative on the $B$-field is rendered zero by the sphere-level three-point S-matrix element.

Furthermore, we demonstrate that the remaining 476 parameters in type II superstring theory are fixed to zero by imposing the requirement that the circular reduction of the couplings remains invariant under $O(1,1,\MZ)$ T-duality transformations. This result is consistent with our expectations and highlights the crucial role played by the $O(1,1,\MZ)$ symmetry in constraining the parameter space of the classical effective actions in string theory.

\end{abstract}
\end{titlepage}

\section{Introduction} \label{intro}
Superstring theory is a comprehensive theory of quantum gravity that provides a framework for understanding the fundamental interactions of particles. Its perturbative construction is based on a finite number of massless fields and an infinite tower of massive fields. When considering low-energy phenomena, where the effects of the massive fields are integrated out, the theory is effectively described by supergravity and its appropriate higher-derivative deformations.
To determine the higher-derivative classical and higher-genus couplings in this theory, several techniques have been developed. A recent review article provides an in-depth exploration of these techniques \cite{Ozkan:2024euj}.

One technique for determining the classical higher-derivative effective action is based on the observation in string field theory that the dimensional reduction of the classical effective action of string theory on a torus $T^d$ should exhibit T-duality symmetry $O(d,d,\mathbb{R})$ at all orders of $\alpha'$ \cite{Sen:1991zi, Hohm:2014sxa}. The $O(d,d,\mathbb{R})$ transformations include both geometrical transformations that leave the parent geometrical couplings invariant and nongeometrical transformations that transform a parent geometrical coupling into other geometrical couplings.
This symmetry has been used in cosmological reduction to constrain the NS-NS coupling constants in the parent theory \cite{David:2021jqn,David:2022jcl,Hyakutake:2023oiy}. Furthermore, a non-geometric aspect of this symmetry, known as $\beta$-transformation, has been employed in \cite{Baron:2022but,Garousi:2022qmk,Garousi:2023diq,Baron:2023qkx,Baron:2023oxk} to determine the coupling constants without relying on dimensional reduction.

To determine coupling constants at a specific order of $\alpha'$ while enforcing T-duality symmetry, a straightforward approach involves cyclic reduction of the most general covariant and gauge-invariant independent couplings at that order \cite{Garousi:2019wgz,Garousi:2019jbq}. In this method, it is assumed that the independent geometric couplings remain invariant under the non-geometric $O(1,1,\mathbb{Z}) = \mathbb{Z}_2$ group or the Buscher transformations \cite{Buscher:1987sk,Buscher:1987qj}, except for certain total derivative terms in the base space. To preserve the standard diffeomorphism symmetry of the couplings in the effective action, modifications to the Buscher transformations in the base space are necessary to incorporate higher derivative corrections at all orders of $\alpha'$ \cite{Kaloper:1997ux,Garousi:2019wgz}. Consequently, T-duality establishes, in general, a relationship between the couplings at order $\alpha'^n$ and all couplings at orders $\alpha'^0, \alpha', \alpha'^2, ..., \alpha'^{n-1}, \alpha'^n$. This enables one to fix all the unknown coupling constants at the order of $\alpha'^n$ in terms of the known coupling constants at orders $\alpha'^0, \alpha', \alpha'^2, ..., \alpha'^{n-1}$, and also one unknown parameter at the order $\alpha'^n$. As a result, there is an infinite number of T-dual multiplet couplings in the effective action at all orders of $\alpha'$ \cite{Garousi:2021yyd}.
 
Furthermore, the original couplings exhibit diffeomorphism symmetry and B-field gauge symmetry, indicating that the circular reduction of the couplings should  possess the anomalous B-field gauge symmetry and a $U(1) \times U(1)$ gauge symmetry in the base space. The first $U(1)$ gauge symmetry arises from the momentum vector field, while the second $U(1)$ gauge symmetry arises from the winding vector field in the base space \cite{Maharana:1992my,Kaloper:1997ux}. Consequently, the higher-derivative corrections to the Buscher rules and the total derivative terms in the base space should also respect these gauge symmetries. The T-duality $\MZ_2$-constraint has been successfully utilized to determine the NS-NS covariant couplings at order $\alpha'^2$ in the bosonic and heterotic theory \cite{Garousi:2019mca,Garousi:2023kxw}, as well as the couplings at order $\alpha'^3$ in the superstring theories \cite{Garousi:2020gio,Garousi:2020lof}, up to an overall factor. Importantly, the determined couplings are fully consistent with the S-matrix elements \cite{Garousi:2022ghs,Gholian:2023kjj}.

Moreover, it is known that the non-perturbative D-brane/O-plane objects in string theory undergo covariant transformations under the  $\mathbb{Z}_2$-group \cite{Bergshoeff:1996cy}. The diffeomorphism and gauge symmetries, as well as the T-duality symmetry, can be utilized to determine the higher-derivative world-volume effective action of these objects \cite{Garousi:2017fbe} . In this context, the T-duality constraint can be stated as follows: the T-duality transformation of the world-volume reduction of a D$_p$-brane/O$_p$-plane should be equivalent to the transverse reduction of a D$_{p-1}$-brane/O$_{p-1}$-plane, up to some gauge-invariant total derivative terms in the base space.

To impose T-duality, however, the static gauge must be utilized to fix the diffeomorphism symmetry. The world-volume and transverse reductions of O-planes in the static gauge reveal a $U(1)\times U(1)$ gauge symmetry, enabling the use of the aforementioned constraint to determine the world-volume effective action of O-planes up to overall factors \cite{Robbins:2014ara,Garousi:2014oya,Mashhadi:2020mzf}. However, in the case of the reduction of a D-brane in the static gauge, it exhibits a $U(1)\times U(1)$ gauge symmetry only when the world-volume components of the momentum vector field in the transverse reduction are zero. It is under this specific background of the base space that the effective action of the D-brane satisfies the aforementioned constraint. Consequently, the T-duality constraint on only the massless closed string fields cannot fully fix the world-volume couplings \cite{Hosseini:2022vrr}\footnote{Note that in \cite{Hosseini:2022vrr}, we were not aware at that time that setting the world-volume components of the momentum vector field to zero in the transverse reduction is equivalent to imposing the $U(1) \times U(1)$ gauge symmetry on the covariant coupling in the transverse reduction.}. On the other hand, the T-duality constraint on only the massless open string fields cannot fully fix the world-volume couplings, up to overall factors, either \cite{Karimi:2018vaf,Garousi:2022rcv}. However, T-duality on both massless open and closed string fields may be able to fully fix the D-brane world-volume couplings, up to overall factors.


When utilizing the T-duality technique described above to determine the space-time effective actions or world-volume effective actions at a specific order of $\alpha'$, it is necessary to establish a basis for the independent covariant couplings at that order of $\alpha'$ with arbitrary coupling constants. In a specific basis, the couplings should not be interconnected through various Bianchi identities, field redefinitions, or integration by parts \cite{Metsaev:1987zx}. At each order of $\alpha'$, there are many different bases; however, they are all related to each other through the utilization of these freedoms.

In this paper, our focus is on obtaining a specific NS-NS basis at order $\alpha'^3$ in the spacetime effective action. This basis incorporates couplings that involve an odd number of $B$-fields, known as odd-parity couplings\footnote{Note that the world-sheet action of a string in the presence of a background B-field incorporates the 2-dimensional antisymmetric Levi-Civita tensor. As a result, the $B$-field exhibits an odd behavior under world-sheet parity.}. In a previous study \cite{Garousi:2020mqn}, a similar basis was established for even-parity couplings, which involve an even number of $B$-fields, and it contained 872 couplings. The $\mathbb{Z}_2$-symmetry is then responsible for fixing all the coupling constants in superstring theory, except for an overall factor, as demonstrated in \cite{Garousi:2020gio}. Our analysis in this paper will reveal that the basis for odd-parity couplings has 477 independent terms. In the specific scheme that we choose for this basis, one of the couplings involves four derivatives on the $B$-field. This term has three field strengths; hence, its coupling constant is zero according to the sphere-level three-point function in string theory. The remaining 476 couplings may be found by T-duality.

From a general perspective, it is anticipated that the couplings in bosonic string theory and the NS-NS couplings in type II superstring theory, where the $B$-field gauge transformations adhere to standard transformations, exhibit parity invariance. Conversely, in the heterotic theory, where the $B$-field gauge transformation deviates from the standard due to the Green-Schwarz mechanism \cite{Green:1984sg}, parity invariance is not expected. This parity invariance may arise in bosonic and type II superstring theories due to the assumption  that  the T-duality transformations may relate  the odd-parity couplings at order $\alpha'^n$ to the  odd-parity couplings at orders $\alpha'^0, \alpha', \alpha'^2, ..., \alpha'^{n-1}$, and the fact  that the Bianchi identities do not allow for the existence of odd-parity couplings at order $\alpha'$ in either theory. The construction of such couplings is impossible without considering the Chern-Simons three-form $\Omega$, which only appears in the heterotic theory due to the anomalous $B$-field gauge transformation.

By observing that there are no odd-parity couplings at orders $\alpha'^0$ and $\alpha'$, and utilizing the assumption  that T-duality relates the odd-parity couplings at order $\alpha'^2$ to the odd-parity couplings at orders $\alpha'^0, \alpha'$, it follows that T-duality does not permit the existence of odd-parity couplings at order $\alpha'^2$. Similarly, with the absence of odd-parity couplings at orders $\alpha'^0, \alpha', \alpha'^2$, T-duality does not permit the existence of odd-parity couplings at order $\alpha'^3$. This logic can be extended to all higher orders of $\alpha'$ to conclude that there are no odd-parity couplings at any order of $\alpha'$ in either bosonic or superstring theories. Therefore, there should be  no odd-parity T-dual multiplets in the effective actions of these theories.

Conversely, since there is an odd-parity coupling at order $\alpha'$ in the heterotic theory due  to the Green-Schwarz mechanism, i.e., $\alpha' H\Omega$, T-duality relates the odd-parity couplings at order $\alpha'^2$ to the odd-parity coupling at order $\alpha'$ \cite{Garousi:2023kxw}. Similarly, since there are odd-parity couplings at orders $\alpha'$ and $\alpha'^2$, T-duality relates the odd-parity couplings at order $\alpha'^3$ to the odd-parity couplings at orders $\alpha'$ and $\alpha'^2$. This line of reasoning can be extended to higher orders of $\alpha'$ in the heterotic theory, leading to the conclusion that there exist odd-parity couplings at all subsequent orders of $\alpha'$. The above couplings, which are related to the coupling $\alpha' H\Omega$, belong to one T-dual multiplet, i.e.,
\beqa
T_1^{(\rm odd)}=a_1\alpha' {\bS}_1^{(\rm odd)}+a_1^2\alpha'^2 {\bS}_2^{(\rm odd)}+a_1^3\alpha'^3{\bS}_3^{(\rm odd)}+\cdots\,,
\eeqa
where ${\bS}_1^{(\rm odd)}$ has the coupling $H\Omega$, and $a_1$ is its coefficient, which is $-1/4$. 

Moreover, there are even-parity couplings at order $\alpha'^3$ in the heterotic theory with coefficient $\zeta(3)$. The anomalous $B$-field gauge transformation then produces odd-parity couplings at order $\alpha'^4$ which have one Chern-Simons three-form $\Omega$. Then T-duality relates the odd-parity couplings at order $\alpha'^5$ to the odd-parity couplings at order $\alpha'^4$. Using the above logic, one finds there is another odd-parity T-dual multiplet with coefficient $\zeta(3)$. Similarly, there are odd-parity T-dual multiplets with coefficients $\zeta(5)$, $\zeta(7)$, and so on. Note that there are no couplings in the effective action of string theory which have coefficient $\zeta(2k)$ \cite{Stieberger:2009rr}.

As a consequence, one would expect the coupling constants of the 476 odd-parity couplings to be zero in both bosonic and superstring theories, while allowing for the possibility of non-zero values in the heterotic theory. The non-zero terms should have the overall factor $a_1^3$. However, the inclusion of the T-duality constraint on these 476 couplings within the heterotic theory, which necessitates T-duality transformations at orders $\alpha'$ and $\alpha'^2$, introduces considerable complexity to the calculations. Therefore, these calculations have been deferred for future research.

In this paper, after finding the basis for the odd-parity couplings at order $\alpha'^3$ in the particular scheme, which involves one coupling with three field strengths and 476 couplings with four or more field strengths, we specifically impose the T-duality constraint on the 476 couplings within superstring theory, where T-duality transformations at orders $\alpha'$ and $\alpha'^2$ are absent. We anticipate that the T-duality constraint will enforce all 476 parameters to be zero. This calculation serves to validate the effectiveness of T-duality in constraining the effective actions in string theory.

The outline of the paper is as follows: In section 2, we present the most general covariant and $B$-field gauge-invariant odd-parity couplings involving the metric, dilaton, and B-field at order $\alpha'^3$. There are 19,996 such couplings. We then incorporate the most general total derivative terms and field redefinitions with arbitrary parameters. To satisfy various Bianchi identities, we rewrite them in the local inertial frame, where the first partial derivative of the metric is zero and its higher derivatives are not zero. Furthermore, we express terms containing derivatives of the B-field strength $H$ in terms of the potential, i.e., $H = dB$. By utilizing the arbitrary parameters in the total derivative terms and field redefinitions, we demonstrate that there are only 477 physical parameters, while all other parameters are unphysical as they can be eliminated through Bianchi identities, field redefinitions, and total derivative terms. The choice of which set of 477 parameters is selected among the 19,996 parameters specifies the basis scheme. We establish that it is possible to select schemes in which there are no terms involving $R, R_{\mu\nu}, \nabla_\mu H^{\mu\alpha\beta}$, or $\nabla_\mu\nabla^\mu\Phi$, and apart from one coupling with a term containing four derivatives, there are no couplings with more than two partial derivatives. Even with these choices, there remain 2,190 couplings. We choose a specific scheme for the 477 physical couplings among these 2,190 couplings. One of these couplings is the one with a term involving four partial derivatives, and its corresponding coupling constant is fixed to zero by the three-point S-matrix element. In this section, we provide the explicit form of the remaining 476 couplings. 
In section 3, we perform circular reduction on these 476 couplings and impose the T-duality $\mathbb{Z}_2$-constraint in superstring theory on the resulting couplings in the base space. We demonstrate that all parameters are forced to be zero by this constraint. 
In Section 4, we offer a brief discussion of our results. For the calculations in this paper, we utilized the "xAct" package \cite{Nutma:2013zea}.

\section{The NS-NS basis  at order  $\alpha'^3$}\label{sec.2}

The effective action of string theory has two expansions. One expansion is the genus expansion, which includes the classical and a tower of quantum effects. The other one is a stringy expansion at each genus order, which is an expansion in terms of higher-derivative couplings. The number of derivatives in each coupling can be accounted for by the order of $\alpha'$. The classical effective action of string theory at the critical dimension $D$ has the following $\alpha'$-expansion in the string frame:
\beqa
{\bf S}_{\rm eff}=\sum_{n=0}^{\infty}\alpha'^n{\bf S}^{(n)}&;& {\bf S}^{(n)}= \frac{\gamma_n}{\kappa^2}\int d^{D} x\sqrt{-g} e^{-2\Phi}\mathcal{L}_n\,,\labell{seff}
\eeqa
where $\kappa$ is related to the $D$-dimensional Newton's constant, and  $\gamma_n$ is the overall normalization of the effective action at order $\alpha'^n$, for example $\gamma_0=-2$. The effective action must be invariant under coordinate transformations and under $B$-field gauge transformations. Therefore, the NS-NS fields must appear in the Lagrangian $\mathcal{L}_n$ through their field strengths and covariant derivatives, i.e., the geometrical coupling. For instance, the effective action at order $\alpha'^0$ has the following three geometrical couplings:
\beqa
\mathcal{L}_0&= R-\frac{1}{12}H_{\alpha\beta\gamma}H^{\alpha\beta\gamma}+4\nabla_\alpha\Phi\nabla^\alpha\Phi\,.\labell{S0}
\eeqa
 It is invariant under the non-geometric $\mathbb{Z}_2$-constraint.

To find the effective action using the $\mathbb{Z}_2$-constraint, one first needs to find all independent couplings with unfixed coupling constants and then impose the $\mathbb{Z}_2$-symmetry on them to find the relations between the coupling constants. To find the independent odd-parity couplings at order $\alpha'^3$, following \cite{Garousi:2019cdn}, one should first write all gauge-invariant NS-NS couplings at the eight-derivative order that have an odd number of $B$-field. Using the package "xAct" \cite{Nutma:2013zea}, one finds that there are 19,996 such couplings, i.e.,
\beqa
\mathcal{L}'_3=& c'_1 R_{\beta}{}^{\delta \epsilon\nu} R_{\gamma\epsilon}{}^{\mu \zeta} \
R_{\delta\nu\mu\zeta} \nabla_\alpha H^{\alpha\beta\gamma}+\cdots\,,\labell{L3}
\eeqa
where $c'_1,\cdots, c'_{19996}$ are some coupling constants. However, the above couplings are not all independent. Some of them are related by total derivative terms, some of them are related by field redefinitions, and some others are related by various Bianchi identities.

To remove the total derivative terms and the field redefinition freedom from the above couplings, we add the following terms to the Lagrangian $\mathcal{L}'_3$:
\beqa
\frac{\alpha'^3\gamma_3}{\kappa^2}\int d^{D}x \sqrt{-g}e^{-2\Phi} \mathcal{J}_3&\!\!\!\!\!\equiv\!\!\!\!\!\!& \frac{\alpha'^3\gamma_3}{\kappa^2}\int d^{D}x\sqrt{-g} \nabla_\alpha (e^{-2\Phi}{\cal I}_3^\alpha)\,,\labell{J3}\nn\\
\frac{\alpha'^3\gamma_3}{\kappa^2}\int d^{D}x\sqrt{-g}e^{-2\Phi}\mathcal{K}_3
&\!\!\!\!\!\equiv\!\!\!\!\!\!& \frac{\alpha'^3\gamma_3}{\kappa^2}\int d^{D} x\sqrt{-g}e^{-2\Phi}\Big[(\frac{1}{2} \nabla_{\gamma}H^{\alpha \beta \gamma} -  H^{\alpha \beta}{}_{\gamma} \nabla^{\gamma}\Phi)\delta B^{(3)}_{\alpha\beta}\nn\\
&& -(  R^{\alpha \beta}-\frac{1}{4} H^{\alpha \gamma \delta} H^{\beta}{}_{\gamma \delta}+ 2 \nabla^{\beta}\nabla^{\alpha}\Phi)\delta g^{(3)}_{\alpha\beta}\labell{eq.13}\\
\nn\\
&&-2( R\! -\!\frac{1}{12} H_{\alpha \beta \gamma} H^{\alpha \beta \gamma} \!+\! 4 \nabla_{\alpha}\nabla^{\alpha}\Phi \!-4 \nabla_{\alpha}\Phi \nabla^{\alpha}\Phi)(\delta\Phi^{(3)}-\frac{1}{4}\delta g^{(3)\mu}{}_\mu) \Big],\nn
\eeqa
where the vector ${\cal I}_3^\alpha$ in the total derivative terms represents all possible covariant and gauge-invariant terms at the seven-derivative level with odd parity, \ie
\beqa
{\cal I}_3^\alpha= &J_1 H^{\alpha\delta\epsilon}H_{\beta\delta\epsilon}H^{\mu\zeta\eta}R^{\beta\gamma}R_{\gamma\mu\zeta\eta}+\cdots\,,
\eeqa
where the coefficients $J_1,\cdots, J_{10310}$ represent 10,310 arbitrary parameters. The field redefinition contribution at order $\alpha'^3$ results from replacing the following field redefinition:
\begin{eqnarray}
g_{\mu\nu}&\rightarrow &g_{\mu\nu}+\alpha'^3 \delta g^{(3)}_{\mu\nu}\,,\nn\\
B_{\mu\nu}&\rightarrow &B_{\mu\nu}+ \alpha'^3\delta B^{(3)}_{\mu\nu}\,,\nn\\
\Phi &\rightarrow &\Phi+ \alpha'^3\delta\Phi^{(3)}\,,\labell{gbp}
\end{eqnarray}
into the leading-order action \reef{S0} and keeping only the terms at linear order of perturbation. Integration by parts has also been used in writing the second equation in \reef{eq.13}. To produce odd-parity couplings, the perturbation $\delta B^{(3)}_{\mu\nu}$ must have even-parity terms, and the perturbations $g^{(3)}_{\mu\nu}$ and $\Phi^{(3)}$ must have odd-parity terms at the six-derivative level. In other words, we require:
\beqa
 \delta B^{(3)}_{\alpha\beta}&=& e_1 H_{[\alpha}{}^{\gamma\delta}H_{\beta]\gamma}{}^{\epsilon}H_{\delta}{}^{\varepsilon\theta}H_{\epsilon\varepsilon}{}^{\eta}H_{\theta}{}^{\mu\nu}H_{\eta\mu\nu}+\cdots\,, \nn\\
 \delta g^{(3)}_{\alpha\beta}&=& g_1 R^{\gamma\delta}R_{\epsilon\delta\mu\{ \alpha}\nabla_{\beta\}}H_{\gamma}{}^{\epsilon\mu}+\cdots \,,\nn\\
 \delta\Phi^{(3)}&=&f_1 \nabla^\alpha \Phi\nabla_{\delta} R_{\alpha\beta\gamma\epsilon}\nabla^\epsilon H^{\beta\gamma\delta}+\cdots\,.\labell{eq.12}
\eeqa
The coefficients $e_1,\cdots, e_{2680}$, $g_1,\cdots, g_{2373}$, and $f_1,\cdots, f_{364}$ represent 5,417 arbitrary parameters. It should be noted that these arbitrary parameters have absorbed a factor of $\gamma_0/\gamma_3$. When adding the total derivative terms and field redefinition terms to $\mathcal{L}'_3$, the resulting Lagrangian is the same, but with different parameters $c_1, c_2, \cdots$. We refer to this new Lagrangian as ${\cal L}_3$. Consequently, we have the equation:
\beqa
\Delta_3-{\cal J}_3-{\cal K}_3&=&0\,.\labell{DLK}
\eeqa
Here, $\Delta_3={ \cal L}_3-\mathcal{L}'_3$ is equivalent to $\mathcal{L}'_3$, but with coefficients $\delta c_1,\delta c_2,\cdots$ where $\delta c_i= c_i-c'_i$. Solving the above equation yields some linear relations between only $\delta c_1,\delta c_2,\cdots$, which indicate how the couplings are related among themselves through the total derivative and field redefinition terms. There are also many relations between $\delta c_1,\delta c_2,\cdots$ and the coefficients of the total derivative terms and field redefinitions, which are not of interest.

To solve the equation \reef{DLK}, it is necessary to express it in terms of independent couplings by imposing the following Bianchi identities:
\beqa
 R_{\alpha[\beta\gamma\delta]}&=&0\,,\nn\\
 \nabla_{[\mu}R_{\alpha\beta]\gamma\delta}&=&0\,,\labell{bian}\\
\nabla_{[\mu}H_{\alpha\beta\gamma]}&=&0\,,\nn\\
{[}\nabla,\nabla{]}\mathcal{O}-R\mathcal{O}&=&0\,.\nn
\eeqa
To impose these Bianchi identities in a gauge-invariant form, one can contract the left-hand side of each Bianchi identity and its derivatives with the NS-NS field strengths and their derivatives to generate terms at order $\alpha'^3$. The coefficients of these terms are arbitrary. By adding these terms to the equation \reef{DLK}, the equation can be solved to find linear relations between only $\delta c_1,\delta c_2,\cdots$. Alternatively, to impose the Bianchi identities in a non-gauge-invariant form, one can rewrite the terms in \reef{DLK} in a local frame where the covariant derivatives are expressed in terms of partial derivatives and the first partial derivative of the metric is zero. Additionally, the terms involving $H$ in \reef{DLK} that have partial derivatives on $H$ can be rewritten in terms of the B-field, i.e., $H=dB$. In this way, all the Bianchi identities are automatically satisfied \cite{Garousi:2019cdn}. In fact, when expressing the couplings in terms of potentials rather than field strengths, there are no Bianchi identities at all. This latter approach is easier for imposing the Bianchi identities using a computer. Furthermore, in this approach, there is no need to introduce another large number of arbitrary parameters to include the Bianchi identities in the equation \reef{DLK}.

Using the above steps, one can rewrite the different terms on the left-hand side of \reef{DLK} in terms of independent but non-gauge invariant couplings. The coefficients of the independent terms must be zero, leading to a set of algebraic equations. The solution to these equations has two parts. The first part consists of 477 relations between only $\delta c_i$'s, while the second part consists of additional relations between the coefficients of total derivative terms, field redefinitions, and $\delta c_i$'s, which are not of interest to us. The number of relations in the first part determines the number of independent couplings in ${\cal L}_3'$. In a particular scheme, it is possible to set some of the coefficients in ${\cal L}_3'$ to zero. However,  after replacing the non-zero terms in \reef{DLK}, the number of relations between only $\delta c_i$'s should remain unchanged. In other words, there must always be 477 relations.
To ensure this, we set the coefficients of the couplings in ${\cal L}_3'$ to zero for each term that contains $R$, $R_{\mu\nu}$, $\nabla_\mu H^{\mu\alpha\beta}$, or $\nabla_\mu\nabla^\mu\Phi$. After setting these coefficients to zero, there are still 477 relations between $\delta c_i$'s, allowing us to remove these terms from consideration.

We then attempt to set zero couplings in ${\cal L}_3'$ that have terms with more than two partial derivatives. By imposing this condition and solving \reef{DLK} again, one would find 476 relations between only $\delta c_i$'s. This means that at least one of the independent couplings has terms with more than two derivatives. We have found this independent coupling to be 
\beqa
[R^2\nabla^3H]_1 &=& c_{477} R_{\alpha\epsilon\beta}{}^\mu R_{\gamma\mu\delta\nu}\nabla^\nu\nabla^\epsilon\nabla^\delta H^{\alpha\beta\gamma}\,.\labell{T53}
\eeqa
We found this coupling by dividing the couplings involving more than two derivatives into two parts and setting the coefficients of one part to zero. If the corresponding equations in \reef{DLK} yield 477 relations between the remaining $\delta c_i$'s, then that choice is allowed; otherwise, the other part is allowed to be set to zero. We then divided the non-zero part into two parts and set half of them to zero. Again, if the corresponding equations in \reef{DLK} give 477 relations between the remaining $\delta c_i$'s, then that choice is allowed; otherwise, the other part is allowed to be set to zero. By repeating this strategy, one finds that the above coupling is one of the independent couplings. All other couplings that have terms with more than two partial derivatives are allowed to be zero\footnote{It is interesting to note that in the basis of even-parity NS-NS couplings at order $\alpha'^3$, which excludes terms that contain $R$, $R_{\mu\nu}$, $\nabla_\mu H^{\mu\alpha\beta}$, or $\nabla_\mu\nabla^\mu\Phi$, there is also one coupling that has a term with more than two derivatives \cite{Garousi:2020mqn}. Moreover, in the basis of M-theory couplings at the eight-derivative level, which excludes terms that contain $R$, $R_{\mu\nu}$,or  $\nabla_\mu F^{\mu\nu\alpha\beta}$, there is also one coupling that has a term with more than two derivatives \cite{Garousi:2021iaw}.}. 

There are still 2,189 couplings that have no terms with more than two derivatives and no terms with structures $R,\,R_{\mu\nu},\,\nabla_\mu H^{\mu\alpha\beta}$, $ \nabla_\mu\nabla^\mu\Phi$. Hence, there are still many choices for the non-zero coefficients that satisfy the 477 relations $\delta c_i=0$. In the particular scheme that we have chosen, the 477 couplings appear in 36 structures.

One structure is \reef{T53}, which has only one coupling. Since it involves three fields, and the sphere-level three-point function in string theory has at most six derivatives, the coefficient of this coupling is zero in string theory. All other couplings appear in the following 35 structures:
\beqa
{\cal L}_3&\!\!\!\!\!=\!\!\!\!\! &[(\nabla H)^3\nabla^2\Phi]_3+[\nabla HR(\nabla^2\Phi)^2]_2+[(\nabla H)^3R]_6+[\nabla HR^2\nabla^2\Phi]_4+[\nabla HR^3]_3\nn\\
&&+[\nabla H(\nabla\Phi)^2(\nabla^2\Phi)^2]_1+[\nabla HR^2(\nabla\Phi)^2]_3+[H^2\nabla H(\nabla^2\Phi)^2]_7+[(\nabla H)^3(\nabla\Phi)^2]_3\nn\\&&+[\nabla HR (\nabla\Phi)^2\nabla^2\Phi]_3+[HR\nabla\Phi(\nabla^2\Phi)^2]_4+[HR^2\nabla\Phi\nabla^2\Phi]_8+[H(\nabla H)^2\nabla\Phi\nabla^2\Phi]_{23}\nn\\&&+[HR^3\nabla\Phi]_6+[H^2\nabla HR\nabla^2\Phi]_{27}+[H(\nabla H)^2R\nabla\Phi]_{20}+[H^2\nabla HR^2]_{42}+[H^2(\nabla H)^3]_{22}\nn\\&&
+[H(\nabla H)^2(\nabla\Phi)^3]_3+[HR(\nabla\Phi)^3\nabla^2\Phi]_3+[H^3\nabla\Phi(\nabla^2\Phi)^2]_3+[H^2\nabla H(\nabla\Phi)^2\nabla^2\Phi]_{16}\nn\\&&
+[H^2\nabla HR(\nabla\Phi)^2]_{12}+[H^4\nabla H\nabla^2\Phi]_{21}+[H^3(\nabla H)^2\nabla\Phi]_{34}+[H^4\nabla H R]_{67}+[H^3R\nabla\Phi\nabla^2\Phi]_{19}\nn\\&&
+[H^3R^2\nabla\Phi]_{25}+[H^3(\nabla\Phi)^3\nabla^2\Phi]_1+[H^2\nabla H(\nabla\Phi)^4]_1+[H^5\nabla\Phi\nabla^2\Phi]_8+[H^5R\nabla\Phi]_{26}\nn\\&&
+[H^4\nabla H(\nabla\Phi)^2]_{23}+[H^6\nabla H]_{24}+[H^7\nabla\Phi]_3\,.\labell{T55}
\eeqa
The notation $[X]_n$ indicates that the structure $X$ has $n$ different contractions with an arbitrary coupling constant for each contraction. The six structures that have no dilaton are as follows:
\beqa
[\nabla HR^3]_3&=&  c_{99}  
    R_{\alpha  }{}^{\epsilon  }{}_{\beta  
}{}^{\varepsilon  } R_{\gamma  }{}^{\mu  }{}_{\epsilon  
}{}^{\zeta  } R_{\delta  \zeta  \varepsilon  \mu  } 
\nabla^{\delta  }H^{\alpha  \beta  \gamma  } +   c_{98} 
     R_{\alpha  }{}^{\epsilon  }{}_{\beta  
}{}^{\varepsilon  } R_{\gamma  }{}^{\mu  }{}_{\epsilon  
}{}^{\zeta  } R_{\delta  \mu  \varepsilon  \zeta  } 
\nabla^{\delta  }H^{\alpha  \beta  \gamma  }\nn\\&& +   c_{100} 
     R_{\alpha  \delta  \beta  }{}^{\epsilon  } 
R_{\gamma  }{}^{\varepsilon  \mu  \zeta  } 
R_{\epsilon  \mu  \varepsilon  \zeta  } \nabla^{\delta 
 }H^{\alpha  \beta  \gamma  }\,,\nn 
\eeqa
\beqa
[(\nabla H)^3R]_6&=&  
   c_{230}  
    R_{\gamma  \mu  \varepsilon  \zeta  } 
\nabla_{\delta  }H_{\alpha  }{}^{\epsilon  \varepsilon  } 
\nabla^{\delta  }H^{\alpha  \beta  \gamma  } \nabla^{\zeta  
}H_{\beta  \epsilon  }{}^{\mu  } +   c_{232}  
    R_{\delta  \zeta  \epsilon  \mu  } \nabla^{\delta 
 }H^{\alpha  \beta  \gamma  } \nabla^{\varepsilon  }H_{\alpha  
\beta  }{}^{\epsilon  } \nabla^{\zeta  }H_{\gamma  \varepsilon  
}{}^{\mu  }  \nn\\&&+   c_{231}  
    R_{\delta  \mu  \epsilon  \zeta  } \nabla^{\delta 
 }H^{\alpha  \beta  \gamma  } \nabla^{\varepsilon  }H_{\alpha  
\beta  }{}^{\epsilon  } \nabla^{\zeta  }H_{\gamma  \varepsilon  
}{}^{\mu  } +   c_{233}  
    R_{\epsilon  \zeta  \varepsilon  \mu  } 
\nabla_{\delta  }H_{\alpha  \beta  }{}^{\epsilon  } 
\nabla^{\delta  }H^{\alpha  \beta  \gamma  } \nabla^{\zeta  
}H_{\gamma  }{}^{\varepsilon  \mu  } \nn\\&& +   c_{234}  
    R_{\gamma  \mu  \epsilon  \zeta  } \nabla^{\delta 
 }H^{\alpha  \beta  \gamma  } \nabla^{\varepsilon  }H_{\alpha  
\beta  }{}^{\epsilon  } \nabla^{\zeta  }H_{\delta  \varepsilon  
}{}^{\mu  } +   c_{235}  
    R_{\gamma  \zeta  \varepsilon  \mu  } 
\nabla^{\delta  }H^{\alpha  \beta  \gamma  } \nabla^{\epsilon  
}H_{\alpha  \beta  \delta  } \nabla^{\zeta  }H_{\epsilon  }{}^{
\varepsilon  \mu  },\nn
\eeqa
\beqa
[H^6\nabla H]_{24}&\!\!\!\!\!=\!\!\!\!\!& c_{454}  
    H_{\alpha  \beta  }{}^{\delta  } H^{\alpha  \beta  \gamma  
} H_{\gamma  }{}^{\epsilon  \varepsilon  } H_{\epsilon  }{}^{\mu 
 \zeta  } H_{\varepsilon  \mu  }{}^{\eta  } H_{\zeta  
}{}^{\theta  \iota  } \nabla_{\eta  }H_{\delta  \theta  \iota  
} +   c_{458}  
    H_{\alpha  \beta  }{}^{\delta  } H^{\alpha  \beta  \gamma  
} H_{\gamma  }{}^{\epsilon  \varepsilon  } H_{\delta  }{}^{\mu  
\zeta  } H_{\epsilon  \mu  }{}^{\eta  } H_{\varepsilon  
}{}^{\theta  \iota  } \nabla_{\eta  }H_{\zeta  \theta  \iota  
} \nn\\&\!\!\!\!\!\!\!\!\!\!& +   c_{459}  
    H_{\alpha  \beta  }{}^{\delta  } H^{\alpha  \beta  \gamma  
} H_{\gamma  }{}^{\epsilon  \varepsilon  } H_{\delta  }{}^{\mu  
\zeta  } H_{\epsilon  \varepsilon  }{}^{\eta  } H_{\mu  
}{}^{\theta  \iota  } \nabla_{\eta  }H_{\zeta  \theta  \iota  
} +   c_{455}  
    H_{\alpha  }{}^{\delta  \epsilon  } H^{\alpha  \beta  
\gamma  } H_{\beta  \delta  }{}^{\varepsilon  } H_{\gamma  
\epsilon  }{}^{\mu  } H_{\varepsilon  }{}^{\zeta  \eta  } 
H_{\zeta  }{}^{\theta  \iota  } \nabla_{\eta  }H_{\mu  \theta  
\iota  } \nn\\&\!\!\!\!\!\!\!\!\!\!&+   c_{456}  
    H_{\alpha  \beta  }{}^{\delta  } H^{\alpha  \beta  \gamma  
} H_{\gamma  }{}^{\epsilon  \varepsilon  } H_{\delta  \epsilon  
}{}^{\mu  } H_{\varepsilon  }{}^{\zeta  \eta  } H_{\zeta  
}{}^{\theta  \iota  } \nabla_{\eta  }H_{\mu  \theta  \iota  } 
+   c_{457}  
    H_{\alpha  \beta  \gamma  } H^{\alpha  \beta  \gamma  } 
H_{\delta  \epsilon  }{}^{\mu  } H^{\delta  \epsilon  
\varepsilon  } H_{\varepsilon  }{}^{\zeta  \eta  } H_{\zeta  
}{}^{\theta  \iota  } \nabla_{\eta  }H_{\mu  \theta  \iota  } 
\nn\\&\!\!\!\!\!\!\!\!\!\!&+   c_{460}  
    H_{\alpha  \beta  }{}^{\delta  } H^{\alpha  \beta  \gamma  
} H_{\gamma  }{}^{\epsilon  \varepsilon  } H_{\epsilon  }{}^{\mu 
 \zeta  } H_{\varepsilon  }{}^{\eta  \theta  } H_{\mu  \eta  
}{}^{\iota  } \nabla_{\theta  }H_{\delta  \zeta  \iota  } +  
    c_{461}  
    H_{\alpha  }{}^{\delta  \epsilon  } H^{\alpha  \beta  
\gamma  } H_{\beta  \delta  }{}^{\varepsilon  } H_{\gamma  }{}^{
\mu  \zeta  } H_{\epsilon  }{}^{\eta  \theta  } H_{\mu  \eta  
}{}^{\iota  } \nabla_{\theta  }H_{\varepsilon  \zeta  \iota  } 
\nn\\&\!\!\!\!\!\!\!\!\!\!&+   c_{462}  
    H_{\alpha  \beta  }{}^{\delta  } H^{\alpha  \beta  \gamma  
} H_{\gamma  }{}^{\epsilon  \varepsilon  } H_{\epsilon  }{}^{\mu 
 \zeta  } H_{\eta  \theta  }{}^{\iota  } H_{\mu  }{}^{\eta  
\theta  } \nabla_{\iota  }H_{\delta  \varepsilon  \zeta  } +  
    c_{463}  
    H_{\alpha  \beta  }{}^{\delta  } H^{\alpha  \beta  \gamma  
} H_{\gamma  }{}^{\epsilon  \varepsilon  } H_{\epsilon  }{}^{\mu 
 \zeta  } H_{\zeta  \eta  }{}^{\iota  } H_{\mu  }{}^{\eta  
\theta  } \nabla_{\iota  }H_{\delta  \varepsilon  \theta  }\nn\\&\!\!\!\!\!\!\!\!\!\!& +  
 c_{465}  
    H_{\alpha  \beta  }{}^{\delta  } H^{\alpha  \beta  \gamma  
} H_{\gamma  }{}^{\epsilon  \varepsilon  } H_{\epsilon  }{}^{\mu 
 \zeta  } H_{\varepsilon  \mu  }{}^{\eta  } H_{\zeta  
}{}^{\theta  \iota  } \nabla_{\iota  }H_{\delta  \eta  \theta  
} +   c_{464}  
    H_{\alpha  \beta  }{}^{\delta  } H^{\alpha  \beta  \gamma  
} H_{\gamma  }{}^{\epsilon  \varepsilon  } H_{\epsilon  }{}^{\mu 
 \zeta  } H_{\varepsilon  }{}^{\eta  \theta  } H_{\mu  \zeta  
}{}^{\iota  } \nabla_{\iota  }H_{\delta  \eta  \theta  } \nn\\&\!\!\!\!\!\!\!\!\!\!& +   
   c_{468}  
    H_{\alpha  \beta  }{}^{\delta  } H^{\alpha  \beta  \gamma  
} H_{\gamma  }{}^{\epsilon  \varepsilon  } H_{\delta  }{}^{\mu  
\zeta  } H_{\epsilon  \mu  }{}^{\eta  } H_{\eta  }{}^{\theta  
\iota  } \nabla_{\iota  }H_{\varepsilon  \zeta  \theta  } +  
    c_{467}  
    H_{\alpha  \beta  }{}^{\delta  } H^{\alpha  \beta  \gamma  
} H_{\gamma  }{}^{\epsilon  \varepsilon  } H_{\delta  }{}^{\mu  
\zeta  } H_{\epsilon  }{}^{\eta  \theta  } H_{\mu  \eta  
}{}^{\iota  } \nabla_{\iota  }H_{\varepsilon  \zeta  \theta  }\nn\\&\!\!\!\!\!\!\!\!\!\!&
+   c_{469}  
    H_{\alpha  }{}^{\delta  \epsilon  } H^{\alpha  \beta  
\gamma  } H_{\beta  \delta  }{}^{\varepsilon  } H_{\gamma  }{}^{
\mu  \zeta  } H_{\epsilon  \mu  }{}^{\eta  } H_{\zeta  
}{}^{\theta  \iota  } \nabla_{\iota  }H_{\varepsilon  \eta  
\theta  } +   c_{466}  
    H_{\alpha  \beta  }{}^{\delta  } H^{\alpha  \beta  \gamma  
} H_{\gamma  }{}^{\epsilon  \varepsilon  } H_{\delta  }{}^{\mu  
\zeta  } H_{\epsilon  }{}^{\eta  \theta  } H_{\eta  \theta  
}{}^{\iota  } \nabla_{\iota  }H_{\varepsilon  \mu  \zeta  }\nn\\&\!\!\!\!\!\!\!\!\!\!& +  
 c_{474}  
    H_{\alpha  }{}^{\delta  \epsilon  } H^{\alpha  \beta  
\gamma  } H_{\beta  \delta  }{}^{\varepsilon  } H_{\gamma  }{}^{
\mu  \zeta  } H_{\epsilon  \mu  }{}^{\eta  } H_{\varepsilon  
}{}^{\theta  \iota  } \nabla_{\iota  }H_{\zeta  \eta  \theta  
} +   c_{475}  
    H_{\alpha  \beta  }{}^{\delta  } H^{\alpha  \beta  \gamma  
} H_{\gamma  }{}^{\epsilon  \varepsilon  } H_{\delta  }{}^{\mu  
\zeta  } H_{\epsilon  \mu  }{}^{\eta  } H_{\varepsilon  
}{}^{\theta  \iota  } \nabla_{\iota  }H_{\zeta  \eta  \theta  
}\nn\\&\!\!\!\!\!\!\!\!\!\!& +   c_{476}  
    H_{\alpha  \beta  \gamma  } H^{\alpha  \beta  \gamma  } 
H_{\delta  }{}^{\mu  \zeta  } H^{\delta  \epsilon  \varepsilon  
} H_{\epsilon  \mu  }{}^{\eta  } H_{\varepsilon  }{}^{\theta  
\iota  } \nabla_{\iota  }H_{\zeta  \eta  \theta  } +   
   c_{159}  
    H_{\alpha  \beta  }{}^{\delta  } H^{\alpha  \beta  \gamma  
} H_{\gamma  }{}^{\epsilon  \varepsilon  } H_{\delta  }{}^{\mu  
\zeta  } H_{\epsilon  \varepsilon  }{}^{\eta  } H_{\mu  
}{}^{\theta  \iota  } \nabla_{\iota  }H_{\zeta  \eta  \theta  
}\nn\\&\!\!\!\!\!\!\!\!\!\!& +   c_{470}  
    H_{\alpha  \beta  }{}^{\delta  } H^{\alpha  \beta  \gamma  
} H_{\gamma  }{}^{\epsilon  \varepsilon  } H_{\delta  }{}^{\mu  
\zeta  } H_{\epsilon  }{}^{\eta  \theta  } H_{\varepsilon  \eta 
 }{}^{\iota  } \nabla_{\iota  }H_{\mu  \zeta  \theta  } +   
   c_{471}  
    H_{\alpha  }{}^{\delta  \epsilon  } H^{\alpha  \beta  
\gamma  } H_{\beta  \delta  }{}^{\varepsilon  } H_{\gamma  
\epsilon  }{}^{\mu  } H_{\varepsilon  }{}^{\zeta  \eta  } 
H_{\zeta  }{}^{\theta  \iota  } \nabla_{\iota  }H_{\mu  \eta  
\theta  } \nn\\&\!\!\!\!\!\!\!\!\!\!&+   c_{472}  
    H_{\alpha  \beta  }{}^{\delta  } H^{\alpha  \beta  \gamma  
} H_{\gamma  }{}^{\epsilon  \varepsilon  } H_{\delta  \epsilon  
}{}^{\mu  } H_{\varepsilon  }{}^{\zeta  \eta  } H_{\zeta  
}{}^{\theta  \iota  } \nabla_{\iota  }H_{\mu  \eta  \theta  } 
+   c_{473}  
    H_{\alpha  \beta  \gamma  } H^{\alpha  \beta  \gamma  } 
H_{\delta  \epsilon  }{}^{\mu  } H^{\delta  \epsilon  
\varepsilon  } H_{\varepsilon  }{}^{\zeta  \eta  } H_{\zeta  
}{}^{\theta  \iota  } \nabla_{\iota  }H_{\mu  \eta  \theta  },
\nn 
\eeqa
\beqa
[H^4\nabla H R]_{67}&=&  
   c_{142}  
    H_{\alpha  }{}^{\delta  \epsilon  } H^{\alpha  \beta  
\gamma  } H_{\beta  }{}^{\varepsilon  \mu  } H_{\delta  
}{}^{\zeta  \eta  } R_{\epsilon  \eta  \mu  \theta  } 
\nabla_{\varepsilon  }H_{\gamma  \zeta  }{}^{\theta  } +   
   c_{143}  
    H_{\alpha  }{}^{\delta  \epsilon  } H^{\alpha  \beta  
\gamma  } H_{\beta  }{}^{\varepsilon  \mu  } H_{\delta  
}{}^{\zeta  \eta  } R_{\epsilon  \theta  \mu  \eta  } 
\nabla_{\varepsilon  }H_{\gamma  \zeta  }{}^{\theta  }\nn\\&& +   
   c_{148}  
    H_{\alpha  }{}^{\delta  \epsilon  } H^{\alpha  \beta  
\gamma  } H_{\beta  }{}^{\varepsilon  \mu  } H^{\zeta  \eta  
\theta  } R_{\gamma  \eta  \mu  \theta  } 
\nabla_{\varepsilon  }H_{\delta  \epsilon  \zeta  } +   
   c_{149}  
    H_{\alpha  \beta  }{}^{\delta  } H^{\alpha  \beta  \gamma  
} H_{\gamma  }{}^{\epsilon  \varepsilon  } H_{\epsilon  }{}^{\mu 
 \zeta  } R_{\mu  \eta  \zeta  \theta  } 
\nabla_{\varepsilon  }H_{\delta  }{}^{\eta  \theta  } \nn\\&&+   
   c_{400}  
    H_{\alpha  }{}^{\delta  \epsilon  } H^{\alpha  \beta  
\gamma  } H_{\beta  }{}^{\varepsilon  \mu  } H^{\zeta  \eta  
\theta  } R_{\varepsilon  \eta  \mu  \theta  } 
\nabla_{\zeta  }H_{\gamma  \delta  \epsilon  } +   
   c_{401}  
    H_{\alpha  \beta  }{}^{\delta  } H^{\alpha  \beta  \gamma  
} H^{\epsilon  \varepsilon  \mu  } H^{\zeta  \eta  \theta  } R
_{\delta  \eta  \mu  \theta  } \nabla_{\zeta  
}H_{\gamma  \epsilon  \varepsilon  } \nn\\&&+   c_{402}  
    H_{\alpha  \beta  }{}^{\delta  } H^{\alpha  \beta  \gamma  
} H_{\epsilon  }{}^{\zeta  \eta  } H^{\epsilon  \varepsilon  
\mu  } R_{\delta  \eta  \mu  \theta  } \nabla_{\zeta  
}H_{\gamma  \varepsilon  }{}^{\theta  } +   c_{403}  
    H_{\alpha  \beta  }{}^{\delta  } H^{\alpha  \beta  \gamma  
} H_{\epsilon  }{}^{\zeta  \eta  } H^{\epsilon  \varepsilon  
\mu  } R_{\delta  \theta  \mu  \eta  } \nabla_{\zeta  
}H_{\gamma  \varepsilon  }{}^{\theta  }\nn\\&& +   c_{404}  
    H_{\alpha  }{}^{\delta  \epsilon  } H^{\alpha  \beta  
\gamma  } H_{\beta  }{}^{\varepsilon  \mu  } H_{\delta  
}{}^{\zeta  \eta  } R_{\epsilon  \eta  \mu  \theta  } 
\nabla_{\zeta  }H_{\gamma  \varepsilon  }{}^{\theta  } +   
   c_{405}  
    H_{\alpha  }{}^{\delta  \epsilon  } H^{\alpha  \beta  
\gamma  } H_{\beta  }{}^{\varepsilon  \mu  } H_{\delta  
}{}^{\zeta  \eta  } R_{\epsilon  \theta  \mu  \eta  } 
\nabla_{\zeta  }H_{\gamma  \varepsilon  }{}^{\theta  } \nn\\&&+   
   c_{406}  
    H_{\alpha  }{}^{\delta  \epsilon  } H^{\alpha  \beta  
\gamma  } H_{\beta  \delta  }{}^{\varepsilon  } H^{\mu  \zeta  
\eta  } R_{\epsilon  \eta  \varepsilon  \theta  } 
\nabla_{\zeta  }H_{\gamma  \mu  }{}^{\theta  } +   
   c_{407}  
    H_{\alpha  }{}^{\delta  \epsilon  } H^{\alpha  \beta  
\gamma  } H_{\beta  }{}^{\varepsilon  \mu  } H^{\zeta  \eta  
\theta  } R_{\gamma  \eta  \mu  \theta  } 
\nabla_{\zeta  }H_{\delta  \epsilon  \varepsilon  }\nn\\&& +   
   c_{408}  
    H_{\alpha  }{}^{\delta  \epsilon  } H^{\alpha  \beta  
\gamma  } H_{\beta  }{}^{\varepsilon  \mu  } H_{\gamma  
}{}^{\zeta  \eta  } R_{\epsilon  \eta  \mu  \theta  } 
\nabla_{\zeta  }H_{\delta  \varepsilon  }{}^{\theta  } +   
   c_{409}  
    H_{\alpha  \beta  }{}^{\delta  } H^{\alpha  \beta  \gamma  
} H_{\gamma  }{}^{\epsilon  \varepsilon  } H^{\mu  \zeta  \eta  
} R_{\epsilon  \eta  \varepsilon  \theta  } 
\nabla_{\zeta  }H_{\delta  \mu  }{}^{\theta  } \nn\\&&+   
   c_{410}  
    H_{\alpha  \beta  }{}^{\delta  } H^{\alpha  \beta  \gamma  
} H_{\gamma  }{}^{\epsilon  \varepsilon  } H^{\mu  \zeta  \eta  
} R_{\delta  \eta  \varepsilon  \theta  } 
\nabla_{\zeta  }H_{\epsilon  \mu  }{}^{\theta  } +   
   c_{411}  
    H_{\alpha  \beta  }{}^{\delta  } H^{\alpha  \beta  \gamma  
} H_{\gamma  }{}^{\epsilon  \varepsilon  } H^{\mu  \zeta  \eta  
} R_{\delta  \theta  \varepsilon  \eta  } 
\nabla_{\zeta  }H_{\epsilon  \mu  }{}^{\theta  }\nn\\&& +   
   c_{412}  
    H_{\alpha  \beta  }{}^{\delta  } H^{\alpha  \beta  \gamma  
} H_{\epsilon  }{}^{\zeta  \eta  } H^{\epsilon  \varepsilon  
\mu  } R_{\gamma  \eta  \delta  \theta  } 
\nabla_{\zeta  }H_{\varepsilon  \mu  }{}^{\theta  } +   
   c_{413}  
    H_{\alpha  }{}^{\delta  \epsilon  } H^{\alpha  \beta  
\gamma  } H_{\beta  }{}^{\varepsilon  \mu  } H_{\delta  
}{}^{\zeta  \eta  } R_{\gamma  \eta  \epsilon  \theta  
} \nabla_{\zeta  }H_{\varepsilon  \mu  }{}^{\theta  }\nn\\&& +   
   c_{414}  
    H_{\alpha  }{}^{\delta  \epsilon  } H^{\alpha  \beta  
\gamma  } H_{\beta  }{}^{\varepsilon  \mu  } H_{\delta  
}{}^{\zeta  \eta  } R_{\gamma  \theta  \epsilon  \eta  
} \nabla_{\zeta  }H_{\varepsilon  \mu  }{}^{\theta  } +   
   c_{415}  
    H_{\alpha  }{}^{\delta  \epsilon  } H^{\alpha  \beta  
\gamma  } H_{\varepsilon  }{}^{\eta  \theta  } H^{\varepsilon  
\mu  \zeta  } R_{\delta  \zeta  \epsilon  \theta  } 
\nabla_{\eta  }H_{\beta  \gamma  \mu  } \nn\\&&+   c_{416}  
    H_{\alpha  }{}^{\delta  \epsilon  } H^{\alpha  \beta  
\gamma  } H_{\beta  }{}^{\varepsilon  \mu  } H^{\zeta  \eta  
\theta  } R_{\epsilon  \theta  \varepsilon  \mu  } 
\nabla_{\eta  }H_{\gamma  \delta  \zeta  } +   c_{417}  
    H_{\alpha  }{}^{\delta  \epsilon  } H^{\alpha  \beta  
\gamma  } H_{\beta  }{}^{\varepsilon  \mu  } H_{\delta  
}{}^{\zeta  \eta  } R_{\epsilon  \theta  \varepsilon  
\mu  } \nabla_{\eta  }H_{\gamma  \zeta  }{}^{\theta  } \nn\\&&+   
   c_{418}  
    H_{\alpha  }{}^{\delta  \epsilon  } H^{\alpha  \beta  
\gamma  } H_{\beta  }{}^{\varepsilon  \mu  } H^{\zeta  \eta  
\theta  } R_{\gamma  \theta  \varepsilon  \mu  } 
\nabla_{\eta  }H_{\delta  \epsilon  \zeta  } +   c_{419} 
     H_{\alpha  }{}^{\delta  \epsilon  } H^{\alpha  \beta  
\gamma  } H_{\beta  }{}^{\varepsilon  \mu  } H^{\zeta  \eta  
\theta  } R_{\gamma  \epsilon  \mu  \theta  } \nabla_{
\eta  }H_{\delta  \varepsilon  \zeta  } \nn\\&&+   c_{420}  
    H_{\alpha  }{}^{\delta  \epsilon  } H^{\alpha  \beta  
\gamma  } H_{\beta  }{}^{\varepsilon  \mu  } H^{\zeta  \eta  
\theta  } R_{\gamma  \varepsilon  \epsilon  \mu  } 
\nabla_{\theta  }H_{\delta  \zeta  \eta  } +   c_{421}  
    H_{\alpha  }{}^{\delta  \epsilon  } H^{\alpha  \beta  
\gamma  } H_{\beta  \delta  }{}^{\varepsilon  } H^{\mu  \zeta  
\eta  } R_{\varepsilon  \theta  \zeta  \eta  } 
\nabla^{\theta  }H_{\gamma  \epsilon  \mu  } \nn\\&&+   c_{423} 
     H_{\alpha  }{}^{\delta  \epsilon  } H^{\alpha  \beta  
\gamma  } H_{\beta  }{}^{\varepsilon  \mu  } H_{\delta  
}{}^{\zeta  \eta  } R_{\epsilon  \eta  \mu  \theta  } 
\nabla^{\theta  }H_{\gamma  \varepsilon  \zeta  } +   
   c_{424}  
    H_{\alpha  }{}^{\delta  \epsilon  } H^{\alpha  \beta  
\gamma  } H_{\beta  }{}^{\varepsilon  \mu  } H_{\delta  
}{}^{\zeta  \eta  } R_{\epsilon  \theta  \mu  \eta  } 
\nabla^{\theta  }H_{\gamma  \varepsilon  \zeta  }\nn\\&& +   
   c_{422}  
    H_{\alpha  }{}^{\delta  \epsilon  } H^{\alpha  \beta  
\gamma  } H_{\beta  }{}^{\varepsilon  \mu  } H_{\delta  
}{}^{\zeta  \eta  } R_{\epsilon  \theta  \zeta  \eta  
} \nabla^{\theta  }H_{\gamma  \varepsilon  \mu  } +   
   c_{425}  
    H_{\alpha  }{}^{\delta  \epsilon  } H^{\alpha  \beta  
\gamma  } H_{\beta  \delta  }{}^{\varepsilon  } H^{\mu  \zeta  
\eta  } R_{\epsilon  \eta  \varepsilon  \theta  } 
\nabla^{\theta  }H_{\gamma  \mu  \zeta  } \nn\\&&+   c_{426}  
    H_{\alpha  }{}^{\delta  \epsilon  } H^{\alpha  \beta  
\gamma  } H_{\beta  }{}^{\varepsilon  \mu  } H_{\gamma  
}{}^{\zeta  \eta  } R_{\mu  \theta  \zeta  \eta  } 
\nabla^{\theta  }H_{\delta  \epsilon  \varepsilon  } +   
   c_{427}  
    H_{\alpha  \beta  }{}^{\delta  } H^{\alpha  \beta  \gamma  
} H_{\gamma  }{}^{\epsilon  \varepsilon  } H^{\mu  \zeta  \eta  
} R_{\varepsilon  \theta  \zeta  \eta  } 
\nabla^{\theta  }H_{\delta  \epsilon  \mu  }\nn\\&& +   c_{428} 
     H_{\alpha  \beta  }{}^{\delta  } H^{\alpha  \beta  \gamma 
 } H_{\gamma  }{}^{\epsilon  \varepsilon  } H_{\epsilon  
}{}^{\mu  \zeta  } R_{\mu  \eta  \zeta  \theta  } 
\nabla^{\theta  }H_{\delta  \varepsilon  }{}^{\eta  } +   
   c_{431}  
    H_{\alpha  \beta  }{}^{\delta  } H^{\alpha  \beta  \gamma  
} H_{\gamma  }{}^{\epsilon  \varepsilon  } H_{\epsilon  
\varepsilon  }{}^{\mu  } R_{\mu  \theta  \zeta  \eta  
} \nabla^{\theta  }H_{\delta  }{}^{\zeta  \eta  }\nn\\&& +   
   c_{429}  
    H_{\alpha  \beta  }{}^{\delta  } H^{\alpha  \beta  \gamma  
} H_{\gamma  }{}^{\epsilon  \varepsilon  } H^{\mu  \zeta  \eta  
} R_{\epsilon  \eta  \varepsilon  \theta  } 
\nabla^{\theta  }H_{\delta  \mu  \zeta  } +   c_{430}  
    H_{\alpha  \beta  }{}^{\delta  } H^{\alpha  \beta  \gamma  
} H_{\gamma  }{}^{\epsilon  \varepsilon  } H_{\epsilon  }{}^{\mu 
 \zeta  } R_{\varepsilon  \theta  \zeta  \eta  } 
\nabla^{\theta  }H_{\delta  \mu  }{}^{\eta  }\nn\\&& +   
   c_{433}  
    H_{\alpha  }{}^{\delta  \epsilon  } H^{\alpha  \beta  
\gamma  } H_{\beta  \delta  }{}^{\varepsilon  } H_{\gamma  }{}^{
\mu  \zeta  } R_{\mu  \eta  \zeta  \theta  } \nabla^{
\theta  }H_{\epsilon  \varepsilon  }{}^{\eta  } +   
   c_{434}  
    H_{\alpha  \beta  }{}^{\delta  } H^{\alpha  \beta  \gamma  
} H_{\gamma  }{}^{\epsilon  \varepsilon  } H_{\delta  }{}^{\mu  
\zeta  } R_{\mu  \eta  \zeta  \theta  } 
\nabla^{\theta  }H_{\epsilon  \varepsilon  }{}^{\eta  } \nn\\&&+   
   c_{435}  
    H_{\alpha  \beta  \gamma  } H^{\alpha  \beta  \gamma  } 
H_{\delta  }{}^{\mu  \zeta  } H^{\delta  \epsilon  \varepsilon  
} R_{\mu  \eta  \zeta  \theta  } \nabla^{\theta  
}H_{\epsilon  \varepsilon  }{}^{\eta  } +   c_{432}  
    H_{\alpha  \beta  }{}^{\delta  } H^{\alpha  \beta  \gamma  
} H_{\gamma  }{}^{\epsilon  \varepsilon  } H^{\mu  \zeta  \eta  
} R_{\delta  \theta  \zeta  \eta  } \nabla^{\theta  
}H_{\epsilon  \varepsilon  \mu  } \nn\\&&+   c_{436}  
    H_{\alpha  \beta  }{}^{\delta  } H^{\alpha  \beta  \gamma  
} H_{\gamma  }{}^{\epsilon  \varepsilon  } H^{\mu  \zeta  \eta  
} R_{\delta  \eta  \varepsilon  \theta  } 
\nabla^{\theta  }H_{\epsilon  \mu  \zeta  } +   c_{437} 
     H_{\alpha  \beta  }{}^{\delta  } H^{\alpha  \beta  \gamma 
 } H_{\gamma  }{}^{\epsilon  \varepsilon  } H^{\mu  \zeta  \eta 
 } R_{\delta  \theta  \varepsilon  \eta  } 
\nabla^{\theta  }H_{\epsilon  \mu  \zeta  }\nn\\&& +   c_{438} 
     H_{\alpha  }{}^{\delta  \epsilon  } H^{\alpha  \beta  
\gamma  } H_{\beta  \delta  }{}^{\varepsilon  } H_{\gamma  }{}^{
\mu  \zeta  } R_{\varepsilon  \eta  \zeta  \theta  } 
\nabla^{\theta  }H_{\epsilon  \mu  }{}^{\eta  } +   
   c_{439}  
    H_{\alpha  \beta  }{}^{\delta  } H^{\alpha  \beta  \gamma  
} H_{\gamma  }{}^{\epsilon  \varepsilon  } H_{\delta  }{}^{\mu  
\zeta  } R_{\varepsilon  \eta  \zeta  \theta  } 
\nabla^{\theta  }H_{\epsilon  \mu  }{}^{\eta  } \nn\\&&+   
   c_{440}  
    H_{\alpha  \beta  \gamma  } H^{\alpha  \beta  \gamma  } 
H_{\delta  }{}^{\mu  \zeta  } H^{\delta  \epsilon  \varepsilon  
} R_{\varepsilon  \eta  \zeta  \theta  } 
\nabla^{\theta  }H_{\epsilon  \mu  }{}^{\eta  } +   
   c_{441}  
    H_{\alpha  }{}^{\delta  \epsilon  } H^{\alpha  \beta  
\gamma  } H_{\beta  \delta  }{}^{\varepsilon  } H_{\gamma  }{}^{
\mu  \zeta  } R_{\varepsilon  \theta  \zeta  \eta  } 
\nabla^{\theta  }H_{\epsilon  \mu  }{}^{\eta  }\nn\\&& +   
   c_{447}  
    H_{\alpha  }{}^{\delta  \epsilon  } H^{\alpha  \beta  
\gamma  } H_{\beta  \delta  }{}^{\varepsilon  } H_{\gamma  
\epsilon  }{}^{\mu  } R_{\mu  \theta  \zeta  \eta  } 
\nabla^{\theta  }H_{\varepsilon  }{}^{\zeta  \eta  } +   
   c_{448}  
    H_{\alpha  \beta  }{}^{\delta  } H^{\alpha  \beta  \gamma  
} H_{\gamma  }{}^{\epsilon  \varepsilon  } H_{\delta  \epsilon  
}{}^{\mu  } R_{\mu  \theta  \zeta  \eta  } 
\nabla^{\theta  }H_{\varepsilon  }{}^{\zeta  \eta  } \nn\\&&+   
   c_{449}  
    H_{\alpha  \beta  \gamma  } H^{\alpha  \beta  \gamma  } 
H_{\delta  \epsilon  }{}^{\mu  } H^{\delta  \epsilon  
\varepsilon  } R_{\mu  \theta  \zeta  \eta  } \nabla^{
\theta  }H_{\varepsilon  }{}^{\zeta  \eta  } +   c_{442} 
     H_{\alpha  \beta  }{}^{\delta  } H^{\alpha  \beta  \gamma 
 } H_{\epsilon  }{}^{\zeta  \eta  } H^{\epsilon  \varepsilon  
\mu  } R_{\gamma  \eta  \delta  \theta  } 
\nabla^{\theta  }H_{\varepsilon  \mu  \zeta  }\nn\\&& +   
   c_{443}  
    H_{\alpha  }{}^{\delta  \epsilon  } H^{\alpha  \beta  
\gamma  } H_{\beta  }{}^{\varepsilon  \mu  } H_{\delta  
}{}^{\zeta  \eta  } R_{\gamma  \eta  \epsilon  \theta  
} \nabla^{\theta  }H_{\varepsilon  \mu  \zeta  } +   
   c_{444}  
    H_{\alpha  }{}^{\delta  \epsilon  } H^{\alpha  \beta  
\gamma  } H_{\beta  }{}^{\varepsilon  \mu  } H_{\delta  
}{}^{\zeta  \eta  } R_{\gamma  \theta  \epsilon  \eta  
} \nabla^{\theta  }H_{\varepsilon  \mu  \zeta  } \nn\\&&+   
   c_{445}  
    H_{\alpha  \beta  }{}^{\delta  } H^{\alpha  \beta  \gamma  
} H_{\gamma  }{}^{\epsilon  \varepsilon  } H_{\epsilon  }{}^{\mu 
 \zeta  } R_{\delta  \eta  \zeta  \theta  } 
\nabla^{\theta  }H_{\varepsilon  \mu  }{}^{\eta  } +   
   c_{446}  
    H_{\alpha  \beta  }{}^{\delta  } H^{\alpha  \beta  \gamma  
} H_{\gamma  }{}^{\epsilon  \varepsilon  } H_{\epsilon  }{}^{\mu 
 \zeta  } R_{\delta  \theta  \zeta  \eta  } 
\nabla^{\theta  }H_{\varepsilon  \mu  }{}^{\eta  } \nn\\&&+   
   c_{450}  
    H_{\alpha  \beta  }{}^{\delta  } H^{\alpha  \beta  \gamma  
} H_{\gamma  }{}^{\epsilon  \varepsilon  } H^{\mu  \zeta  \eta  
} R_{\delta  \theta  \epsilon  \varepsilon  } 
\nabla^{\theta  }H_{\mu  \zeta  \eta  } +   c_{451}  
    H_{\alpha  \beta  }{}^{\delta  } H^{\alpha  \beta  \gamma  
} H_{\gamma  }{}^{\epsilon  \varepsilon  } H_{\epsilon  }{}^{\mu 
 \zeta  } R_{\delta  \eta  \varepsilon  \theta  } 
\nabla^{\theta  }H_{\mu  \zeta  }{}^{\eta  } \nn\\&&+   c_{452} 
     H_{\alpha  \beta  }{}^{\delta  } H^{\alpha  \beta  \gamma 
 } H_{\gamma  }{}^{\epsilon  \varepsilon  } H_{\epsilon  
}{}^{\mu  \zeta  } R_{\delta  \theta  \varepsilon  
\eta  } \nabla^{\theta  }H_{\mu  \zeta  }{}^{\eta  } +   
   c_{453}  
    H_{\alpha  }{}^{\delta  \epsilon  } H^{\alpha  \beta  
\gamma  } H_{\beta  \delta  }{}^{\varepsilon  } H_{\gamma  }{}^{
\mu  \zeta  } R_{\epsilon  \eta  \varepsilon  \theta  
} \nabla^{\theta  }H_{\mu  \zeta  }{}^{\eta  } \nn\\&&+   
   c_{160}  
    H_{\alpha  }{}^{\delta  \epsilon  } H^{\alpha  \beta  
\gamma  } H_{\varepsilon  }{}^{\eta  \theta  } H^{\varepsilon  
\mu  \zeta  } R_{\epsilon  \eta  \zeta  \theta  } 
\nabla_{\mu  }H_{\beta  \gamma  \delta  } +   c_{174}  
    H_{\alpha  }{}^{\delta  \epsilon  } H^{\alpha  \beta  
\gamma  } H_{\beta  \delta  }{}^{\varepsilon  } H^{\mu  \zeta  
\eta  } R_{\varepsilon  \theta  \zeta  \eta  } 
\nabla_{\mu  }H_{\gamma  \epsilon  }{}^{\theta  }\nn\\&& +   
   c_{175}  
    H_{\alpha  \beta  }{}^{\delta  } H^{\alpha  \beta  \gamma  
} H_{\epsilon  }{}^{\zeta  \eta  } H^{\epsilon  \varepsilon  
\mu  } R_{\delta  \theta  \zeta  \eta  } \nabla_{\mu  
}H_{\gamma  \varepsilon  }{}^{\theta  } +   c_{176}  
    H_{\alpha  }{}^{\delta  \epsilon  } H^{\alpha  \beta  
\gamma  } H_{\beta  }{}^{\varepsilon  \mu  } H_{\delta  
}{}^{\zeta  \eta  } R_{\epsilon  \theta  \zeta  \eta  
} \nabla_{\mu  }H_{\gamma  \varepsilon  }{}^{\theta  } \nn\\&&+   
   c_{182}  
    H_{\alpha  \beta  }{}^{\delta  } H^{\alpha  \beta  \gamma  
} H_{\gamma  }{}^{\epsilon  \varepsilon  } H^{\mu  \zeta  \eta  
} R_{\varepsilon  \theta  \zeta  \eta  } \nabla_{\mu  
}H_{\delta  \epsilon  }{}^{\theta  } +   c_{183}  
    H_{\alpha  \beta  }{}^{\delta  } H^{\alpha  \beta  \gamma  
} H_{\gamma  }{}^{\epsilon  \varepsilon  } H_{\epsilon  }{}^{\mu 
 \zeta  } R_{\varepsilon  \eta  \zeta  \theta  } 
\nabla_{\mu  }H_{\delta  }{}^{\eta  \theta  }\nn\\&& +   
   c_{184}  
    H_{\alpha  \beta  }{}^{\delta  } H^{\alpha  \beta  \gamma  
} H_{\gamma  }{}^{\epsilon  \varepsilon  } H^{\mu  \zeta  \eta  
} R_{\delta  \theta  \zeta  \eta  } \nabla_{\mu  
}H_{\epsilon  \varepsilon  }{}^{\theta  } +   c_{188}  
    H_{\alpha  }{}^{\delta  \epsilon  } H^{\alpha  \beta  
\gamma  } H_{\beta  \delta  }{}^{\varepsilon  } H_{\gamma  }{}^{
\mu  \zeta  } R_{\varepsilon  \eta  \zeta  \theta  } 
\nabla_{\mu  }H_{\epsilon  }{}^{\eta  \theta  } \nn\\&&+   
   c_{193}  
    H_{\alpha  \beta  }{}^{\delta  } H^{\alpha  \beta  \gamma  
} H_{\gamma  }{}^{\epsilon  \varepsilon  } H_{\epsilon  }{}^{\mu 
 \zeta  } R_{\delta  \eta  \zeta  \theta  } 
\nabla_{\mu  }H_{\varepsilon  }{}^{\eta  \theta  },\nn 
\eeqa
\beqa
[H^2\nabla HR^2]_{42}&=&
   c_{1}  
    H^{\alpha  \beta  \gamma  } H^{\delta  \epsilon  
\varepsilon  } R_{\beta  \delta  \mu  \zeta  } 
R_{\gamma  \eta  \epsilon  \varepsilon  } 
\nabla_{\alpha  }H^{\mu  \zeta  \eta  } +   c_{2}  
    H^{\alpha  \beta  \gamma  } H^{\delta  \epsilon  
\varepsilon  } R_{\beta  \gamma  \delta  \mu  } 
R_{\epsilon  \zeta  \varepsilon  \eta  } 
\nabla_{\alpha  }H^{\mu  \zeta  \eta  } \nn\\&&+   c_{152}  
    H_{\alpha  }{}^{\delta  \epsilon  } H^{\alpha  \beta  
\gamma  } R_{\epsilon  }{}^{\mu  \zeta  \eta  } 
R_{\varepsilon  \zeta  \mu  \eta  } 
\nabla^{\varepsilon  }H_{\beta  \gamma  \delta  } +   
   c_{375}  
    H^{\alpha  \beta  \gamma  } H^{\delta  \epsilon  
\varepsilon  } R_{\gamma  }{}^{\eta  }{}_{\delta  \mu  
} R_{\epsilon  \zeta  \varepsilon  \eta  } 
\nabla^{\zeta  }H_{\alpha  \beta  }{}^{\mu  }\nn\\&& +   
   c_{374}  
    H^{\alpha  \beta  \gamma  } H^{\delta  \epsilon  
\varepsilon  } R_{\gamma  \mu  \delta  }{}^{\eta  } 
R_{\epsilon  \zeta  \varepsilon  \eta  } \nabla^{\zeta 
 }H_{\alpha  \beta  }{}^{\mu  } +   c_{373}  
    H^{\alpha  \beta  \gamma  } H^{\delta  \epsilon  
\varepsilon  } R_{\gamma  }{}^{\eta  }{}_{\delta  \zeta 
 } R_{\epsilon  \mu  \varepsilon  \eta  } 
\nabla^{\zeta  }H_{\alpha  \beta  }{}^{\mu  } \nn\\&&+   
   c_{376}  
    H^{\alpha  \beta  \gamma  } H^{\delta  \epsilon  
\varepsilon  } R_{\gamma  }{}^{\eta  }{}_{\delta  
\epsilon  } R_{\varepsilon  \zeta  \mu  \eta  } 
\nabla^{\zeta  }H_{\alpha  \beta  }{}^{\mu  } +   
   c_{377}  
    H^{\alpha  \beta  \gamma  } H^{\delta  \epsilon  
\varepsilon  } R_{\gamma  }{}^{\eta  }{}_{\delta  
\epsilon  } R_{\varepsilon  \eta  \mu  \zeta  } 
\nabla^{\zeta  }H_{\alpha  \beta  }{}^{\mu  }\nn\\&& +   
   c_{378}  
    H^{\alpha  \beta  \gamma  } H^{\delta  \epsilon  
\varepsilon  } R_{\beta  \mu  \epsilon  }{}^{\eta  } R
_{\gamma  \zeta  \varepsilon  \eta  } \nabla^{\zeta  
}H_{\alpha  \delta  }{}^{\mu  } +   c_{379}  
    H^{\alpha  \beta  \gamma  } H^{\delta  \epsilon  
\varepsilon  } R_{\beta  \mu  \epsilon  }{}^{\eta  } R
_{\gamma  \eta  \varepsilon  \zeta  } \nabla^{\zeta  
}H_{\alpha  \delta  }{}^{\mu  }\nn\\&& +   c_{380}  
    H^{\alpha  \beta  \gamma  } H^{\delta  \epsilon  
\varepsilon  } R_{\beta  \mu  \gamma  }{}^{\eta  } 
R_{\epsilon  \zeta  \varepsilon  \eta  } \nabla^{\zeta 
 }H_{\alpha  \delta  }{}^{\mu  } +   c_{381}  
    H^{\alpha  \beta  \gamma  } H^{\delta  \epsilon  
\varepsilon  } R_{\beta  \epsilon  \gamma  }{}^{\eta  } 
R_{\varepsilon  \zeta  \mu  \eta  } \nabla^{\zeta  
}H_{\alpha  \delta  }{}^{\mu  } \nn\\&&+   c_{382}  
    H^{\alpha  \beta  \gamma  } H^{\delta  \epsilon  
\varepsilon  } R_{\beta  \epsilon  \gamma  }{}^{\eta  } 
R_{\varepsilon  \eta  \mu  \zeta  } \nabla^{\zeta  
}H_{\alpha  \delta  }{}^{\mu  } +   c_{385}  
    H_{\alpha  }{}^{\delta  \epsilon  } H^{\alpha  \beta  
\gamma  } R_{\gamma  }{}^{\eta  }{}_{\varepsilon  \mu  
} R_{\delta  \zeta  \epsilon  \eta  } \nabla^{\zeta  
}H_{\beta  }{}^{\varepsilon  \mu  } \nn\\&&+   c_{383}  
    H_{\alpha  }{}^{\delta  \epsilon  } H^{\alpha  \beta  
\gamma  } R_{\gamma  \zeta  \varepsilon  }{}^{\eta  } 
R_{\delta  \mu  \epsilon  \eta  } \nabla^{\zeta  
}H_{\beta  }{}^{\varepsilon  \mu  } +   c_{384}  
    H_{\alpha  }{}^{\delta  \epsilon  } H^{\alpha  \beta  
\gamma  } R_{\gamma  }{}^{\eta  }{}_{\varepsilon  \zeta 
 } R_{\delta  \mu  \epsilon  \eta  } \nabla^{\zeta  
}H_{\beta  }{}^{\varepsilon  \mu  }\nn\\&& +   c_{386}  
    H_{\alpha  }{}^{\delta  \epsilon  } H^{\alpha  \beta  
\gamma  } R_{\gamma  \varepsilon  \delta  }{}^{\eta  } 
R_{\epsilon  \zeta  \mu  \eta  } \nabla^{\zeta  
}H_{\beta  }{}^{\varepsilon  \mu  } +   c_{387}  
    H_{\alpha  }{}^{\delta  \epsilon  } H^{\alpha  \beta  
\gamma  } R_{\gamma  }{}^{\eta  }{}_{\delta  
\varepsilon  } R_{\epsilon  \zeta  \mu  \eta  } 
\nabla^{\zeta  }H_{\beta  }{}^{\varepsilon  \mu  } \nn\\&&+   
   c_{388}  
    H_{\alpha  }{}^{\delta  \epsilon  } H^{\alpha  \beta  
\gamma  } R_{\gamma  \zeta  \delta  }{}^{\eta  } 
R_{\epsilon  \eta  \varepsilon  \mu  } \nabla^{\zeta  
}H_{\beta  }{}^{\varepsilon  \mu  }+   c_{389}  
    H_{\alpha  }{}^{\delta  \epsilon  } H^{\alpha  \beta  
\gamma  } R_{\gamma  }{}^{\eta  }{}_{\delta  \zeta  } 
R_{\epsilon  \eta  \varepsilon  \mu  } \nabla^{\zeta  
}H_{\beta  }{}^{\varepsilon  \mu  } \nn\\&& +   c_{390}  
    H_{\alpha  }{}^{\delta  \epsilon  } H^{\alpha  \beta  
\gamma  } R_{\gamma  \varepsilon  \delta  }{}^{\eta  } 
R_{\epsilon  \eta  \mu  \zeta  } \nabla^{\zeta  
}H_{\beta  }{}^{\varepsilon  \mu  } +   c_{391}  
    H_{\alpha  }{}^{\delta  \epsilon  } H^{\alpha  \beta  
\gamma  } R_{\gamma  }{}^{\eta  }{}_{\delta  
\varepsilon  } R_{\epsilon  \eta  \mu  \zeta  } 
\nabla^{\zeta  }H_{\beta  }{}^{\varepsilon  \mu  }\nn\\&& +   
   c_{392}  
    H_{\alpha  }{}^{\delta  \epsilon  } H^{\alpha  \beta  
\gamma  } R_{\gamma  }{}^{\eta  }{}_{\delta  \epsilon  
} R_{\varepsilon  \zeta  \mu  \eta  } \nabla^{\zeta  
}H_{\beta  }{}^{\varepsilon  \mu  }+   c_{393}  
    H_{\alpha  \beta  }{}^{\delta  } H^{\alpha  \beta  \gamma  
} R_{\gamma  \zeta  \epsilon  }{}^{\eta  } 
R_{\delta  \eta  \varepsilon  \mu  } \nabla^{\zeta  
}H^{\epsilon  \varepsilon  \mu  } \nn\\&& +   c_{394}  
    H_{\alpha  \beta  }{}^{\delta  } H^{\alpha  \beta  \gamma  
} R_{\gamma  \epsilon  \delta  }{}^{\eta  } R_{
\varepsilon  \zeta  \mu  \eta  } \nabla^{\zeta  }H^{\epsilon  
\varepsilon  \mu  } +   c_{395}  
    H^{\alpha  \beta  \gamma  } H^{\delta  \epsilon  
\varepsilon  } R_{\beta  \mu  \delta  \zeta  } 
R_{\gamma  \eta  \epsilon  \varepsilon  } \nabla^{\eta 
 }H_{\alpha  }{}^{\mu  \zeta  } \nn\\&&+   c_{396}  
    H^{\alpha  \beta  \gamma  } H^{\delta  \epsilon  
\varepsilon  } R_{\beta  \mu  \delta  \epsilon  } 
R_{\gamma  \eta  \varepsilon  \zeta  } \nabla^{\eta  
}H_{\alpha  }{}^{\mu  \zeta  }+   c_{397}  
    H^{\alpha  \beta  \gamma  } H^{\delta  \epsilon  
\varepsilon  } R_{\beta  \delta  \gamma  \mu  } 
R_{\epsilon  \zeta  \varepsilon  \eta  } \nabla^{\eta  
}H_{\alpha  }{}^{\mu  \zeta  } \nn\\&& +   c_{398}  
    H^{\alpha  \beta  \gamma  } H^{\delta  \epsilon  
\varepsilon  } R_{\beta  \delta  \gamma  \epsilon  } R
_{\varepsilon  \eta  \mu  \zeta  } \nabla^{\eta  
}H_{\alpha  }{}^{\mu  \zeta  } +   c_{399}  
    H_{\alpha  }{}^{\delta  \epsilon  } H^{\alpha  \beta  
\gamma  } R_{\beta  \delta  \gamma  \varepsilon  } 
R_{\epsilon  \eta  \mu  \zeta  } \nabla^{\eta  
}H^{\varepsilon  \mu  \zeta  } \nn\\&&+   c_{194}  
    H^{\alpha  \beta  \gamma  } H^{\delta  \epsilon  
\varepsilon  } R_{\gamma  }{}^{\zeta  }{}_{\mu  
}{}^{\eta  } R_{\epsilon  \zeta  \varepsilon  \eta  } 
\nabla^{\mu  }H_{\alpha  \beta  \delta  } +   c_{195}  
    H^{\alpha  \beta  \gamma  } H^{\delta  \epsilon  
\varepsilon  } R_{\gamma  }{}^{\zeta  }{}_{\epsilon  
}{}^{\eta  } R_{\varepsilon  \zeta  \mu  \eta  } 
\nabla^{\mu  }H_{\alpha  \beta  \delta  } \nn\\&&+   c_{196}  
    H^{\alpha  \beta  \gamma  } H^{\delta  \epsilon  
\varepsilon  } R_{\gamma  }{}^{\zeta  }{}_{\epsilon  
}{}^{\eta  } R_{\varepsilon  \eta  \mu  \zeta  } 
\nabla^{\mu  }H_{\alpha  \beta  \delta  } +   c_{206}  
    H_{\alpha  }{}^{\delta  \epsilon  } H^{\alpha  \beta  
\gamma  } R_{\delta  }{}^{\zeta  }{}_{\varepsilon  }{}^{
\eta  } R_{\epsilon  \zeta  \mu  \eta  } \nabla^{\mu  
}H_{\beta  \gamma  }{}^{\varepsilon  } \nn\\&&+   c_{207}  
    H_{\alpha  }{}^{\delta  \epsilon  } H^{\alpha  \beta  
\gamma  } R_{\delta  }{}^{\zeta  }{}_{\varepsilon  }{}^{
\eta  } R_{\epsilon  \eta  \mu  \zeta  } \nabla^{\mu  
}H_{\beta  \gamma  }{}^{\varepsilon  } +   c_{208}  
    H_{\alpha  }{}^{\delta  \epsilon  } H^{\alpha  \beta  
\gamma  } R_{\delta  }{}^{\zeta  }{}_{\epsilon  
}{}^{\eta  } R_{\varepsilon  \zeta  \mu  \eta  } 
\nabla^{\mu  }H_{\beta  \gamma  }{}^{\varepsilon  } \nn\\&&+   
   c_{215}  
    H_{\alpha  }{}^{\delta  \epsilon  } H^{\alpha  \beta  
\gamma  } R_{\gamma  }{}^{\zeta  }{}_{\varepsilon  }{}^{
\eta  } R_{\epsilon  \zeta  \mu  \eta  } \nabla^{\mu  
}H_{\beta  \delta  }{}^{\varepsilon  }+   c_{216}  
    H_{\alpha  }{}^{\delta  \epsilon  } H^{\alpha  \beta  
\gamma  } R_{\gamma  }{}^{\zeta  }{}_{\varepsilon  }{}^{
\eta  } R_{\epsilon  \eta  \mu  \zeta  } \nabla^{\mu  
}H_{\beta  \delta  }{}^{\varepsilon  }  \nn\\&&+   c_{217}  
    H_{\alpha  }{}^{\delta  \epsilon  } H^{\alpha  \beta  
\gamma  } R_{\gamma  }{}^{\zeta  }{}_{\epsilon  
}{}^{\eta  } R_{\varepsilon  \zeta  \mu  \eta  } 
\nabla^{\mu  }H_{\beta  \delta  }{}^{\varepsilon  } +   
   c_{225}  
    H_{\alpha  \beta  }{}^{\delta  } H^{\alpha  \beta  \gamma  
} R_{\delta  }{}^{\zeta  }{}_{\mu  }{}^{\eta  } 
R_{\epsilon  \zeta  \varepsilon  \eta  } \nabla^{\mu  
}H_{\gamma  }{}^{\epsilon  \varepsilon  }\nn\\&& +   c_{226}  
    H_{\alpha  \beta  }{}^{\delta  } H^{\alpha  \beta  \gamma  
} R_{\delta  }{}^{\zeta  }{}_{\epsilon  }{}^{\eta  } 
R_{\varepsilon  \zeta  \mu  \eta  } \nabla^{\mu  
}H_{\gamma  }{}^{\epsilon  \varepsilon  } +   c_{227}  
    H_{\alpha  \beta  }{}^{\delta  } H^{\alpha  \beta  \gamma  
} R_{\delta  }{}^{\zeta  }{}_{\epsilon  }{}^{\eta  } 
R_{\varepsilon  \eta  \mu  \zeta  } \nabla^{\mu  
}H_{\gamma  }{}^{\epsilon  \varepsilon  },\nn
\eeqa
\beqa
[H^2(\nabla H)^3]_{22}&\!\!\!\!\!=\!\!\!\!\!&
 c_{238}  
    H^{\alpha  \beta  \gamma  } H^{\delta  \epsilon  
\varepsilon  } \nabla_{\delta  }H_{\alpha  \beta  }{}^{\mu  } 
\nabla_{\zeta  }H_{\epsilon  \varepsilon  \eta  } \nabla^{\eta 
 }H_{\gamma  \mu  }{}^{\zeta  } +   c_{239}  
    H^{\alpha  \beta  \gamma  } H^{\delta  \epsilon  
\varepsilon  } \nabla_{\delta  }H_{\alpha  \beta  }{}^{\mu  } 
\nabla_{\eta  }H_{\epsilon  \varepsilon  \zeta  } \nabla^{\eta 
 }H_{\gamma  \mu  }{}^{\zeta  }\nn\\&\!\!\!\!\!\!\!\!\!\!& +   c_{248}  
    H^{\alpha  \beta  \gamma  } H^{\delta  \epsilon  
\varepsilon  } \nabla_{\gamma  }H_{\mu  \zeta  \eta  } 
\nabla_{\delta  }H_{\alpha  \beta  }{}^{\mu  } \nabla^{\eta  
}H_{\epsilon  \varepsilon  }{}^{\zeta  } +   c_{249}  
    H_{\alpha  }{}^{\delta  \epsilon  } H^{\alpha  \beta  
\gamma  } \nabla_{\gamma  }H_{\mu  \zeta  \eta  } 
\nabla_{\delta  }H_{\beta  }{}^{\varepsilon  \mu  } 
\nabla^{\eta  }H_{\epsilon  \varepsilon  }{}^{\zeta  }\nn\\&\!\!\!\!\!\!\!\!\!\!&+   
   c_{251}  
    H^{\alpha  \beta  \gamma  } H^{\delta  \epsilon  
\varepsilon  } \nabla_{\gamma  }H_{\varepsilon  \zeta  \eta  } 
\nabla_{\delta  }H_{\alpha  \beta  }{}^{\mu  } \nabla^{\eta  
}H_{\epsilon  \mu  }{}^{\zeta  } +   c_{254}  
    H_{\alpha  }{}^{\delta  \epsilon  } H^{\alpha  \beta  
\gamma  } \nabla_{\delta  }H_{\beta  \gamma  }{}^{\varepsilon  
} \nabla_{\eta  }H_{\varepsilon  \mu  \zeta  } \nabla^{\eta  
}H_{\epsilon  }{}^{\mu  \zeta  }\nn\\&\!\!\!\!\!\!\!\!\!\!& +   c_{255}  
    H_{\alpha  }{}^{\delta  \epsilon  } H^{\alpha  \beta  
\gamma  } \nabla^{\varepsilon  }H_{\beta  \gamma  \delta  } 
\nabla_{\eta  }H_{\varepsilon  \mu  \zeta  } \nabla^{\eta  
}H_{\epsilon  }{}^{\mu  \zeta  } +   c_{256}  
    H_{\alpha  }{}^{\delta  \epsilon  } H^{\alpha  \beta  
\gamma  } \nabla_{\delta  }H_{\beta  \gamma  }{}^{\varepsilon  
} \nabla_{\epsilon  }H_{\mu  \zeta  \eta  } \nabla^{\eta  }H_{
\varepsilon  }{}^{\mu  \zeta  } \nn\\&\!\!\!\!\!\!\!\!\!\!&+   c_{257}  
    H_{\alpha  }{}^{\delta  \epsilon  } H^{\alpha  \beta  
\gamma  } \nabla_{\epsilon  }H_{\mu  \zeta  \eta  } 
\nabla^{\varepsilon  }H_{\beta  \gamma  \delta  } \nabla^{\eta 
 }H_{\varepsilon  }{}^{\mu  \zeta  } +   c_{244}  
    H^{\alpha  \beta  \gamma  } H^{\delta  \epsilon  
\varepsilon  } \nabla_{\varepsilon  }H_{\epsilon  \zeta  \eta  
} \nabla^{\eta  }H_{\delta  \mu  }{}^{\zeta  } \nabla^{\mu  
}H_{\alpha  \beta  \gamma  }\nn\\&\!\!\!\!\!\!\!\!\!\!&+   c_{240}  
    H^{\alpha  \beta  \gamma  } H^{\delta  \epsilon  
\varepsilon  } \nabla_{\eta  }H_{\epsilon  \varepsilon  \zeta  
} \nabla^{\eta  }H_{\gamma  \mu  }{}^{\zeta  } \nabla^{\mu  
}H_{\alpha  \beta  \delta  } +   c_{250}  
    H^{\alpha  \beta  \gamma  } H^{\delta  \epsilon  
\varepsilon  } \nabla_{\gamma  }H_{\mu  \zeta  \eta  } 
\nabla^{\eta  }H_{\epsilon  \varepsilon  }{}^{\zeta  } \nabla^{
\mu  }H_{\alpha  \beta  \delta  }\nn\\&\!\!\!\!\!\!\!\!\!\!& +   c_{252}  
    H^{\alpha  \beta  \gamma  } H^{\delta  \epsilon  
\varepsilon  } \nabla_{\gamma  }H_{\varepsilon  \zeta  \eta  } 
\nabla^{\eta  }H_{\epsilon  \mu  }{}^{\zeta  } \nabla^{\mu  
}H_{\alpha  \beta  \delta  } +   c_{243}  
    H_{\alpha  }{}^{\delta  \epsilon  } H^{\alpha  \beta  
\gamma  } \nabla_{\eta  }H_{\epsilon  \mu  \zeta  } 
\nabla^{\eta  }H_{\delta  \varepsilon  }{}^{\zeta  } 
\nabla^{\mu  }H_{\beta  \gamma  }{}^{\varepsilon  } \nn\\&\!\!\!\!\!\!\!\!\!\!&+   
   c_{245}  
    H_{\alpha  }{}^{\delta  \epsilon  } H^{\alpha  \beta  
\gamma  } \nabla_{\epsilon  }H_{\varepsilon  \zeta  \eta  } 
\nabla^{\eta  }H_{\delta  \mu  }{}^{\zeta  } \nabla^{\mu  }H_{
\beta  \gamma  }{}^{\varepsilon  } +   c_{246}  
    H_{\alpha  }{}^{\delta  \epsilon  } H^{\alpha  \beta  
\gamma  } \nabla_{\varepsilon  }H_{\epsilon  \zeta  \eta  } 
\nabla^{\eta  }H_{\delta  \mu  }{}^{\zeta  } \nabla^{\mu  }H_{
\beta  \gamma  }{}^{\varepsilon  }\nn\\&\!\!\!\!\!\!\!\!\!\!&+   c_{242}  
    H_{\alpha  }{}^{\delta  \epsilon  } H^{\alpha  \beta  
\gamma  } \nabla^{\eta  }H_{\delta  \varepsilon  }{}^{\zeta  } 
\nabla_{\mu  }H_{\epsilon  \zeta  \eta  } \nabla^{\mu  
}H_{\beta  \gamma  }{}^{\varepsilon  } +   c_{236}  
    H_{\alpha  }{}^{\delta  \epsilon  } H^{\alpha  \beta  
\gamma  } \nabla_{\epsilon  }H_{\varepsilon  \zeta  \eta  } 
\nabla^{\eta  }H_{\gamma  \mu  }{}^{\zeta  } \nabla^{\mu  }H_{
\beta  \delta  }{}^{\varepsilon  }\nn\\&\!\!\!\!\!\!\!\!\!\!& +   c_{237}  
    H_{\alpha  }{}^{\delta  \epsilon  } H^{\alpha  \beta  
\gamma  } \nabla_{\varepsilon  }H_{\epsilon  \zeta  \eta  } 
\nabla^{\eta  }H_{\gamma  \mu  }{}^{\zeta  } \nabla^{\mu  }H_{
\beta  \delta  }{}^{\varepsilon  } +   c_{241}  
    H_{\alpha  \beta  }{}^{\delta  } H^{\alpha  \beta  \gamma  
} \nabla_{\eta  }H_{\varepsilon  \mu  \zeta  } \nabla^{\eta  
}H_{\delta  \epsilon  }{}^{\zeta  } \nabla^{\mu  }H_{\gamma  
}{}^{\epsilon  \varepsilon  }\nn\\&\!\!\!\!\!\!\!\!\!\!& +   c_{247}  
    H_{\alpha  \beta  }{}^{\delta  } H^{\alpha  \beta  \gamma  
} \nabla_{\eta  }H_{\epsilon  \varepsilon  \zeta  } 
\nabla^{\eta  }H_{\delta  \mu  }{}^{\zeta  } \nabla^{\mu  }H_{
\gamma  }{}^{\epsilon  \varepsilon  } +   c_{253}  
    H_{\alpha  \beta  }{}^{\delta  } H^{\alpha  \beta  \gamma  
} \nabla_{\delta  }H_{\varepsilon  \zeta  \eta  } \nabla^{\eta 
 }H_{\epsilon  \mu  }{}^{\zeta  } \nabla^{\mu  }H_{\gamma  
}{}^{\epsilon  \varepsilon  }.\nn 
\eeqa
The 29 additional structures that include the dilaton can be found in the Appendix.

Although the total number of couplings in the basis at order $\alpha'^3$ is fixed at 477, the number of couplings in each structure is not fixed. In different basis schemes, one may encounter different structures and a different number of couplings in each structure. The aforementioned structures and the number of terms in each structure are fixed in the scheme that we have chosen.

The parameters $c_1, \cdots, c_{477}$ are coupling constants that are independent of the background. Once determined for a specific background, they remain valid for any other background. In the next section, we consider a background with a circular dimension, and thus, the dimensional reduction of the aforementioned couplings must be invariant under the T-duality $\mathbb{Z}_2$-group. The parameter $c_{477}$ is fixed to zero by the three-point S-matrix element. The T-duality constraint must also reproduce $c_{477}=0$. However, since this term involves three covariant derivatives on $H$, the calculation of its circular reduction is complicated. Therefore, we do not reproduce the result $c_{477}=0$ using T-duality. On the other hand, the remaining 476 couplings involve four to eight fields, and the S-matrix calculation required to fix these couplings is very complex. Moreover, these couplings involve only the tensors $R_{\mu\nu\alpha\beta}$, $H_{\mu\nu\alpha},\nabla_{\beta}H_{\mu\nu\alpha}$, $\nabla_\alpha\Phi, \nabla_\alpha\nabla_\beta\Phi$ for which their circular reduction is easy to perform. Thus, these couplings are determined by the T-duality constraint.

 \section{T-duality  constraint}\label{sec.3}

The observation that the dimensional reduction of the classical effective action of string theory on a torus $T^{(d)}$ must be invariant under the $O(d,d, \mathbb{R})$ transformations \cite{Sen:1991zi,Hohm:2014sxa} indicates that the circular reduction of the couplings in \reef{T55} should be invariant under the discrete group $O(1,1, \mathbb{Z})$, which consists only of non-geometrical transformations. These discrete transformations transform a geometrical coupling, which is a covariant and gauge-invariant coupling, into other geometrical couplings. These discrete transformations at the leading order of $\alpha'$ are known as the Buscher rules \cite{Buscher:1987sk,Buscher:1987qj}.

To express the Buscher rules in their simplest form, it is convenient to use the following circular reduction for the metric, $B$-field, and dilaton \cite{Maharana:1992my}:
 \beqa
g_{\mu\nu}=\left(\matrix{\bg_{ab}+e^{\varphi}g_{a }g_{b }& e^{\varphi}g_{a }&\cr e^{\varphi}g_{b }&e^{\varphi}&}\right),\, B_{\mu\nu}= \left(\matrix{\bb_{ab}+\frac{1}{2}b_{a }g_{b }- \frac{1}{2}b_{b }g_{a }&b_{a }\cr - b_{b }&0&}\right),\,  \Phi=\bar{\phi}+\varphi/4\,.\labell{reduc}\eeqa
Here, $\bg_{ab}$, $\bb_{ab}$, and $\bphi$ represent the metric, B-field, and dilaton in the base space, respectively. $g_a$ and $b_b$ denote the momentum and winding vectors, respectively, while $\varphi$ represents a scalar within this space. The base space fields $\bb_{ab}$ and $b_a$ are of odd parity, and all other fields are of even parity.
The Buscher rules in this parametrization are as follows:
\beqa
\varphi'= -\varphi
\,\,\,,\,\,g'_{a }= b_{a }\,\,\,,\,\, b'_{a }= g_{a } \,\,\,,\,\,\bg_{ab}'=\bg_{ab} \,\,\,,\,\,\bb_{ab}'=\bb_{ab} \,\,\,,\,\,  \bar{\phi}'= \bar{\phi}\,.\labell{T2}
\eeqa
These transformations form a $\mathbb{Z}_2$-group, meaning that $(\psi')' = \psi$ for any field $\psi$ in the base space. Assuming that the diffeomorphism transformations receive no $\alpha'$-deformations, the Buscher rules must be deformed at each order of $\alpha'$, \ie
\beqa
\psi'&=&\psi_0'+\sum_{n=1}^{\infty}\frac{\alpha'^n}{n!}\psi_n'\,. \labell{gBuch}
\eeqa
Here, $\psi_0'$ represents the Buscher rules \reef{T2}, and $\psi_n'$ represents their deformations at order $\alpha'^n$. The deformed Buscher rules must form the $\mathbb{Z}_2$-group.

To impose the T-duality constraint on the classical effective action ${\bf S}_{\rm eff}$, one needs to reduce the theory on a circle using the reduction \reef{reduc} to obtain the $(D-1)$-dimensional effective action $S_{\rm eff}(\psi)$. We then transform this action under the aforementioned $\mathbb{Z}_2$-transformations to produce $S_{\rm eff}(\psi')$. The $\mathbb{Z}_2$-constraint on the effective action is given by
\beqa
S_{\rm eff}(\psi)-S_{\rm eff}(\psi')&=&\int d^{D-1}x \sqrt{-\bg}\nabla_a\Big[e^{-2\bphi}J^a(\psi)\Big]\,,\labell{TS}
\eeqa
where $J^a$ is an arbitrary covariant vector composed of the $(D-1)$-dimensional base space fields. Using a similar $\alpha'$-expansion as in \reef{seff} for the $(D-1)$-dimensional effective action $S_{\rm eff}$ and for the vector $J^a$, and employing the following Taylor expansion for the T-duality transformation of the $(D-1)$-dimensional effective action at order $\alpha'^n$ around the Buscher rule $\psi_0'$:
\begin{equation}
S^{(n)}(\psi')=\sum_{m=0}^{\infty}\alpha'^mS^{(n,m)}(\psi_0')\,,
\end{equation}
where $S^{(n,0)}=S^{(n)}$, we find that the $\mathbb{Z}_2$-constraint in \reef{TS} becomes
\beqa
\sum^\infty_{n=0}\alpha'^nS^{(n)}(\psi)-\!\!\!\!\sum^\infty_{n=0,m=0}\!\!\alpha'^{n+m}S^{(n,m)}(\psi_0')-\!\sum^\infty_{n=0}\alpha'^n\!\!\int d^{D-1}x\, \prt_a\Big[e^{-2\bphi}J^a_{(n)}(\psi)\Big]=0, \labell{TSn}
\eeqa
where $J^a_{(n)}$ is an arbitrary covariant vector composed of the $(D-1)$-dimensional base space fields at order $\alpha'^n$. In the above equation, we have also used the observation made in \cite{Garousi:2019mca} that the $\mathbb{Z}_2$-constraint for curved base space produces exactly the same constraint for the coupling constants as for flat base space. Hence, for simplicity of the calculation, we have assumed the base space is flat. To find the appropriate constraints on the effective actions, one must set the terms at each order of $\alpha'$ to be zero.

The T-duality constraint \reef{TSn} at order $\alpha'^0$ is
\beqa
S^{(0)}(\psi)-S^{(0)}(\psi'_0)- \int d^{D-1}x\, \prt_a\Big[e^{-2\bphi}J_{(0)}^a(\psi)\Big]&=&0\,.\labell{TS0}
\eeqa
Using the basic reductions \reef{reduc}, one finds that the circular reduction of $\bS^{(0)}$ becomes \cite{Kaloper:1997ux,Garousi:2019wgz}.
\beqa
S^{(0)}(\psi)&=&-\frac{2}{\kappa'^2}\int d^{D-1}x\,\sqrt{-\bg}e^{-2\bphi}\Big[\bar{R}-\bar{\nabla}^a\bar{\nabla}_a\vp-\frac{1}{4}\bar{\nabla}_a\vp \bar{\nabla}^a\vp+4\bar{\nabla}_a\bphi\bar{\nabla}^a \bphi+2\bar{\nabla}_a\bphi\bar{\nabla}^a\vp\nn\\
&&\qquad\qquad\qquad\qquad\qquad-\frac{1}{4}e^{\vp}V^2 -\frac{1}{4}e^{-\vp}W^2 -\frac{1}{12}\bH_{abc}\bH^{abc}\Big]\,,\labell{S0psi}
\eeqa
where $\kappa'$ is related to the $(D-1)$-dimensional Newton's constant, the scalar curvature and covariant derivatives in the above equation are defined in terms of the base space metric. The field strengths are defined as
\beqa
V_{ab}&=&\prt_a g_b-\prt_b g_a, \nn\\
W_{ab}&=&\prt_a b_b-\prt_b b_a,\nn\\
\bH_{abc}&=&\prt_{[a}\bb_{bc]}-\frac{1}{2}g_{[a}W_{bc]}-\frac{1}{2}b_{[a}V_{bc]}\,.\labell{potential}
\eeqa
Since $\bH$ is not the exterior derivative of a two-form, it satisfies an anomalous Bianchi identity, whereas $W$ and $V$ satisfy the ordinary Bianchi identity, i.e.,
 \beqa
 \prt_{[a} \bH_{bcd]}&=&-V_{[ab}W_{cd]}\,,\labell{anB}\\
 \prt_{[a} W_{bc]}&=&0\,,\nn\\
  \prt_{[a} V_{bc]}&=&0\,.\nn
 \eeqa
 Our notation for antisymmetry is such that, for example, $g_{[a}W_{bc]}=g_aW_{bc}-g_bW_{ac}-g_cW_{ba}$.
For flat base space, the scalar curvature in \reef{S0psi}  is zero and the covariant derivatives become partial derivatives. It is evident that the T-duality constraint \reef{TS0} is satisfied for the vector $J^a_{(0)}=-2\partial^a\varphi$.

Using the fact that in superstring theory, there is no effective action at orders $\alpha'$ and $\alpha'^2$, one finds that the constraint in \reef{TSn} at order $\alpha'^3$ in superstring theory becomes
\beqa
S^{(3)}(\psi)-S^{(3)}(\psi_0')-S^{(0,3)}(\psi_0')- \int d^{D-1}x\, \prt_a\Big[e^{-2\bphi}J^a_{(3)}(\psi)\Big]=0. \labell{TS3}
\eeqa
Using the reductions in \reef{reduc}, it is straightforward to find the circular reduction of the couplings in  \reef{T55} to obtain $S^{(3)}(\psi)$. We refer interested readers to \cite{Garousi:2019mca} for a useful trick to greatly simplify the calculations in this part. Then, using the Buscher transformations in  \reef{T2}, one can calculate $S^{(3)}(\psi_0')$. 

To determine the third term above, we note that there are no corrections to the Buscher rules at order $\alpha'$ and $\alpha'^2$ in superstring theory, i.e.,
\beqa
&&\varphi'= -\varphi+\frac{\alpha'^3}{3!}\Delta\vp^{(3)}(\psi)+\cdots
\,,\,g'_{a }= b_{a }+\frac{\alpha'^3}{3!}e^{\vp/2}\Delta g^{(3)}_a(\psi)+\cdots
\,,\,\nn\\&&b'_{a }= g_{a }+\frac{\alpha'^3}{3!}e^{-\vp/2}\Delta b^{(3)}_a(\psi)+\cdots
\,,\,\bg_{ab}'=\eta_{ab}+\frac{\alpha'^3}{3!}\Delta \bg^{(3)}_{ab}(\psi)+\cdots
\,,\,\nn\\&&\bH_{abc}'=\bH_{abc}+\frac{\alpha'^3}{3!}\Delta\bH^{(3)}_{abc}(\psi)+\cdots
\,,\,\bar{\phi}'= \bar{\phi}+\frac{\alpha'^3}{3!}\Delta\bphi^{(3)}(\psi)+\cdots.
\labell{T22}
\eeqa
Hence, the Taylor expansion of the T-duality transformation of the leading-order action in \reef{S0psi} around the Buscher transformations in superstring theory is
\beqa
 S^{(0,3)}(\psi_0')
 &\!\!\!\!\!=\!\!\!\!\!& -\frac{2}{3!\kappa'^2}\int d^{D-1}x e^{-2\bphi} \,  \Big[ \Big(\frac{1}{4}\prt^a\vp\prt^b\vp-2\prt^a\prt^b\bphi+\frac{1}{4}\bH^{acd}\bH^b{}_{cd} +\frac{1}{2}e^\vp V^{ac}V^b{}_{c}\nn\\
 &&+\frac{1}{2}e^{-\vp} W^{ac}W^b{}_{c}\Big)\Delta\bar{g}_{ab}^{(3)}+\Big(2\prt_c\prt^c\bphi-2\prt_c\bphi\prt^c\bphi -\frac{1}{8}\prt_c\vp\prt^c\vp-\frac{1}{24}\bH^2-\frac{1}{8}e^\vp V^2\nn\\
 &&-\frac{1}{8}e^{-\vp}W^2\Big)(\eta^{ab}\Delta\bar{g}_{ab}^{(3)}-4\Delta\bphi^{(3)})-\Big(\frac{1}{2}\prt_a\prt^a\vp-\prt_a\bphi\prt^a\vp-\frac{1}{4}e^\vp V^2+\frac{1}{4}e^{-\vp}W^2\Big)\Delta\vp^{(3)}\nn\\
 &&+e^{-\vp/2}\Big(2\prt_b\bphi W^{ab}- \prt_b W^{ab}+\prt_b\vp W^{ab}\Big)\Delta g_a^{(3)} \nn\\
 &&+e^{\vp/2}\Big(2\prt_b\bphi V^{ab}- \prt_b V^{ab}-\prt_b\vp V^{ab}\Big)\Delta b_a^{(3)}-\frac{1}{6}\bH^{abc}\Delta\bH_{abc}^{(3)}\Big] \,,\labell{d3S}
 \eeqa
where we have also removed some total derivative terms.

In order for the Taylor expansion term $S^{(0,3)}(\psi_0')$ to produce an odd-parity contribution, the deformations $\Delta g^{(3)}_a(\psi)$ and $\Delta\bH^{(3)}_{abc}(\psi)$ must be even under parity, while all other deformations must be odd under parity. Since the T-duality transformations \reef{T22} must satisfy the $\mathbb{Z}_2$-group condition, i.e., $(\psi')'=\psi$, the deformations at order $\alpha'^3$ in superstring theory should satisfy the following constraint:
\beqa
-\Delta\vp^{(3)}(\psi)+\Delta\vp^{(3)}(\psi_0') &=&0\,,\nn\\
\Delta b_a^{(3)}(\psi)+\Delta g_a^{(3)}(\psi'_0) &=&0\,,\nn\\
\Delta g_a^{(3)}(\psi)+\Delta b_a^{(3)}(\psi'_0)&=&0\,,\nn\\
\Delta \bg_{ab}^{(3)}(\psi)+\Delta \bg_{ab}^{(3)}(\psi'_0)&=&0\,,\nn\\
\Delta\bphi^{(3)}(\psi)+\Delta\bphi^{(3)}(\psi_0') &=&0\,.\labell{Z22}
\eeqa
The deformation $\Delta \bH_{abc}^{(3)}$ should also satisfy a similar relation, i.e.,
\beqa
\Delta \bH_{abc}^{(3)}(\psi)+\Delta \bH_{abc}^{(3)}(\psi'_0) &=&0\,.\labell{Hdeform}
\eeqa
However, the deformations $\Delta b_a^{(3)},\,\Delta g_a^{(3)},$ and $\Delta \bH_{abc}^{(3)}$ must satisfy the Bianchi identity $\reef{anB}$. In terms of form, this identity is given by
\beqa
d(\bH+\frac{\alpha'^3}{3!}\Delta\bH^{(3)}+\cdots)&=-d(b+\frac{\alpha'^3}{3!}e^{\vp/2}\Delta g^{(3)}+\cdots)\wedge d( g+\frac{\alpha'^3}{3!}e^{-\vp/2}\Delta b^{(3)}  +\cdots)\label{aa},
\eeqa
where $d$ denotes the exterior derivative. This relation at order $\alpha'^0$ gives rise to the Bianchi identity \reef{anB}. At order $\alpha'^3$, it yields the following relation between the deformations at that order:
\beqa
\Delta\bH^{(3)}_{abc}&=&\prt_{[a}\tilde B^{(3)}_{bc]}-e^{-\vp/2}W_{[ab}\Delta b^{(3)}_{c]}-e^{\vp/2}V_{[ab}\Delta g^{(3)}_{c]}\,. \labell{DH}
\eeqa
Here, $\tilde B^{(3)}$ is an arbitrary gauge-invariant 2-form at order $\alpha'^3$ that is even under parity  \cite{Garousi:2023kxw}. By substituting the above deformation into the constraint \reef{Hdeform}, one obtains the following constraint on the tensor $\tilde B^{(3)}$:
\beqa
\tilde B^{(3)}_{ab}(\psi)+\tilde B^{(3)}_{ab}(\psi'_0) &=&0\,.\labell{Bdeform}
\eeqa
This tensor and all deformations, including $\Delta g_a^{(3)}$, $\Delta \varphi^{(3)}$, $\Delta \bar{\varphi}^{(3)}$, $\Delta b_a^{(3)}$, and $\Delta \bar{g}_{ab}^{(3)}$, should be constructed from all contractions of $\partial \varphi$, $\partial \bar{\varphi}$, $e^{\varphi/2}V$, $e^{-\varphi/2}W$, $\bar{H}$, and their derivatives at order $\alpha'^3$ with arbitrary coefficients.
 
Using the package ''xAct,'' one finds that $\tilde{B}^{(3)}$ has 15,209 terms, $\Delta \varphi^{(3)}$ has 2,537 terms, $\Delta \bar{\varphi}^{(3)}$ has 2,537 terms, $\Delta b_a^{(3)}$ has 9,054 terms, $\Delta g_a^{(3)}$ has 9,054 terms, and $\Delta \bar{g}_{ab}^{(3)}$ has 15,217 terms. They must satisfy the $\mathbb{Z}_2$-constraints \reef{Z22} and \reef{Bdeform}. To impose these constraints and find relations between the parameters of the deformations, it is necessary to use the Bianchi identities \reef{anB}. To impose the Bianchi identities, we express the field strengths $V$, $W$, and $\bar{H}$ and their derivatives in terms of the corresponding potentials \reef{potential}. Note that since the first derivative of the vectors $g_a, b_a$ appears in the derivatives of $\bar{H}$, one has to write the field strengths $V, W$ as well as their derivatives in terms of potentials $g_a, b_a$. This is unlike the case in the parent theory, in which in order to impose the $B$-field Bianchi identity, only the derivatives of the field strength $H$ should be written in terms of the potential $B_{\mu\nu}$. Writing the field strengths in terms of the potentials allows us to write the constraints \reef{Z22} and \reef{Bdeform} in non-gauge-invariant but independent terms. The coefficients of the independent terms must be zero, which gives rise to relations between the parameters in the deformations. Since the vector potentials $g_a$ and $b_a$ without derivatives appear only in $\bar{H}$ and its partial derivatives, and the two-form $\bar{b}_{ab}$ also appears only in $\bar{H}$ and its partial derivatives, the constraint between the parameters resulting from terms involving $g_a$ and $b_a$ without derivatives on them should be the same as the constraint produced by $\bar{b}_{ab}$. Therefore, for simplicity of the calculations, in the non-gauge-invariant independent terms, we set the terms that have $g_a$ and $b_a$ without derivatives on them to zero. We then solve the algebraic equations that result from setting the coefficients of all other independent terms to zero. The corresponding solution must be inserted into the deformations $\tilde B^{(3)}$, $\Delta g_a^{(3)}$, $\Delta \varphi^{(3)}$, $\Delta \bar{\varphi}^{(3)}$, $\Delta b_a^{(3)}$, and $\Delta \bar{g}_{ab}^{(3)}$ to ensure that they satisfy the $\mathbb{Z}_2$-group.
 
Having discussed how to calculate $S^{(0,3)}(\psi_0')$, we now consider the total derivative term in the T-duality constraint \reef{TS3}. The vector $J^a_{(3)}$ should have odd parity and be constructed from contractions of $\partial \varphi$, $\partial \bar{\varphi}$, $e^{\varphi/2}V$, $e^{-\varphi/2}W$, $\bar{H}$, and their derivatives at the seven-derivative order, with arbitrary coefficients. Using the package ''xAct,'' one finds that $J^a_{(3)}$ has 68,022 terms.

Since $S^{(0,3)}(\psi'_0)$ in  \reef{d3S}, as well as $S^{(3)}(\psi)-S^{(3)}(\psi_0')$, are odd under the Buscher rules, the last term in  \reef{TS3} should also be odd under the Buscher transformations. Considering the fact that the base space dilaton $\bar{\phi}$ is invariant under the Buscher rules, one concludes that the vector $J^a_{(3)}$ should also be odd under the Buscher rules. In other words, it must satisfy the following relation:
\begin{equation}
J_{(3)}^a(\psi) + J_{(3)}^a(\psi_0') = 0\,.\labell{JJ'}
\end{equation}
One should impose the Bianchi identities on the above constraint to derive relations between the 68,022 parameters in $J^a_{(3)}$. However, if one imposes the Bianchi identities by writing the field strengths in terms of the potentials in \reef{potential}, it would produce too many non-gauge-invariant independent terms at seven-derivative order. It would be extremely difficult to separate the independent terms and determine the coefficients that must be set to zero to find the relations between the 68,022 parameters. In this case, we prefer to impose the Bianchi identities in a gauge-invariant way, as follows.

First, we introduce the totally antisymmetric gauge-invariant odd-parity tensors ${\bf H}_{abcd}$, ${\bf W}_{abc}$, and the even-parity tensor ${\bf V}_{abc}$. These tensors are at order $\alpha'$. Then, we consider all contractions of $\partial \varphi$, $\partial \bar{\varphi}$, $e^{\varphi/2}V$, $e^{-\varphi/2}W$, $\bar{H}$, ${\bf H}$, $e^{-\varphi/2}{\bf W}$, $e^{\varphi/2}{\bf V}$, and their derivatives at the seven-derivative order with one free index. They must be odd under parity and must linearly include the tensors ${\bf H}$, ${\bf W}$, ${\bf V}$, and their derivatives. There are 31,936 such terms. The coefficients of these gauge-invariant terms are arbitrary.

Next, we insert the following gauge-invariant relations for the newly defined tensors:
\beqa
{\bf H}_{abcd} &= &\partial_{[a} \bar{H}_{bcd]} + V_{[ab}W_{cd]} \labell{HWV}\,, \\
{\bf W}_{abc} &=& \partial_{[a} W_{bc]} \,,\nonumber \\
{\bf V}_{abc} &=& \partial_{[a} V_{bc]}\,. \nn
\eeqa
We denote the resulting gauge-invariant vector as ${\bf BI}^a_{(3)}(\psi)$. Then, we add this vector to  \reef{JJ'}, i.e.,
\begin{equation}
J_{(3)}^a(\psi) + J_{(3)}^a(\psi_0') + {\bf BI}^a_{(3)}(\psi) = 0\,. \label{Jz}
\end{equation}
The above equation includes the Bianchi identities in gauge-invariant form. This equation can be expressed in terms of gauge-invariant independent terms, and the coefficients of these terms must be zero. The resulting algebraic equations can be solved, yielding 21,253 relations between the 68,022 parameters in $J^a_{(3)}$. We subsequently substitute these relations into $J^a_{(3)}$. The resulting $J^a_{(3)}$, now with 46,796 parameters, is then inserted into the T-duality constraint in equation \reef{TS3}.

To solve the constraint in \reef{TS3} and find the relations between the coupling constants $c_1, \cdots, c_{476}$, it is also necessary to impose the Bianchi identities \reef{anB}. However, if we were to impose all the Bianchi identities in a gauge-invariant form, it would require introducing 83,280 new parameters in the calculations. This would result in the final algebraic equations having an excessive number of parameters. To reduce the complexity of the final algebraic equations, we choose to impose only the $\bH$-Bianchi identity in a gauge-invariant form and impose the $V$- and $W$-Bianchi identities by expressing the derivatives of these field strengths in terms of their corresponding potentials. In this case, it is not necessary to express the field strengths $V, W$ themselves in terms of their corresponding potentials. This approach helps to avoid an excessive number of independent structures.

To impose the Bianchi identities using the aforementioned mixed method, we consider all contractions of $\partial \varphi$, $\partial \bar{\varphi}$, $e^{\varphi/2}V$, $e^{-\varphi/2}W$, $\bar{H}$, ${\bf H}$, and their derivatives at the eight-derivative order with no free index. These terms must be odd under parity and must linearly include the tensor ${\bf H}$ and its derivatives. There are 26,306 such terms. The coefficients of these gauge-invariant terms are arbitrary.

Next, we insert the first relation in \reef{HWV} into these gauge-invariant terms, and we denote the resulting scalar as ${\bf BI}_{(3)}(\psi)$. This scalar is then added to \reef{TS3}, resulting in the equation \reef{TSB3}:
\begin{equation}
-S^{(0,3)}(\psi_0')-\int d^{D-1}x\, \partial_a\left[e^{-2\bar{\phi}}J^a_{(3)}(\psi)\right]+{\bf BI}_{(3)}(\psi)=S^{(3)}(\psi_0')-S^{(3)}(\psi). \label{TSB3}
\end{equation}
This equation can be expressed in terms of $\bH$-, $V$-, and $W$-gauge-invariant terms. Then, we impose the $V$- and $W$-Bianchi identities by expressing the derivatives of their field strengths in terms of the corresponding potentials. This allows us to find independent terms in which the $\bH$-Bianchi identity is imposed in a gauge-invariant form, while the $V$- and $W$-Bianchi identities are imposed in a non-gauge-invariant form. There are 136,392 such independent terms. The coefficients of these terms, which include the coupling constants $c_1, \cdots, c_{476}$ as well as 109,818 parameters in the deformations of the Buscher rules, total derivative terms, and ${\bf BI}_{(3)}(\psi)$, must be zero. This results in 122,392 algebraic equations involving these parameters.

Solving the 122,392 algebraic equations yields two sets of solutions for these parameters. One set of solutions represents the relations between only the 109,818 parameters in the deformations of the Buscher rules, total derivative terms, and ${\bf BI}_{(3)}(\psi)$. These solutions satisfy the following homogeneous equation:
\begin{equation}
-S^{(0,3)}(\psi_0')-\int d^{D-1}x\, \partial_a\left[e^{-2\bar{\phi}}J^a_{(3)}(\psi)\right]+{\bf BI}_{(3)}(\psi)=0. \label{TSB3h}
\end{equation}
However, we are not interested in the solutions of the above equation. 

The other solution, which is a particular solution of the non-homogeneous equation \reef{TSB3} and the one of interest, expresses the 109,818 parameters in terms of the coupling constants $c_1, \cdots, c_{476}$. This solution also determines the relations between the coupling constants.
We have found the particular solution to be as follows:
\beqa
c_1 = c_2 = \cdots = c_{476} = 0\,.
\eeqa
Hence, all 109,818 parameters corresponding to the particular solution are also zero. This concludes our illustration that T-duality fixes the coupling constants of the odd-parity couplings at order $\alpha'^3$ in superstring theory to be zero.

The above result indicates that the coupling constants in the heterotic theory, which may be non-zero, should depend on the effective actions at orders $\alpha'$ and $\alpha'^2$. Otherwise, the calculation would yield the same result as above, resulting in a zero value. Therefore, the above coupling constants in the heterotic theory should not be proportional to $\zeta(3)$; instead, they should be proportional to $a_1^3$, where $a_1$ and $a_1^2$ are the overall factors of the effective actions at order $\alpha'$ and $\alpha'^2$, respectively.

In the bosonic string theory, which has non-zero even-parity effective actions at orders $\alpha'$ and $\alpha'^2$, the coupling constants should still be zero. However, the bosonic theory does not have any odd-parity couplings at these orders. Therefore, the odd-parity couplings at order $\alpha'^3$, which should be related to the odd-parity couplings at orders $\alpha'$ and $\alpha'^2$, should be zero as well. 

\newpage

\section{Discussion}

In this paper, we have discovered that the basis of odd-parity NS-NS couplings at order $\alpha'^3$ comprises 477 independent couplings. Notably, we have observed that certain bases exclude couplings involving $R$, $R_{\mu\nu}$, $\nabla_\mu H^{\mu}{}_{\alpha\beta}$, $\nabla_\mu\nabla^\mu\Phi$, or terms with more than two partial derivatives, except for a single term with four partial derivatives on $B$. The basis we have selected is presented in equations \reef{T53} and \reef{T55}. However, the coupling \reef{T53} is found to be zero in string theory due to the restriction that the three-point function in string theory can involve at most six momenta, while the coupling \reef{T53} contains eight derivatives.

Furthermore, we have demonstrated that the 476 coupling constants in \reef{T55} are fixed at zero in superstring theory by the T-duality $\MZ_2$-constraint. We anticipate that a similar scenario holds for couplings in bosonic string theory and other odd-parity couplings at higher orders of $\alpha'$. This observation may stem from the conventional $B$-field gauge transformations in both bosonic and superstring theories. However, it is worth mentioning that in the heterotic theory, where the $B$-field gauge transformation deviates from the standard transformations due to the Green-Schwarz mechanism \cite{Green:1984sg}, the coupling constants in \reef{T55} could potentially be non-zero. Nonetheless, due to the complexity arising from corrections at orders $\alpha',\alpha'^2$ in the effective action and corresponding Buscher rules of the heterotic theory, determining the coupling constants by imposing the $\MZ_2$-constraint becomes highly intricate. Therefore, we have postponed this calculation for future investigations.

The absence of odd-parity couplings in bosonic and superstring theories implies that the S-matrix elements in these theories for an odd number of $B$-field vertex operators are zero, i.e.,
\beqa
{\cal A}^{(\rm odd)}=0. 
\eeqa
This observation aligns with the fact that the four-point sphere-level S-matrix elements involving one $B$-field and three gravitons, or three $B$-fields and one graviton, are  zero in superstring theory (see, for example, \cite{Garousi:2013zca}). On the other hand, it is well-established that the graviton, $B$-field, and dilaton are described by a single vertex operator in string theory. For instance, in the case of bosonic string theory, this vertex operator can be expressed as
\beqa
V\sim\epsilon_{\mu\nu}\bar{\partial} X^\mu\partial X^\nu e^{ip\cdot X}\,,
\eeqa
where the polarization tensor $\epsilon_{\mu\nu}$ exhibits antisymmetry for the $B$-field and symmetry for the graviton and dilaton. By exploiting the constraint that the S-matrix element of an odd number of $B$-field operators is zero, one may be able to  find some constraints on the S-matrix element of an even number of $B$-field operators.

Moreover, it is known that closed string S-matrix elements can be expressed in terms of open string S-matrix elements \cite{Kawai:1985xq}. Hence, the fact that the S-matrix element for an odd number of closed string $B$-fields is zero in both bosonic and superstring theories suggests the existence of constraints on the open string S-matrix elements. Unraveling these constraints on the open string S-matrix elements would be a captivating endeavor for future exploration.

According to explicit calculations presented in \cite{Garousi:2023kxw}, it has been discovered that at order $\alpha'^2$, there exist 13 odd-parity NS-NS couplings in the basis. Furthermore, in our recent paper, we have identified a total of 477 such couplings in the basis at order $\alpha'^3$. Additionally, the explicit calculations have revealed the presence of even-parity NS-NS couplings in the bases at different orders, namely $\alpha'$, $\alpha'^2$,   $\alpha'^3$, and so on, with the respective counts of 8, 60, 872, and so forth \cite{Metsaev:1987zx, Garousi:2019cdn, Garousi:2020mqn}. It is worth noting that various schemes exist for selecting the couplings within each basis. Consequently, an intriguing research direction would involve investigating whether there exists an underlying structure that governs the dimensions and selection schemes of these bases.

\vskip 0.5 cm
{\huge \bf Appendix: Dilaton couplings}
\vskip 0.5 cm
The basis scheme that we have chosen for the NS-NS odd-parity couplings consists of 36 structures. In this appendix, we present the 29 structures that include the dilaton. They are as follows:
\beqa
[H^7\nabla\Phi]_3&=&  c_{3}  
    H_{\alpha  }{}^{\beta  \gamma  } H_{\beta  }{}^{\delta  
\epsilon  } H_{\gamma  }{}^{\varepsilon  \mu  } H_{\delta  
\epsilon  }{}^{\zeta  } H_{\varepsilon  }{}^{\eta  \theta  } H_{
\zeta  \theta  \iota  } H_{\mu  \eta  }{}^{\iota  } 
\nabla^{\alpha  }\Phi  +   c_{4}  
    H_{\alpha  }{}^{\beta  \gamma  } H_{\beta  }{}^{\delta  
\epsilon  } H_{\gamma  }{}^{\varepsilon  \mu  } H_{\delta  
\varepsilon  }{}^{\zeta  } H_{\epsilon  \zeta  }{}^{\eta  } 
H_{\eta  \theta  \iota  } H_{\mu  }{}^{\theta  \iota  } 
\nabla^{\alpha  }\Phi  \nn\\&&+   c_{5}  
    H_{\alpha  }{}^{\beta  \gamma  } H_{\beta  }{}^{\delta  
\epsilon  } H_{\gamma  }{}^{\varepsilon  \mu  } H_{\delta  
\epsilon  }{}^{\zeta  } H_{\varepsilon  \zeta  }{}^{\eta  } 
H_{\eta  \theta  \iota  } H_{\mu  }{}^{\theta  \iota  } 
\nabla^{\alpha  }\Phi,\nn
\eeqa
\beqa
[HR^3\nabla\Phi]_6 &=&c_{26}  
    H^{\beta  \gamma  \delta  } R_{\alpha  
}{}^{\epsilon  \varepsilon  \mu  } R_{\beta  \epsilon  
\gamma  }{}^{\zeta  } R_{\delta  \zeta  \varepsilon  
\mu  } \nabla^{\alpha  }\Phi  +   c_{27}  
    H^{\beta  \gamma  \delta  } R_{\alpha  
}{}^{\epsilon  }{}_{\beta  }{}^{\varepsilon  } R_{\gamma 
 }{}^{\mu  }{}_{\epsilon  }{}^{\zeta  } R_{\delta  
\zeta  \varepsilon  \mu  } \nabla^{\alpha  }\Phi \nn\\&& +   
   c_{24}  
    H^{\beta  \gamma  \delta  } R_{\alpha  
}{}^{\epsilon  \varepsilon  \mu  } R_{\beta  
\varepsilon  \gamma  }{}^{\zeta  } R_{\delta  \mu  
\epsilon  \zeta  } \nabla^{\alpha  }\Phi  +   c_{25}  
    H^{\beta  \gamma  \delta  } R_{\alpha  
}{}^{\epsilon  }{}_{\beta  }{}^{\varepsilon  } R_{\gamma 
 }{}^{\mu  }{}_{\epsilon  }{}^{\zeta  } R_{\delta  \mu  
\varepsilon  \zeta  } \nabla^{\alpha  }\Phi  \nn\\&&+   c_{36} 
     H^{\beta  \gamma  \delta  } R_{\alpha  
}{}^{\epsilon  }{}_{\beta  }{}^{\varepsilon  } R_{\gamma 
 }{}^{\mu  }{}_{\delta  }{}^{\zeta  } R_{\epsilon  \mu  
\varepsilon  \zeta  } \nabla^{\alpha  }\Phi  +   c_{37} 
     H^{\beta  \gamma  \delta  } R_{\alpha  
}{}^{\epsilon  }{}_{\beta  \gamma  } R_{\delta  
}{}^{\varepsilon  \mu  \zeta  } R_{\epsilon  \mu  
\varepsilon  \zeta  } \nabla^{\alpha  }\Phi , \nn
\eeqa
\beqa
[H^3R^2\nabla\Phi]_{25}&=& c_{28}  
    H_{\beta  \gamma  }{}^{\epsilon  } H^{\beta  \gamma  
\delta  } H^{\varepsilon  \mu  \zeta  } R_{\alpha  
}{}^{\eta  }{}_{\varepsilon  \mu  } R_{\delta  \zeta  
\epsilon  \eta  } \nabla^{\alpha  }\Phi  +   c_{29}  
    H_{\alpha  }{}^{\beta  \gamma  } H_{\beta  }{}^{\delta  
\epsilon  } H^{\varepsilon  \mu  \zeta  } R_{\gamma  
}{}^{\eta  }{}_{\varepsilon  \mu  } R_{\delta  \zeta  
\epsilon  \eta  } \nabla^{\alpha  }\Phi \nn\\&& +   c_{30}  
    H_{\beta  }{}^{\epsilon  \varepsilon  } H^{\beta  \gamma  
\delta  } H_{\gamma  }{}^{\mu  \zeta  } R_{\alpha  
\epsilon  \mu  }{}^{\eta  } R_{\delta  \zeta  
\varepsilon  \eta  } \nabla^{\alpha  }\Phi  +   c_{31}  
    H_{\beta  }{}^{\epsilon  \varepsilon  } H^{\beta  \gamma  
\delta  } H^{\mu  \zeta  \eta  } R_{\alpha  \mu  
\gamma  \epsilon  } R_{\delta  \zeta  \varepsilon  
\eta  } \nabla^{\alpha  }\Phi \nn\\&& +   c_{32}  
    H_{\beta  }{}^{\epsilon  \varepsilon  } H^{\beta  \gamma  
\delta  } H^{\mu  \zeta  \eta  } R_{\alpha  \mu  
\gamma  \zeta  } R_{\delta  \eta  \epsilon  
\varepsilon  } \nabla^{\alpha  }\Phi  +   c_{33}  
    H_{\beta  }{}^{\epsilon  \varepsilon  } H^{\beta  \gamma  
\delta  } H_{\gamma  }{}^{\mu  \zeta  } R_{\alpha  
\epsilon  \mu  }{}^{\eta  } R_{\delta  \eta  
\varepsilon  \zeta  } \nabla^{\alpha  }\Phi \nn\\&& +   c_{34} 
     H_{\beta  }{}^{\epsilon  \varepsilon  } H^{\beta  \gamma  
\delta  } H_{\gamma  }{}^{\mu  \zeta  } R_{\alpha  
}{}^{\eta  }{}_{\epsilon  \varepsilon  } R_{\delta  
\eta  \mu  \zeta  } \nabla^{\alpha  }\Phi  +   c_{38}  
    H_{\beta  }{}^{\epsilon  \varepsilon  } H^{\beta  \gamma  
\delta  } H^{\mu  \zeta  \eta  } R_{\alpha  \mu  
\gamma  \delta  } R_{\epsilon  \zeta  \varepsilon  
\eta  } \nabla^{\alpha  }\Phi \nn\\&& +   c_{39}  
    H_{\beta  \gamma  }{}^{\epsilon  } H^{\beta  \gamma  
\delta  } H_{\delta  }{}^{\varepsilon  \mu  } R_{\alpha 
 }{}^{\zeta  }{}_{\varepsilon  }{}^{\eta  } R_{\epsilon  
\zeta  \mu  \eta  } \nabla^{\alpha  }\Phi  +   c_{40}  
    H_{\alpha  }{}^{\beta  \gamma  } H_{\beta  }{}^{\delta  
\epsilon  } H_{\delta  }{}^{\varepsilon  \mu  } 
R_{\gamma  }{}^{\zeta  }{}_{\varepsilon  }{}^{\eta  } R
_{\epsilon  \zeta  \mu  \eta  } \nabla^{\alpha  }\Phi  
\nn\\&&+   c_{41}  
    H_{\beta  \gamma  }{}^{\epsilon  } H^{\beta  \gamma  
\delta  } H^{\varepsilon  \mu  \zeta  } R_{\alpha  
\varepsilon  \delta  }{}^{\eta  } R_{\epsilon  \eta  
\mu  \zeta  } \nabla^{\alpha  }\Phi  +   c_{42}  
    H_{\beta  \gamma  }{}^{\epsilon  } H^{\beta  \gamma  
\delta  } H_{\delta  }{}^{\varepsilon  \mu  } R_{\alpha 
 }{}^{\zeta  }{}_{\varepsilon  }{}^{\eta  } R_{\epsilon  
\eta  \mu  \zeta  } \nabla^{\alpha  }\Phi \nn\\&& +   c_{43}  
    H_{\beta  \gamma  }{}^{\epsilon  } H^{\beta  \gamma  
\delta  } H^{\varepsilon  \mu  \zeta  } R_{\alpha  
}{}^{\eta  }{}_{\delta  \varepsilon  } R_{\epsilon  
\eta  \mu  \zeta  } \nabla^{\alpha  }\Phi  +   c_{44}  
    H_{\alpha  }{}^{\beta  \gamma  } H_{\beta  }{}^{\delta  
\epsilon  } H^{\varepsilon  \mu  \zeta  } R_{\gamma  
\varepsilon  \delta  }{}^{\eta  } R_{\epsilon  \eta  
\mu  \zeta  } \nabla^{\alpha  }\Phi  \nn\\&&+   c_{45}  
    H_{\alpha  }{}^{\beta  \gamma  } H_{\beta  }{}^{\delta  
\epsilon  } H_{\delta  }{}^{\varepsilon  \mu  } 
R_{\gamma  }{}^{\zeta  }{}_{\varepsilon  }{}^{\eta  } R
_{\epsilon  \eta  \mu  \zeta  } \nabla^{\alpha  }\Phi  
+   c_{46}  
    H_{\alpha  }{}^{\beta  \gamma  } H_{\beta  }{}^{\delta  
\epsilon  } H^{\varepsilon  \mu  \zeta  } R_{\gamma  
}{}^{\eta  }{}_{\delta  \varepsilon  } R_{\epsilon  
\eta  \mu  \zeta  } \nabla^{\alpha  }\Phi \nn\\&& +   c_{48}  
    H_{\beta  }{}^{\epsilon  \varepsilon  } H^{\beta  \gamma  
\delta  } H_{\gamma  \epsilon  }{}^{\mu  } R_{\alpha  
}{}^{\zeta  }{}_{\delta  }{}^{\eta  } R_{\varepsilon  
\zeta  \mu  \eta  } \nabla^{\alpha  }\Phi  +   c_{49}  
    H_{\beta  \gamma  }{}^{\epsilon  } H^{\beta  \gamma  
\delta  } H_{\delta  }{}^{\varepsilon  \mu  } R_{\alpha 
 }{}^{\zeta  }{}_{\epsilon  }{}^{\eta  } R_{\varepsilon  
\zeta  \mu  \eta  } \nabla^{\alpha  }\Phi  \nn\\&&+   c_{50}  
    H_{\beta  \gamma  \delta  } H^{\beta  \gamma  \delta  } 
H^{\epsilon  \varepsilon  \mu  } R_{\alpha  }{}^{\zeta  
}{}_{\epsilon  }{}^{\eta  } R_{\varepsilon  \zeta  \mu  
\eta  } \nabla^{\alpha  }\Phi  +   c_{52}  
    H_{\alpha  }{}^{\beta  \gamma  } H_{\beta  }{}^{\delta  
\epsilon  } H_{\delta  }{}^{\varepsilon  \mu  } 
R_{\gamma  }{}^{\zeta  }{}_{\epsilon  }{}^{\eta  } 
R_{\varepsilon  \zeta  \mu  \eta  } \nabla^{\alpha  
}\Phi \nn\\&& +   c_{51}  
    H_{\alpha  }{}^{\beta  \gamma  } H_{\beta  }{}^{\delta  
\epsilon  } H_{\delta  \epsilon  }{}^{\varepsilon  } R_{
\gamma  }{}^{\mu  \zeta  \eta  } R_{\varepsilon  \zeta 
 \mu  \eta  } \nabla^{\alpha  }\Phi  +   c_{53}  
    H_{\alpha  }{}^{\beta  \gamma  } H_{\beta  \gamma  
}{}^{\delta  } H^{\epsilon  \varepsilon  \mu  } 
R_{\delta  }{}^{\zeta  }{}_{\epsilon  }{}^{\eta  } 
R_{\varepsilon  \zeta  \mu  \eta  } \nabla^{\alpha  
}\Phi \nn\\&& +   c_{54}  
    H_{\beta  }{}^{\epsilon  \varepsilon  } H^{\beta  \gamma  
\delta  } H_{\gamma  }{}^{\mu  \zeta  } R_{\alpha  
\epsilon  \delta  }{}^{\eta  } R_{\varepsilon  \eta  
\mu  \zeta  } \nabla^{\alpha  }\Phi  +   c_{55}  
    H_{\beta  }{}^{\epsilon  \varepsilon  } H^{\beta  \gamma  
\delta  } H_{\gamma  }{}^{\mu  \zeta  } R_{\alpha  
}{}^{\eta  }{}_{\delta  \epsilon  } R_{\varepsilon  
\eta  \mu  \zeta  } \nabla^{\alpha  }\Phi \nn\\&& +   c_{56}  
    H_{\alpha  }{}^{\beta  \gamma  } H_{\delta  }{}^{\mu  
\zeta  } H^{\delta  \epsilon  \varepsilon  } R_{\beta  
\epsilon  \gamma  }{}^{\eta  } R_{\varepsilon  \eta  
\mu  \zeta  } \nabla^{\alpha  }\Phi, \nn
\eeqa
\beqa
[H^5R\nabla\Phi]_{26}&=& 
    c_{7}  
    H_{\beta  \gamma  }{}^{\epsilon  } H^{\beta  \gamma  
\delta  } H_{\delta  }{}^{\varepsilon  \mu  } H_{\varepsilon  
}{}^{\zeta  \eta  } H_{\mu  \zeta  }{}^{\theta  } 
R_{\alpha  \eta  \epsilon  \theta  } \nabla^{\alpha  }
\Phi  +   c_{9}  
    H_{\beta  }{}^{\epsilon  \varepsilon  } H^{\beta  \gamma  
\delta  } H_{\gamma  \epsilon  }{}^{\mu  } H_{\delta  
\varepsilon  }{}^{\zeta  } H_{\mu  }{}^{\eta  \theta  } 
R_{\alpha  \eta  \zeta  \theta  } \nabla^{\alpha  
}\Phi \nn\\&& +   c_{10}  
    H_{\beta  \gamma  }{}^{\epsilon  } H^{\beta  \gamma  
\delta  } H_{\delta  }{}^{\varepsilon  \mu  } H_{\epsilon  
\varepsilon  }{}^{\zeta  } H_{\mu  }{}^{\eta  \theta  } 
R_{\alpha  \eta  \zeta  \theta  } \nabla^{\alpha  
}\Phi  +   c_{11}  
    H_{\beta  \gamma  \delta  } H^{\beta  \gamma  \delta  } 
H_{\epsilon  \varepsilon  }{}^{\zeta  } H^{\epsilon  \varepsilon 
 \mu  } H_{\mu  }{}^{\eta  \theta  } R_{\alpha  \eta  
\zeta  \theta  } \nabla^{\alpha  }\Phi \nn\\&& +   c_{8}  
    H_{\beta  }{}^{\epsilon  \varepsilon  } H^{\beta  \gamma  
\delta  } H_{\gamma  \epsilon  }{}^{\mu  } H_{\delta  
}{}^{\zeta  \eta  } H_{\varepsilon  \zeta  }{}^{\theta  } 
R_{\alpha  \eta  \mu  \theta  } \nabla^{\alpha  }\Phi 
 +   c_{13}  
    H_{\beta  \gamma  }{}^{\epsilon  } H^{\beta  \gamma  
\delta  } H_{\delta  }{}^{\varepsilon  \mu  } H_{\epsilon  }{}^{
\zeta  \eta  } H_{\varepsilon  \mu  }{}^{\theta  } 
R_{\alpha  \theta  \zeta  \eta  } \nabla^{\alpha  
}\Phi  \nn\\&&+   c_{12}  
    H_{\beta  \gamma  }{}^{\epsilon  } H^{\beta  \gamma  
\delta  } H_{\delta  }{}^{\varepsilon  \mu  } H_{\epsilon  }{}^{
\zeta  \eta  } H_{\varepsilon  \zeta  }{}^{\theta  } 
R_{\alpha  \theta  \mu  \eta  } \nabla^{\alpha  }\Phi 
 +   c_{6}  
    H_{\beta  \gamma  }{}^{\epsilon  } H^{\beta  \gamma  
\delta  } H_{\delta  }{}^{\varepsilon  \mu  } H_{\varepsilon  
}{}^{\zeta  \eta  } H_{\zeta  \eta  }{}^{\theta  } 
R_{\alpha  \mu  \epsilon  \theta  } \nabla^{\alpha  
}\Phi \nn\\&& +   c_{14}  
    H_{\alpha  }{}^{\beta  \gamma  } H_{\beta  }{}^{\delta  
\epsilon  } H_{\delta  }{}^{\varepsilon  \mu  } H_{\zeta  \eta  
\theta  } H^{\zeta  \eta  \theta  } R_{\gamma  
\varepsilon  \epsilon  \mu  } \nabla^{\alpha  }\Phi  +   
   c_{18}  
    H_{\alpha  }{}^{\beta  \gamma  } H_{\beta  }{}^{\delta  
\epsilon  } H_{\delta  }{}^{\varepsilon  \mu  } H_{\varepsilon  
\mu  }{}^{\zeta  } H_{\zeta  }{}^{\eta  \theta  } 
R_{\gamma  \eta  \epsilon  \theta  } \nabla^{\alpha  }
\Phi \nn\\&& +   c_{17}  
    H_{\alpha  }{}^{\beta  \gamma  } H_{\beta  }{}^{\delta  
\epsilon  } H_{\delta  }{}^{\varepsilon  \mu  } H_{\varepsilon  
}{}^{\zeta  \eta  } H_{\mu  \zeta  }{}^{\theta  } 
R_{\gamma  \eta  \epsilon  \theta  } \nabla^{\alpha  }
\Phi  +   c_{19}  
    H_{\alpha  }{}^{\beta  \gamma  } H_{\beta  }{}^{\delta  
\epsilon  } H_{\delta  \epsilon  }{}^{\varepsilon  } H_{\mu  
\zeta  }{}^{\theta  } H^{\mu  \zeta  \eta  } R_{\gamma 
 \eta  \varepsilon  \theta  } \nabla^{\alpha  }\Phi \nn\\&& +   
   c_{20}  
    H_{\alpha  }{}^{\beta  \gamma  } H_{\beta  }{}^{\delta  
\epsilon  } H_{\delta  }{}^{\varepsilon  \mu  } H_{\epsilon  
\varepsilon  }{}^{\zeta  } H_{\mu  }{}^{\eta  \theta  } 
R_{\gamma  \eta  \zeta  \theta  } \nabla^{\alpha  
}\Phi  +   c_{21}  
    H_{\alpha  }{}^{\beta  \gamma  } H_{\beta  }{}^{\delta  
\epsilon  } H_{\delta  \epsilon  }{}^{\varepsilon  } 
H_{\varepsilon  }{}^{\mu  \zeta  } H_{\mu  }{}^{\eta  \theta  } 
R_{\gamma  \eta  \zeta  \theta  } \nabla^{\alpha  
}\Phi \nn\\&& +   c_{22}  
    H_{\alpha  }{}^{\beta  \gamma  } H_{\beta  }{}^{\delta  
\epsilon  } H_{\delta  }{}^{\varepsilon  \mu  } H_{\varepsilon  
}{}^{\zeta  \eta  } H_{\zeta  \eta  }{}^{\theta  } 
R_{\gamma  \theta  \epsilon  \mu  } \nabla^{\alpha  
}\Phi  +   c_{23}  
    H_{\alpha  }{}^{\beta  \gamma  } H_{\beta  }{}^{\delta  
\epsilon  } H_{\delta  }{}^{\varepsilon  \mu  } H_{\epsilon  
}{}^{\zeta  \eta  } H_{\varepsilon  \mu  }{}^{\theta  } 
R_{\gamma  \theta  \zeta  \eta  } \nabla^{\alpha  
}\Phi  \nn\\&&+   c_{15}  
    H_{\alpha  }{}^{\beta  \gamma  } H_{\beta  }{}^{\delta  
\epsilon  } H_{\delta  }{}^{\varepsilon  \mu  } H_{\varepsilon  
}{}^{\zeta  \eta  } H_{\zeta  \eta  }{}^{\theta  } 
R_{\gamma  \mu  \epsilon  \theta  } \nabla^{\alpha  
}\Phi  +   c_{16}  
    H_{\alpha  }{}^{\beta  \gamma  } H_{\beta  }{}^{\delta  
\epsilon  } H_{\delta  }{}^{\varepsilon  \mu  } H_{\epsilon  
}{}^{\zeta  \eta  } H_{\varepsilon  \zeta  }{}^{\theta  } 
R_{\gamma  \mu  \eta  \theta  } \nabla^{\alpha  }\Phi 
\nn\\&& +   c_{35}  
    H_{\alpha  }{}^{\beta  \gamma  } H_{\beta  \gamma  
}{}^{\delta  } H_{\epsilon  \varepsilon  }{}^{\zeta  } 
H^{\epsilon  \varepsilon  \mu  } H_{\mu  }{}^{\eta  \theta  } 
R_{\delta  \eta  \zeta  \theta  } \nabla^{\alpha  
}\Phi  +   c_{47}  
    H_{\alpha  }{}^{\beta  \gamma  } H_{\beta  }{}^{\delta  
\epsilon  } H_{\gamma  }{}^{\varepsilon  \mu  } H_{\delta  }{}^{
\zeta  \eta  } H_{\zeta  \eta  }{}^{\theta  } 
R_{\epsilon  \theta  \varepsilon  \mu  } 
\nabla^{\alpha  }\Phi  \nn\\&&+   c_{59}  
    H_{\alpha  }{}^{\beta  \gamma  } H_{\beta  }{}^{\delta  
\epsilon  } H_{\gamma  \delta  }{}^{\varepsilon  } H_{\epsilon  
}{}^{\mu  \zeta  } H_{\mu  }{}^{\eta  \theta  } 
R_{\varepsilon  \eta  \zeta  \theta  } \nabla^{\alpha  
}\Phi  +   c_{60}  
    H_{\alpha  }{}^{\beta  \gamma  } H_{\beta  \gamma  
}{}^{\delta  } H_{\delta  }{}^{\epsilon  \varepsilon  } 
H_{\epsilon  }{}^{\mu  \zeta  } H_{\mu  }{}^{\eta  \theta  } R
_{\varepsilon  \eta  \zeta  \theta  } \nabla^{\alpha  }
\Phi \nn\\&& +   c_{57}  
    H_{\alpha  }{}^{\beta  \gamma  } H_{\beta  }{}^{\delta  
\epsilon  } H_{\gamma  }{}^{\varepsilon  \mu  } H_{\delta  }{}^{
\zeta  \eta  } H_{\epsilon  \zeta  }{}^{\theta  } 
R_{\varepsilon  \eta  \mu  \theta  } \nabla^{\alpha  }
\Phi  +   c_{58}  
    H_{\alpha  }{}^{\beta  \gamma  } H_{\beta  }{}^{\delta  
\epsilon  } H_{\gamma  }{}^{\varepsilon  \mu  } H_{\delta  
\epsilon  }{}^{\zeta  } H_{\zeta  }{}^{\eta  \theta  } 
R_{\varepsilon  \eta  \mu  \theta  } \nabla^{\alpha  }
\Phi  \nn\\&&+   c_{61}  
    H_{\alpha  }{}^{\beta  \gamma  } H_{\beta  }{}^{\delta  
\epsilon  } H_{\gamma  }{}^{\varepsilon  \mu  } H_{\delta  
\varepsilon  }{}^{\zeta  } H_{\epsilon  }{}^{\eta  \theta  } 
R_{\mu  \eta  \zeta  \theta  } \nabla^{\alpha  }\Phi  
+   c_{62}  
    H_{\alpha  }{}^{\beta  \gamma  } H_{\beta  }{}^{\delta  
\epsilon  } H_{\gamma  }{}^{\varepsilon  \mu  } H_{\delta  
\epsilon  }{}^{\zeta  } H_{\varepsilon  }{}^{\eta  \theta  } 
R_{\mu  \eta  \zeta  \theta  } \nabla^{\alpha  }\Phi ,
\nn
\eeqa
\beqa
[H^5\nabla\Phi\nabla^2\Phi]_8&\!\!\!\!\!\!=\!\!\!\!\!\!&   c_{64}  
    H_{\alpha  }{}^{\delta  \epsilon  } H_{\beta  \delta  }{}^{
\varepsilon  } H_{\gamma  }{}^{\mu  \zeta  } H_{\epsilon  \mu  
}{}^{\eta  } H_{\varepsilon  \zeta  \eta  } \nabla^{\alpha  
}\Phi  \nabla^{\gamma  }\nabla^{\beta  }\Phi  +   
   c_{65}  
    H_{\alpha  \beta  }{}^{\delta  } H_{\gamma  }{}^{\epsilon  
\varepsilon  } H_{\delta  }{}^{\mu  \zeta  } H_{\epsilon  \mu  
}{}^{\eta  } H_{\varepsilon  \zeta  \eta  } \nabla^{\alpha  
}\Phi  \nabla^{\gamma  }\nabla^{\beta  }\Phi  \nn\\&\!\!\!\!\!\!\!\!\!\!\!\!\!\!& +   
   c_{66}  
    H_{\alpha  }{}^{\delta  \epsilon  } H_{\beta  
}{}^{\varepsilon  \mu  } H_{\gamma  }{}^{\zeta  \eta  } 
H_{\delta  \epsilon  \varepsilon  } H_{\mu  \zeta  \eta  } 
\nabla^{\alpha  }\Phi  \nabla^{\gamma  }\nabla^{\beta  }\Phi  
+   c_{67}  
    H_{\alpha  }{}^{\delta  \epsilon  } H_{\beta  \delta  }{}^{
\varepsilon  } H_{\gamma  }{}^{\mu  \zeta  } H_{\epsilon  
\varepsilon  }{}^{\eta  } H_{\mu  \zeta  \eta  } 
\nabla^{\alpha  }\Phi  \nabla^{\gamma  }\nabla^{\beta  }\Phi  
  \nn\\&\!\!\!\!\!\!\!\!\!\!\!\!\!\!&+   c_{68}  
    H_{\alpha  \beta  }{}^{\delta  } H_{\gamma  }{}^{\epsilon  
\varepsilon  } H_{\delta  }{}^{\mu  \zeta  } H_{\epsilon  
\varepsilon  }{}^{\eta  } H_{\mu  \zeta  \eta  } 
\nabla^{\alpha  }\Phi  \nabla^{\gamma  }\nabla^{\beta  }\Phi  
+   c_{69}  
    H_{\alpha  }{}^{\delta  \epsilon  } H_{\beta  \delta  }{}^{
\varepsilon  } H_{\gamma  \varepsilon  }{}^{\mu  } H_{\epsilon  
}{}^{\zeta  \eta  } H_{\mu  \zeta  \eta  } \nabla^{\alpha  
}\Phi  \nabla^{\gamma  }\nabla^{\beta  }\Phi   \nn\\&\!\!\!\!\!\!\!\!\!\!\!\!\!\!& +   
   c_{70}  
    H_{\alpha  \beta  }{}^{\delta  } H_{\gamma  }{}^{\epsilon  
\varepsilon  } H_{\delta  \epsilon  }{}^{\mu  } H_{\varepsilon  
}{}^{\zeta  \eta  } H_{\mu  \zeta  \eta  } \nabla^{\alpha  
}\Phi  \nabla^{\gamma  }\nabla^{\beta  }\Phi  +   
   c_{71}  
    H_{\alpha  \beta  }{}^{\delta  } H_{\gamma  }{}^{\epsilon  
\varepsilon  } H_{\delta  \epsilon  \varepsilon  } H_{\mu  
\zeta  \eta  } H^{\mu  \zeta  \eta  } \nabla^{\alpha  }\Phi  
\nabla^{\gamma  }\nabla^{\beta  }\Phi,\nn
\eeqa
\beqa
[HR^2\nabla\Phi\nabla^2\Phi]_8&=& 
 c_{63}  
    H^{\gamma  \delta  \epsilon  } R_{\beta  
}{}^{\varepsilon  }{}_{\gamma  }{}^{\mu  } R_{\delta  
\varepsilon  \epsilon  \mu  } \nabla^{\alpha  }\Phi  
\nabla^{\beta  }\nabla_{\alpha  }\Phi  +   c_{80}  
    H^{\delta  \epsilon  \varepsilon  } R_{\alpha  
}{}^{\mu  }{}_{\delta  \epsilon  } R_{\beta  
\varepsilon  \gamma  \mu  } \nabla^{\alpha  }\Phi  
\nabla^{\gamma  }\nabla^{\beta  }\Phi   \nn\\&&+   c_{81}  
    H_{\beta  }{}^{\delta  \epsilon  } R_{\alpha  }{}^{
\varepsilon  }{}_{\delta  }{}^{\mu  } R_{\gamma  
\varepsilon  \epsilon  \mu  } \nabla^{\alpha  }\Phi  
\nabla^{\gamma  }\nabla^{\beta  }\Phi  +   c_{83}  
    H^{\delta  \epsilon  \varepsilon  } R_{\alpha  
\delta  \beta  }{}^{\mu  } R_{\gamma  \mu  \epsilon  
\varepsilon  } \nabla^{\alpha  }\Phi  \nabla^{\gamma  
}\nabla^{\beta  }\Phi  \nn\\&& +   c_{84}  
    H_{\beta  }{}^{\delta  \epsilon  } R_{\alpha  }{}^{
\varepsilon  }{}_{\delta  }{}^{\mu  } R_{\gamma  \mu  
\epsilon  \varepsilon  } \nabla^{\alpha  }\Phi  \nabla^{\gamma 
 }\nabla^{\beta  }\Phi  +   c_{85}  
    H^{\delta  \epsilon  \varepsilon  } R_{\alpha  
}{}^{\mu  }{}_{\beta  \delta  } R_{\gamma  \mu  
\epsilon  \varepsilon  } \nabla^{\alpha  }\Phi  \nabla^{\gamma 
 }\nabla^{\beta  }\Phi   \nn\\&&+   c_{93}  
    H_{\beta  }{}^{\delta  \epsilon  } R_{\alpha  }{}^{
\varepsilon  }{}_{\gamma  }{}^{\mu  } R_{\delta  
\varepsilon  \epsilon  \mu  } \nabla^{\alpha  }\Phi  
\nabla^{\gamma  }\nabla^{\beta  }\Phi  +   c_{94}  
    H_{\alpha  \beta  }{}^{\delta  } R_{\gamma  
}{}^{\epsilon  \varepsilon  \mu  } R_{\delta  
\varepsilon  \epsilon  \mu  } \nabla^{\alpha  }\Phi  
\nabla^{\gamma  }\nabla^{\beta  }\Phi, \nn
\eeqa
\beqa
[H^3R\nabla\Phi\nabla^2\Phi]_{19}&=& 
 c_{72}  
    H_{\beta  }{}^{\delta  \epsilon  } H_{\varepsilon  \mu  
\zeta  } H^{\varepsilon  \mu  \zeta  } R_{\alpha  
\delta  \gamma  \epsilon  } \nabla^{\alpha  }\Phi  
\nabla^{\gamma  }\nabla^{\beta  }\Phi  +   c_{73}  
    H_{\beta  }{}^{\delta  \epsilon  } H_{\delta  
}{}^{\varepsilon  \mu  } H_{\varepsilon  \mu  }{}^{\zeta  } 
R_{\alpha  \epsilon  \gamma  \zeta  } \nabla^{\alpha  
}\Phi  \nabla^{\gamma  }\nabla^{\beta  }\Phi  \nn\\&& +   
   c_{76}  
    H_{\beta  }{}^{\delta  \epsilon  } H_{\delta  
}{}^{\varepsilon  \mu  } H_{\varepsilon  \mu  }{}^{\zeta  } 
R_{\alpha  \zeta  \gamma  \epsilon  } \nabla^{\alpha  
}\Phi  \nabla^{\gamma  }\nabla^{\beta  }\Phi  +   
   c_{77}  
    H_{\beta  }{}^{\delta  \epsilon  } H_{\gamma  
}{}^{\varepsilon  \mu  } H_{\delta  \varepsilon  }{}^{\zeta  } 
R_{\alpha  \zeta  \epsilon  \mu  } \nabla^{\alpha  
}\Phi  \nabla^{\gamma  }\nabla^{\beta  }\Phi  \nn\\&& +   
   c_{78}  
    H_{\beta  }{}^{\delta  \epsilon  } H_{\gamma  
}{}^{\varepsilon  \mu  } H_{\delta  \epsilon  }{}^{\zeta  } 
R_{\alpha  \zeta  \varepsilon  \mu  } \nabla^{\alpha  
}\Phi  \nabla^{\gamma  }\nabla^{\beta  }\Phi  +   
   c_{74}  
    H_{\beta  }{}^{\delta  \epsilon  } H_{\delta  
}{}^{\varepsilon  \mu  } H_{\epsilon  \varepsilon  }{}^{\zeta  } 
R_{\alpha  \mu  \gamma  \zeta  } \nabla^{\alpha  
}\Phi  \nabla^{\gamma  }\nabla^{\beta  }\Phi  \nn\\&& +   
   c_{75}  
    H_{\beta  }{}^{\delta  \epsilon  } H_{\delta  \epsilon  
}{}^{\varepsilon  } H_{\varepsilon  }{}^{\mu  \zeta  } 
R_{\alpha  \mu  \gamma  \zeta  } \nabla^{\alpha  
}\Phi  \nabla^{\gamma  }\nabla^{\beta  }\Phi  +   
   c_{79}  
    H_{\alpha  }{}^{\delta  \epsilon  } H_{\delta  
}{}^{\varepsilon  \mu  } H_{\varepsilon  \mu  }{}^{\zeta  } 
R_{\beta  \epsilon  \gamma  \zeta  } \nabla^{\alpha  }
\Phi  \nabla^{\gamma  }\nabla^{\beta  }\Phi   \nn\\&&+   c_{90} 
     H_{\alpha  }{}^{\delta  \epsilon  } H_{\beta  
}{}^{\varepsilon  \mu  } H_{\varepsilon  \mu  }{}^{\zeta  } 
R_{\gamma  \zeta  \delta  \epsilon  } \nabla^{\alpha  
}\Phi  \nabla^{\gamma  }\nabla^{\beta  }\Phi  +   
   c_{91}  
    H_{\alpha  }{}^{\delta  \epsilon  } H_{\beta  
}{}^{\varepsilon  \mu  } H_{\delta  \varepsilon  }{}^{\zeta  } 
R_{\gamma  \zeta  \epsilon  \mu  } \nabla^{\alpha  
}\Phi  \nabla^{\gamma  }\nabla^{\beta  }\Phi  \nn\\&& +   
   c_{92}  
    H_{\alpha  }{}^{\delta  \epsilon  } H_{\beta  
}{}^{\varepsilon  \mu  } H_{\delta  \epsilon  }{}^{\zeta  } 
R_{\gamma  \zeta  \varepsilon  \mu  } \nabla^{\alpha  
}\Phi  \nabla^{\gamma  }\nabla^{\beta  }\Phi  +   
   c_{82}  
    H_{\alpha  \beta  }{}^{\delta  } H_{\epsilon  \varepsilon  
}{}^{\zeta  } H^{\epsilon  \varepsilon  \mu  } 
R_{\gamma  \mu  \delta  \zeta  } \nabla^{\alpha  
}\Phi  \nabla^{\gamma  }\nabla^{\beta  }\Phi  \nn\\&&+   
   c_{86}  
    H_{\alpha  }{}^{\delta  \epsilon  } H_{\beta  
}{}^{\varepsilon  \mu  } H_{\delta  \varepsilon  }{}^{\zeta  } 
R_{\gamma  \mu  \epsilon  \zeta  } \nabla^{\alpha  
}\Phi  \nabla^{\gamma  }\nabla^{\beta  }\Phi  +   
   c_{87}  
    H_{\alpha  }{}^{\delta  \epsilon  } H_{\beta  \delta  }{}^{
\varepsilon  } H_{\varepsilon  }{}^{\mu  \zeta  } 
R_{\gamma  \mu  \epsilon  \zeta  } \nabla^{\alpha  
}\Phi  \nabla^{\gamma  }\nabla^{\beta  }\Phi  \nn\\&& +   
   c_{88}  
    H_{\alpha  }{}^{\delta  \epsilon  } H_{\beta  \delta  }{}^{
\varepsilon  } H_{\epsilon  }{}^{\mu  \zeta  } 
R_{\gamma  \mu  \varepsilon  \zeta  } \nabla^{\alpha  
}\Phi  \nabla^{\gamma  }\nabla^{\beta  }\Phi  +   
   c_{89}  
    H_{\alpha  \beta  }{}^{\delta  } H_{\delta  }{}^{\epsilon  
\varepsilon  } H_{\epsilon  }{}^{\mu  \zeta  } 
R_{\gamma  \mu  \varepsilon  \zeta  } \nabla^{\alpha  
}\Phi  \nabla^{\gamma  }\nabla^{\beta  }\Phi  \nn\\&& +   
   c_{95}  
    H_{\alpha  \beta  }{}^{\delta  } H_{\gamma  }{}^{\epsilon  
\varepsilon  } H_{\epsilon  }{}^{\mu  \zeta  } 
R_{\delta  \mu  \varepsilon  \zeta  } \nabla^{\alpha  
}\Phi  \nabla^{\gamma  }\nabla^{\beta  }\Phi  +   
   c_{96}  
    H_{\alpha  }{}^{\delta  \epsilon  } H_{\beta  \delta  }{}^{
\varepsilon  } H_{\gamma  }{}^{\mu  \zeta  } 
R_{\epsilon  \mu  \varepsilon  \zeta  } \nabla^{\alpha 
 }\Phi  \nabla^{\gamma  }\nabla^{\beta  }\Phi  \nn\\&& +   
   c_{97}  
    H_{\alpha  \beta  }{}^{\delta  } H_{\gamma  }{}^{\epsilon  
\varepsilon  } H_{\delta  }{}^{\mu  \zeta  } 
R_{\epsilon  \mu  \varepsilon  \zeta  } \nabla^{\alpha 
 }\Phi  \nabla^{\gamma  }\nabla^{\beta  }\Phi,\nn
\eeqa
\beqa
[H^3(\nabla\Phi)^3\nabla^2\Phi]_1&=&   c_{104}  
    H_{\beta  \gamma  }{}^{\epsilon  } H_{\delta  
}{}^{\varepsilon  \mu  } H_{\epsilon  \varepsilon  \mu  } 
\nabla_{\alpha  }\Phi  \nabla^{\alpha  }\Phi  \nabla^{\beta  
}\Phi  \nabla^{\delta  }\nabla^{\gamma  }\Phi,\nn
\eeqa
\beqa
[\nabla H(\nabla\Phi)^2(\nabla^2\Phi)^2]_1&=&  
   c_{130}  
    \nabla^{\alpha  }\Phi  \nabla^{\beta  }\Phi  
\nabla^{\gamma  }\nabla_{\alpha  }\Phi  \nabla_{\epsilon  }H_{
\beta  \gamma  \delta  } \nabla^{\epsilon  }\nabla^{\delta  
}\Phi, \nn
\eeqa
\beqa
[HR(\nabla\Phi)^3\nabla^2\Phi]_3&=&  c_{101}  
    H_{\beta  }{}^{\epsilon  \varepsilon  } R_{\gamma  
\epsilon  \delta  \varepsilon  } \nabla^{\alpha  }\Phi  
\nabla^{\beta  }\Phi  \nabla^{\gamma  }\Phi  \nabla^{\delta  
}\nabla_{\alpha  }\Phi  +   c_{105}  
    H_{\gamma  }{}^{\epsilon  \varepsilon  } R_{\beta  
\epsilon  \delta  \varepsilon  } \nabla_{\alpha  }\Phi  
\nabla^{\alpha  }\Phi  \nabla^{\beta  }\Phi  \nabla^{\delta  
}\nabla^{\gamma  }\Phi  \nn\\&&+   c_{126}  
    H_{\alpha  \delta  }{}^{\varepsilon  } R_{\beta  
\epsilon  \gamma  \varepsilon  } \nabla^{\alpha  }\Phi  
\nabla^{\beta  }\Phi  \nabla^{\gamma  }\Phi  \nabla^{\epsilon 
 }\nabla^{\delta  }\Phi ,\nn
\eeqa
\beqa
[H^3\nabla\Phi(\nabla^2\Phi)^2]_3&\!\!\!\!\!=\!\!\!\!\!&  c_{102} 
     H_{\alpha  \gamma  }{}^{\epsilon  } H_{\delta  
}{}^{\varepsilon  \mu  } H_{\epsilon  \varepsilon  \mu  } 
\nabla^{\alpha  }\Phi  \nabla^{\gamma  }\nabla^{\beta  }\Phi  
\nabla^{\delta  }\nabla_{\beta  }\Phi  +   c_{107}  
    H_{\beta  \gamma  }{}^{\epsilon  } H_{\delta  
}{}^{\varepsilon  \mu  } H_{\epsilon  \varepsilon  \mu  } 
\nabla^{\alpha  }\Phi  \nabla^{\beta  }\nabla_{\alpha  }\Phi  
\nabla^{\delta  }\nabla^{\gamma  }\Phi  \nn\\&\!\!\!\!\!\!\!\!\!\!\!\!\!\!& +   c_{127}  
    H_{\alpha  \beta  }{}^{\varepsilon  } H_{\gamma  \delta  
}{}^{\mu  } H_{\epsilon  \varepsilon  \mu  } \nabla^{\alpha  
}\Phi  \nabla^{\gamma  }\nabla^{\beta  }\Phi  
\nabla^{\epsilon  }\nabla^{\delta  }\Phi,\nn
\eeqa
\beqa
[HR\nabla\Phi(\nabla^2\Phi)^2]_4&=&c_{103}  
    H_{\gamma  }{}^{\epsilon  \varepsilon  } R_{\alpha  
\epsilon  \delta  \varepsilon  } \nabla^{\alpha  }\Phi  
\nabla^{\gamma  }\nabla^{\beta  }\Phi  \nabla^{\delta  
}\nabla_{\beta  }\Phi  +   c_{108}  
    H_{\gamma  }{}^{\epsilon  \varepsilon  } R_{\beta  
\epsilon  \delta  \varepsilon  } \nabla^{\alpha  }\Phi  
\nabla^{\beta  }\nabla_{\alpha  }\Phi  \nabla^{\delta  
}\nabla^{\gamma  }\Phi \nn\\&& +   c_{128}  
    H_{\beta  \delta  }{}^{\varepsilon  } R_{\alpha  
\varepsilon  \gamma  \epsilon  } \nabla^{\alpha  }\Phi  
\nabla^{\gamma  }\nabla^{\beta  }\Phi  \nabla^{\epsilon  
}\nabla^{\delta  }\Phi  +   c_{129}  
    H_{\alpha  \beta  }{}^{\varepsilon  } R_{\gamma  
\delta  \epsilon  \varepsilon  } \nabla^{\alpha  }\Phi  
\nabla^{\gamma  }\nabla^{\beta  }\Phi  \nabla^{\epsilon  
}\nabla^{\delta  }\Phi,\nn
\eeqa
\beqa
[\nabla HR(\nabla^2\Phi)^2]_2&=& c_{151}  
    R_{\beta  \epsilon  \delta  \varepsilon  } 
\nabla^{\beta  }\nabla^{\alpha  }\Phi  \nabla^{\delta  
}\nabla^{\gamma  }\Phi  \nabla^{\varepsilon  }H_{\alpha  
\gamma  }{}^{\epsilon  } +   c_{153}  
    R_{\gamma  \varepsilon  \delta  \epsilon  } 
\nabla^{\beta  }\nabla^{\alpha  }\Phi  \nabla^{\gamma  
}\nabla_{\alpha  }\Phi  \nabla^{\varepsilon  }H_{\beta  
}{}^{\delta  \epsilon  },\nn
\eeqa
\beqa
[\nabla HR (\nabla\Phi)^2\nabla^2\Phi]_3&=&  c_{109} 
     R_{\beta  \epsilon  \delta  \varepsilon  } 
\nabla^{\alpha  }\Phi  \nabla^{\beta  }\Phi  \nabla_{\gamma  
}H_{\alpha  }{}^{\epsilon  \varepsilon  } \nabla^{\delta  
}\nabla^{\gamma  }\Phi  +   c_{150}  
    R_{\beta  \varepsilon  \delta  \epsilon  } 
\nabla^{\alpha  }\Phi  \nabla^{\beta  }\Phi  \nabla^{\delta  
}\nabla^{\gamma  }\Phi  \nabla^{\varepsilon  }H_{\alpha  
\gamma  }{}^{\epsilon  } \nn\\&&+   c_{154}  
    R_{\gamma  \varepsilon  \delta  \epsilon  } 
\nabla_{\alpha  }\Phi  \nabla^{\alpha  }\Phi  \nabla^{\gamma  
}\nabla^{\beta  }\Phi  \nabla^{\varepsilon  }H_{\beta  
}{}^{\delta  \epsilon  },\nn
\eeqa
\beqa
[\nabla HR^2(\nabla\Phi)^2]_3&=&  c_{123} 
     R_{\beta  }{}^{\varepsilon  }{}_{\gamma  }{}^{\mu  
} R_{\delta  \mu  \epsilon  \varepsilon  } 
\nabla^{\alpha  }\Phi  \nabla^{\beta  }\Phi  \nabla^{\epsilon 
 }H_{\alpha  }{}^{\gamma  \delta  } +   c_{155}  
    R_{\alpha  \varepsilon  \gamma  }{}^{\mu  } 
R_{\beta  \mu  \delta  \epsilon  } \nabla^{\alpha  
}\Phi  \nabla^{\beta  }\Phi  \nabla^{\varepsilon  }H^{\gamma  
\delta  \epsilon  } \nn\\&&+   c_{156}  
    R_{\alpha  \gamma  \beta  }{}^{\mu  } 
R_{\delta  \varepsilon  \epsilon  \mu  } 
\nabla^{\alpha  }\Phi  \nabla^{\beta  }\Phi  
\nabla^{\varepsilon  }H^{\gamma  \delta  \epsilon  },\nn
\eeqa
\beqa
[\nabla HR^2\nabla^2\Phi]_4&=&  c_{124}  
    R_{\beta  }{}^{\varepsilon  }{}_{\gamma  }{}^{\mu  
} R_{\delta  \varepsilon  \epsilon  \mu  } 
\nabla^{\beta  }\nabla^{\alpha  }\Phi  \nabla^{\epsilon  
}H_{\alpha  }{}^{\gamma  \delta  } +   c_{125}  
    R_{\beta  }{}^{\varepsilon  }{}_{\gamma  }{}^{\mu  
} R_{\delta  \mu  \epsilon  \varepsilon  } 
\nabla^{\beta  }\nabla^{\alpha  }\Phi  \nabla^{\epsilon  
}H_{\alpha  }{}^{\gamma  \delta  } \nn\\&&+   c_{157}  
    R_{\alpha  \varepsilon  \gamma  }{}^{\mu  } 
R_{\beta  \mu  \delta  \epsilon  } \nabla^{\beta  
}\nabla^{\alpha  }\Phi  \nabla^{\varepsilon  }H^{\gamma  
\delta  \epsilon  } +   c_{158}  
    R_{\alpha  \gamma  \beta  }{}^{\mu  } 
R_{\delta  \varepsilon  \epsilon  \mu  } \nabla^{\beta 
 }\nabla^{\alpha  }\Phi  \nabla^{\varepsilon  }H^{\gamma  
\delta  \epsilon  },\nn
\eeqa
\beqa
 [H^2\nabla HR(\nabla\Phi)^2]_{12}&\!\!\!\!\!=\!\!\!\!\!&  c_{135}  
    H_{\alpha  }{}^{\gamma  \delta  } H^{\epsilon  \varepsilon  
\mu  } R_{\gamma  \mu  \delta  \zeta  } 
\nabla^{\alpha  }\Phi  \nabla^{\beta  }\Phi  
\nabla_{\varepsilon  }H_{\beta  \epsilon  }{}^{\zeta  } +   
   c_{144}  
    H_{\gamma  }{}^{\varepsilon  \mu  } H^{\gamma  \delta  
\epsilon  } R_{\alpha  \mu  \beta  \zeta  } 
\nabla^{\alpha  }\Phi  \nabla^{\beta  }\Phi  
\nabla_{\varepsilon  }H_{\delta  \epsilon  }{}^{\zeta  } \nn\\&\!\!\!\!\!\!\!\!\!\!\!\!\!\!& +   
   c_{312}  
    H_{\alpha  }{}^{\gamma  \delta  } H_{\gamma  }{}^{\epsilon  
\varepsilon  } R_{\epsilon  \mu  \varepsilon  \zeta  } 
\nabla^{\alpha  }\Phi  \nabla^{\beta  }\Phi  \nabla^{\zeta  
}H_{\beta  \delta  }{}^{\mu  } +   c_{313}  
    H_{\alpha  }{}^{\gamma  \delta  } H_{\gamma  \delta  
}{}^{\epsilon  } R_{\epsilon  \zeta  \varepsilon  \mu  
} \nabla^{\alpha  }\Phi  \nabla^{\beta  }\Phi  \nabla^{\zeta  
}H_{\beta  }{}^{\varepsilon  \mu  } \nn\\&\!\!\!\!\!\!\!\!\!\!\!\!\!\!& +   c_{314}  
    H_{\beta  }{}^{\epsilon  \varepsilon  } H^{\beta  \gamma  
\delta  } R_{\epsilon  \mu  \varepsilon  \zeta  } 
\nabla_{\alpha  }\Phi  \nabla^{\alpha  }\Phi  \nabla^{\zeta  
}H_{\gamma  \delta  }{}^{\mu  } +   c_{315}  
    H_{\alpha  }{}^{\gamma  \delta  } H_{\beta  }{}^{\epsilon  
\varepsilon  } R_{\epsilon  \mu  \varepsilon  \zeta  } 
\nabla^{\alpha  }\Phi  \nabla^{\beta  }\Phi  \nabla^{\zeta  
}H_{\gamma  \delta  }{}^{\mu  }   \nn\\&\!\!\!\!\!\!\!\!\!\!\!\!\!\!&+   c_{316}  
    H_{\beta  }{}^{\epsilon  \varepsilon  } H^{\beta  \gamma  
\delta  } R_{\delta  \mu  \varepsilon  \zeta  } 
\nabla_{\alpha  }\Phi  \nabla^{\alpha  }\Phi  \nabla^{\zeta  
}H_{\gamma  \epsilon  }{}^{\mu  } +   c_{317}  
    H_{\alpha  }{}^{\gamma  \delta  } H_{\beta  }{}^{\epsilon  
\varepsilon  } R_{\delta  \mu  \varepsilon  \zeta  } 
\nabla^{\alpha  }\Phi  \nabla^{\beta  }\Phi  \nabla^{\zeta  
}H_{\gamma  \epsilon  }{}^{\mu  } \nn\\&\!\!\!\!\!\!\!\!\!\!\!\!\!\!& +   c_{318}  
    H_{\alpha  }{}^{\gamma  \delta  } H_{\gamma  }{}^{\epsilon  
\varepsilon  } R_{\beta  \zeta  \varepsilon  \mu  } 
\nabla^{\alpha  }\Phi  \nabla^{\beta  }\Phi  \nabla^{\zeta  
}H_{\delta  \epsilon  }{}^{\mu  } +   c_{319}  
    H_{\beta  \gamma  }{}^{\epsilon  } H^{\beta  \gamma  
\delta  } R_{\epsilon  \zeta  \varepsilon  \mu  } 
\nabla_{\alpha  }\Phi  \nabla^{\alpha  }\Phi  \nabla^{\zeta  
}H_{\delta  }{}^{\varepsilon  \mu  }  \nn\\&\!\!\!\!\!\!\!\!\!\!\!\!\!\!&+   c_{320}  
    H_{\alpha  }{}^{\gamma  \delta  } H_{\beta  \gamma  
}{}^{\epsilon  } R_{\epsilon  \zeta  \varepsilon  \mu  
} \nabla^{\alpha  }\Phi  \nabla^{\beta  }\Phi  \nabla^{\zeta  
}H_{\delta  }{}^{\varepsilon  \mu  } +   c_{321}  
    H_{\alpha  }{}^{\gamma  \delta  } H_{\gamma  }{}^{\epsilon  
\varepsilon  } R_{\beta  \zeta  \delta  \mu  } 
\nabla^{\alpha  }\Phi  \nabla^{\beta  }\Phi  \nabla^{\zeta  
}H_{\epsilon  \varepsilon  }{}^{\mu  },\nn
\eeqa
\beqa
[H^2\nabla HR\nabla^2\Phi]_{27}&=&  
   c_{115}  
    H_{\alpha  }{}^{\gamma  \delta  } H^{\epsilon  \varepsilon  
\mu  } R_{\delta  \zeta  \varepsilon  \mu  } 
\nabla^{\beta  }\nabla^{\alpha  }\Phi  \nabla_{\epsilon  
}H_{\beta  \gamma  }{}^{\zeta  } +   c_{117}  
    H_{\alpha  }{}^{\gamma  \delta  } H_{\gamma  }{}^{\epsilon  
\varepsilon  } R_{\delta  \mu  \varepsilon  \zeta  } 
\nabla^{\beta  }\nabla^{\alpha  }\Phi  \nabla_{\epsilon  
}H_{\beta  }{}^{\mu  \zeta  } \nn\\&&+   c_{132}  
    H_{\gamma  }{}^{\varepsilon  \mu  } H^{\gamma  \delta  
\epsilon  } R_{\beta  \zeta  \epsilon  \mu  } \nabla^{
\beta  }\nabla^{\alpha  }\Phi  \nabla_{\varepsilon  }H_{\alpha 
 \delta  }{}^{\zeta  } +   c_{131}  
    H_{\gamma  }{}^{\varepsilon  \mu  } H^{\gamma  \delta  
\epsilon  } R_{\beta  \mu  \epsilon  \zeta  } \nabla^{
\beta  }\nabla^{\alpha  }\Phi  \nabla_{\varepsilon  }H_{\alpha 
 \delta  }{}^{\zeta  } \nn\\&& +   c_{136}  
    H_{\alpha  }{}^{\gamma  \delta  } H^{\epsilon  \varepsilon  
\mu  } R_{\gamma  \mu  \delta  \zeta  } \nabla^{\beta 
 }\nabla^{\alpha  }\Phi  \nabla_{\varepsilon  }H_{\beta  
\epsilon  }{}^{\zeta  } +   c_{141}  
    H_{\alpha  }{}^{\gamma  \delta  } H^{\epsilon  \varepsilon  
\mu  } R_{\beta  \zeta  \delta  \mu  } \nabla^{\beta  
}\nabla^{\alpha  }\Phi  \nabla_{\varepsilon  }H_{\gamma  
\epsilon  }{}^{\zeta  } \nn\\&& +   c_{140}  
    H_{\alpha  }{}^{\gamma  \delta  } H^{\epsilon  \varepsilon  
\mu  } R_{\beta  \mu  \delta  \zeta  } \nabla^{\beta  
}\nabla^{\alpha  }\Phi  \nabla_{\varepsilon  }H_{\gamma  
\epsilon  }{}^{\zeta  } +   c_{146}  
    H_{\gamma  }{}^{\varepsilon  \mu  } H^{\gamma  \delta  
\epsilon  } R_{\alpha  \mu  \beta  \zeta  } 
\nabla^{\beta  }\nabla^{\alpha  }\Phi  \nabla_{\varepsilon  
}H_{\delta  \epsilon  }{}^{\zeta  } \nn\\&& +   c_{339}  
    H_{\gamma  \delta  }{}^{\varepsilon  } H^{\gamma  \delta  
\epsilon  } R_{\beta  \mu  \varepsilon  \zeta  } 
\nabla^{\beta  }\nabla^{\alpha  }\Phi  \nabla^{\zeta  
}H_{\alpha  \epsilon  }{}^{\mu  } +   c_{340}  
    H_{\alpha  }{}^{\gamma  \delta  } H^{\epsilon  \varepsilon  
\mu  } R_{\delta  \zeta  \varepsilon  \mu  } 
\nabla^{\beta  }\nabla^{\alpha  }\Phi  \nabla^{\zeta  
}H_{\beta  \gamma  \epsilon  }  \nn\\&&+   c_{341}  
    H_{\alpha  }{}^{\gamma  \delta  } H_{\gamma  }{}^{\epsilon  
\varepsilon  } R_{\epsilon  \mu  \varepsilon  \zeta  } 
\nabla^{\beta  }\nabla^{\alpha  }\Phi  \nabla^{\zeta  
}H_{\beta  \delta  }{}^{\mu  } +   c_{342}  
    H_{\alpha  }{}^{\gamma  \delta  } H^{\epsilon  \varepsilon  
\mu  } R_{\gamma  \mu  \delta  \zeta  } \nabla^{\beta 
 }\nabla^{\alpha  }\Phi  \nabla^{\zeta  }H_{\beta  \epsilon  
\varepsilon  } \nn\\&& +   c_{344}  
    H_{\alpha  }{}^{\gamma  \delta  } H_{\gamma  }{}^{\epsilon  
\varepsilon  } R_{\delta  \zeta  \varepsilon  \mu  } 
\nabla^{\beta  }\nabla^{\alpha  }\Phi  \nabla^{\zeta  
}H_{\beta  \epsilon  }{}^{\mu  } +   c_{343}  
    H_{\alpha  }{}^{\gamma  \delta  } H_{\gamma  }{}^{\epsilon  
\varepsilon  } R_{\delta  \mu  \varepsilon  \zeta  } 
\nabla^{\beta  }\nabla^{\alpha  }\Phi  \nabla^{\zeta  
}H_{\beta  \epsilon  }{}^{\mu  }  \nn\\&&+   c_{345}  
    H_{\alpha  }{}^{\gamma  \delta  } H_{\gamma  \delta  
}{}^{\epsilon  } R_{\epsilon  \zeta  \varepsilon  \mu  
} \nabla^{\beta  }\nabla^{\alpha  }\Phi  \nabla^{\zeta  
}H_{\beta  }{}^{\varepsilon  \mu  } +   c_{346}  
    H_{\alpha  }{}^{\gamma  \delta  } H^{\epsilon  \varepsilon  
\mu  } R_{\beta  \zeta  \varepsilon  \mu  } 
\nabla^{\beta  }\nabla^{\alpha  }\Phi  \nabla^{\zeta  
}H_{\gamma  \delta  \epsilon  } \nn\\&& +   c_{347}  
    H_{\alpha  }{}^{\gamma  \delta  } H_{\beta  }{}^{\epsilon  
\varepsilon  } R_{\epsilon  \mu  \varepsilon  \zeta  } 
\nabla^{\beta  }\nabla^{\alpha  }\Phi  \nabla^{\zeta  
}H_{\gamma  \delta  }{}^{\mu  } +   c_{349}  
    H_{\alpha  }{}^{\gamma  \delta  } H^{\epsilon  \varepsilon  
\mu  } R_{\beta  \zeta  \delta  \mu  } \nabla^{\beta  
}\nabla^{\alpha  }\Phi  \nabla^{\zeta  }H_{\gamma  \epsilon  
\varepsilon  } \nn\\&& +   c_{348}  
    H_{\alpha  }{}^{\gamma  \delta  } H^{\epsilon  \varepsilon  
\mu  } R_{\beta  \mu  \delta  \zeta  } \nabla^{\beta  
}\nabla^{\alpha  }\Phi  \nabla^{\zeta  }H_{\gamma  \epsilon  
\varepsilon  } +   c_{350}  
    H_{\alpha  }{}^{\gamma  \delta  } H_{\beta  }{}^{\epsilon  
\varepsilon  } R_{\delta  \mu  \varepsilon  \zeta  } 
\nabla^{\beta  }\nabla^{\alpha  }\Phi  \nabla^{\zeta  
}H_{\gamma  \epsilon  }{}^{\mu  }  \nn\\&&+   c_{351}  
    H_{\gamma  }{}^{\varepsilon  \mu  } H^{\gamma  \delta  
\epsilon  } R_{\alpha  \mu  \beta  \zeta  } 
\nabla^{\beta  }\nabla^{\alpha  }\Phi  \nabla^{\zeta  
}H_{\delta  \epsilon  \varepsilon  } +   c_{353}  
    H_{\alpha  }{}^{\gamma  \delta  } H_{\gamma  }{}^{\epsilon  
\varepsilon  } R_{\beta  \zeta  \varepsilon  \mu  } 
\nabla^{\beta  }\nabla^{\alpha  }\Phi  \nabla^{\zeta  
}H_{\delta  \epsilon  }{}^{\mu  }  \nn\\&&+   c_{352}  
    H_{\alpha  }{}^{\gamma  \delta  } H_{\gamma  }{}^{\epsilon  
\varepsilon  } R_{\beta  \mu  \varepsilon  \zeta  } 
\nabla^{\beta  }\nabla^{\alpha  }\Phi  \nabla^{\zeta  
}H_{\delta  \epsilon  }{}^{\mu  } +   c_{354}  
    H_{\alpha  }{}^{\gamma  \delta  } H_{\beta  \gamma  
}{}^{\epsilon  } R_{\epsilon  \zeta  \varepsilon  \mu  
} \nabla^{\beta  }\nabla^{\alpha  }\Phi  \nabla^{\zeta  
}H_{\delta  }{}^{\varepsilon  \mu  } \nn\\&& +   c_{356}  
    H_{\alpha  }{}^{\gamma  \delta  } H_{\gamma  }{}^{\epsilon  
\varepsilon  } R_{\beta  \zeta  \delta  \mu  } 
\nabla^{\beta  }\nabla^{\alpha  }\Phi  \nabla^{\zeta  
}H_{\epsilon  \varepsilon  }{}^{\mu  } +   c_{355}  
    H_{\alpha  }{}^{\gamma  \delta  } H_{\gamma  }{}^{\epsilon  
\varepsilon  } R_{\beta  \mu  \delta  \zeta  } 
\nabla^{\beta  }\nabla^{\alpha  }\Phi  \nabla^{\zeta  
}H_{\epsilon  \varepsilon  }{}^{\mu  } \nn\\&& +   c_{357}  
    H_{\alpha  }{}^{\gamma  \delta  } H_{\gamma  \delta  
}{}^{\epsilon  } R_{\beta  \zeta  \varepsilon  \mu  } 
\nabla^{\beta  }\nabla^{\alpha  }\Phi  \nabla^{\zeta  
}H_{\epsilon  }{}^{\varepsilon  \mu  },\nn
\eeqa

\beqa
[H^2\nabla H(\nabla\Phi)^4]_1&=&  c_{169}  
    H_{\beta  }{}^{\delta  \epsilon  } H_{\delta  
}{}^{\varepsilon  \mu  } \nabla_{\alpha  }\Phi  \nabla^{\alpha 
 }\Phi  \nabla^{\beta  }\Phi  \nabla^{\gamma  }\Phi  \nabla_{
\mu  }H_{\gamma  \epsilon  \varepsilon  },\nn
\eeqa
\beqa
&&[H^2\nabla H(\nabla\Phi)^2\nabla^2\Phi]_{16}=\nn\\&&  
   c_{106}  
    H_{\gamma  }{}^{\epsilon  \varepsilon  } H_{\epsilon  
\varepsilon  }{}^{\mu  } \nabla^{\alpha  }\Phi  \nabla_{\beta  
}H_{\alpha  \delta  \mu  } \nabla^{\beta  }\Phi  
\nabla^{\delta  }\nabla^{\gamma  }\Phi  +   c_{110}  
    H_{\alpha  }{}^{\epsilon  \varepsilon  } H_{\epsilon  
\varepsilon  }{}^{\mu  } \nabla^{\alpha  }\Phi  \nabla^{\beta  
}\Phi  \nabla_{\delta  }H_{\beta  \gamma  \mu  } 
\nabla^{\delta  }\nabla^{\gamma  }\Phi  \nn\\&\!\!\!\!\!\!\!\!\!\!\!\!&+   c_{112}  
    H_{\alpha  }{}^{\epsilon  \varepsilon  } H_{\gamma  
\epsilon  }{}^{\mu  } \nabla^{\alpha  }\Phi  \nabla^{\beta  
}\Phi  \nabla_{\delta  }H_{\beta  \varepsilon  \mu  } \nabla^{
\delta  }\nabla^{\gamma  }\Phi  +   c_{113}  
    H_{\alpha  \gamma  }{}^{\epsilon  } H_{\epsilon  
}{}^{\varepsilon  \mu  } \nabla^{\alpha  }\Phi  \nabla^{\beta  
}\Phi  \nabla_{\delta  }H_{\beta  \varepsilon  \mu  } \nabla^{
\delta  }\nabla^{\gamma  }\Phi \nn\\&\!\!\!\!\!\!\!\!\!\!\!\!& +   c_{116}  
    H_{\alpha  \gamma  }{}^{\epsilon  } H_{\delta  
}{}^{\varepsilon  \mu  } \nabla^{\alpha  }\Phi  \nabla^{\beta  
}\Phi  \nabla^{\delta  }\nabla^{\gamma  }\Phi  
\nabla_{\epsilon  }H_{\beta  \varepsilon  \mu  } +   
   c_{119}  
    H_{\beta  }{}^{\delta  \epsilon  } H_{\delta  
}{}^{\varepsilon  \mu  } \nabla_{\alpha  }\Phi  \nabla^{\alpha 
 }\Phi  \nabla^{\gamma  }\nabla^{\beta  }\Phi  
\nabla_{\epsilon  }H_{\gamma  \varepsilon  \mu  }  \nn\\&\!\!\!\!\!\!\!\!\!\!\!\!&+   
   c_{121}  
    H_{\alpha  \gamma  }{}^{\epsilon  } H_{\beta  
}{}^{\varepsilon  \mu  } \nabla^{\alpha  }\Phi  \nabla^{\beta  
}\Phi  \nabla^{\delta  }\nabla^{\gamma  }\Phi  
\nabla_{\epsilon  }H_{\delta  \varepsilon  \mu  } +   
   c_{134}  
    H_{\alpha  }{}^{\epsilon  \varepsilon  } H_{\gamma  
\epsilon  }{}^{\mu  } \nabla^{\alpha  }\Phi  \nabla^{\beta  
}\Phi  \nabla^{\delta  }\nabla^{\gamma  }\Phi  
\nabla_{\varepsilon  }H_{\beta  \delta  \mu  } \nn\\&\!\!\!\!\!\!\!\!\!\!\!\!& +   
   c_{161}  
    H_{\delta  \epsilon  }{}^{\mu  } H^{\delta  \epsilon  
\varepsilon  } \nabla^{\alpha  }\Phi  \nabla^{\beta  }\Phi  
\nabla^{\gamma  }\nabla_{\alpha  }\Phi  \nabla_{\mu  
}H_{\beta  \gamma  \varepsilon  } +   c_{162}  
    H_{\alpha  }{}^{\epsilon  \varepsilon  } H_{\gamma  
\epsilon  }{}^{\mu  } \nabla^{\alpha  }\Phi  \nabla^{\beta  
}\Phi  \nabla^{\delta  }\nabla^{\gamma  }\Phi  \nabla_{\mu  
}H_{\beta  \delta  \varepsilon  }  \nn\\&\!\!\!\!\!\!\!\!\!\!\!\!&+   c_{163}  
    H_{\alpha  \gamma  }{}^{\epsilon  } H_{\epsilon  
}{}^{\varepsilon  \mu  } \nabla^{\alpha  }\Phi  \nabla^{\beta  
}\Phi  \nabla^{\delta  }\nabla^{\gamma  }\Phi  \nabla_{\mu  
}H_{\beta  \delta  \varepsilon  } +   c_{166}  
    H_{\alpha  \gamma  }{}^{\epsilon  } H_{\delta  
}{}^{\varepsilon  \mu  } \nabla^{\alpha  }\Phi  \nabla^{\beta  
}\Phi  \nabla^{\delta  }\nabla^{\gamma  }\Phi  \nabla_{\mu  
}H_{\beta  \epsilon  \varepsilon  }  \nn\\&\!\!\!\!\!\!\!\!\!\!\!\!&+   c_{170}  
    H_{\beta  }{}^{\delta  \epsilon  } H_{\delta  
}{}^{\varepsilon  \mu  } \nabla^{\alpha  }\Phi  \nabla^{\beta  
}\Phi  \nabla^{\gamma  }\nabla_{\alpha  }\Phi  \nabla_{\mu  
}H_{\gamma  \epsilon  \varepsilon  } +   c_{172}  
    H_{\beta  }{}^{\delta  \epsilon  } H_{\delta  
}{}^{\varepsilon  \mu  } \nabla_{\alpha  }\Phi  \nabla^{\alpha 
 }\Phi  \nabla^{\gamma  }\nabla^{\beta  }\Phi  \nabla_{\mu  
}H_{\gamma  \epsilon  \varepsilon  } \nn\\&\!\!\!\!\!\!\!\!\!\!\!\!& +   c_{177}  
    H_{\beta  }{}^{\delta  \epsilon  } H_{\gamma  
}{}^{\varepsilon  \mu  } \nabla^{\alpha  }\Phi  \nabla^{\beta  
}\Phi  \nabla^{\gamma  }\nabla_{\alpha  }\Phi  \nabla_{\mu  
}H_{\delta  \epsilon  \varepsilon  } +   c_{180}  
    H_{\alpha  \gamma  }{}^{\epsilon  } H_{\beta  
}{}^{\varepsilon  \mu  } \nabla^{\alpha  }\Phi  \nabla^{\beta  
}\Phi  \nabla^{\delta  }\nabla^{\gamma  }\Phi  \nabla_{\mu  
}H_{\delta  \epsilon  \varepsilon  },\nn 
\eeqa
\beqa
&&[H^2\nabla H(\nabla^2\Phi)^2]_7=\nn\\&& 
   c_{111}  
    H_{\alpha  }{}^{\epsilon  \varepsilon  } H_{\epsilon  
\varepsilon  }{}^{\mu  } \nabla^{\beta  }\nabla^{\alpha  }\Phi 
 \nabla_{\delta  }H_{\beta  \gamma  \mu  } \nabla^{\delta  
}\nabla^{\gamma  }\Phi  +   c_{118}  
    H_{\beta  }{}^{\delta  \epsilon  } H_{\delta  
}{}^{\varepsilon  \mu  } \nabla^{\beta  }\nabla^{\alpha  }\Phi 
 \nabla^{\gamma  }\nabla_{\alpha  }\Phi  \nabla_{\epsilon  
}H_{\gamma  \varepsilon  \mu  } \nn\\&&+   c_{122}  
    H_{\alpha  \gamma  }{}^{\epsilon  } H_{\beta  
}{}^{\varepsilon  \mu  } \nabla^{\beta  }\nabla^{\alpha  }\Phi 
 \nabla^{\delta  }\nabla^{\gamma  }\Phi  \nabla_{\epsilon  
}H_{\delta  \varepsilon  \mu  } +   c_{164}  
    H_{\alpha  }{}^{\epsilon  \varepsilon  } H_{\gamma  
\epsilon  }{}^{\mu  } \nabla^{\beta  }\nabla^{\alpha  }\Phi  
\nabla^{\delta  }\nabla^{\gamma  }\Phi  \nabla_{\mu  
}H_{\beta  \delta  \varepsilon  } \nn\\&&+   c_{165}  
    H_{\alpha  \gamma  }{}^{\epsilon  } H_{\epsilon  
}{}^{\varepsilon  \mu  } \nabla^{\beta  }\nabla^{\alpha  }\Phi 
 \nabla^{\delta  }\nabla^{\gamma  }\Phi  \nabla_{\mu  
}H_{\beta  \delta  \varepsilon  } +   c_{171}  
    H_{\beta  }{}^{\delta  \epsilon  } H_{\delta  
}{}^{\varepsilon  \mu  } \nabla^{\beta  }\nabla^{\alpha  }\Phi 
 \nabla^{\gamma  }\nabla_{\alpha  }\Phi  \nabla_{\mu  
}H_{\gamma  \epsilon  \varepsilon  } \nn\\&&+   c_{181}  
    H_{\alpha  \gamma  }{}^{\epsilon  } H_{\beta  
}{}^{\varepsilon  \mu  } \nabla^{\beta  }\nabla^{\alpha  }\Phi 
 \nabla^{\delta  }\nabla^{\gamma  }\Phi  \nabla_{\mu  
}H_{\delta  \epsilon  \varepsilon  },\nn
\eeqa
\beqa
&&[H^4\nabla H(\nabla\Phi)^2]_{23}=\nn \\&&
   c_{145}  
    H_{\alpha  }{}^{\gamma  \delta  } H_{\beta  }{}^{\epsilon  
\varepsilon  } H_{\gamma  }{}^{\mu  \zeta  } H_{\mu  \zeta  
}{}^{\eta  } \nabla^{\alpha  }\Phi  \nabla^{\beta  }\Phi  
\nabla_{\varepsilon  }H_{\delta  \epsilon  \eta  } +   
   c_{322}  
    H_{\alpha  }{}^{\gamma  \delta  } H_{\gamma  }{}^{\epsilon  
\varepsilon  } H_{\delta  }{}^{\mu  \zeta  } H_{\epsilon  
\varepsilon  }{}^{\eta  } \nabla^{\alpha  }\Phi  \nabla^{\beta 
 }\Phi  \nabla_{\zeta  }H_{\beta  \mu  \eta  } \nn\\&\!\!\!\!\!\!\!\!\!\!&+   
   c_{323}  
    H_{\alpha  }{}^{\gamma  \delta  } H_{\gamma  }{}^{\epsilon  
\varepsilon  } H_{\epsilon  }{}^{\mu  \zeta  } H_{\mu  \zeta  
}{}^{\eta  } \nabla^{\alpha  }\Phi  \nabla^{\beta  }\Phi  
\nabla_{\eta  }H_{\beta  \delta  \varepsilon  } +   
   c_{324}  
    H_{\alpha  }{}^{\gamma  \delta  } H_{\gamma  }{}^{\epsilon  
\varepsilon  } H_{\epsilon  }{}^{\mu  \zeta  } H_{\varepsilon  
\mu  }{}^{\eta  } \nabla^{\alpha  }\Phi  \nabla^{\beta  }\Phi 
 \nabla_{\eta  }H_{\beta  \delta  \zeta  } \nn\\&\!\!\!\!\!\!\!\!\!\!&+   c_{325}  
    H_{\alpha  }{}^{\gamma  \delta  } H_{\gamma  }{}^{\epsilon  
\varepsilon  } H_{\epsilon  \varepsilon  }{}^{\mu  } H_{\mu  
}{}^{\zeta  \eta  } \nabla^{\alpha  }\Phi  \nabla^{\beta  
}\Phi  \nabla_{\eta  }H_{\beta  \delta  \zeta  } +   
   c_{326}  
    H_{\alpha  }{}^{\gamma  \delta  } H_{\gamma  }{}^{\epsilon  
\varepsilon  } H_{\delta  }{}^{\mu  \zeta  } H_{\epsilon  
\varepsilon  }{}^{\eta  } \nabla^{\alpha  }\Phi  \nabla^{\beta 
 }\Phi  \nabla_{\eta  }H_{\beta  \mu  \zeta  }\nn\\&\!\!\!\!\!\!\!\!\!\!& +   
   c_{327}  
    H_{\alpha  }{}^{\gamma  \delta  } H_{\gamma  }{}^{\epsilon  
\varepsilon  } H_{\delta  \epsilon  }{}^{\mu  } H_{\varepsilon  
}{}^{\zeta  \eta  } \nabla^{\alpha  }\Phi  \nabla^{\beta  
}\Phi  \nabla_{\eta  }H_{\beta  \mu  \zeta  } +   
   c_{328}  
    H_{\alpha  }{}^{\gamma  \delta  } H_{\gamma  \delta  
}{}^{\epsilon  } H_{\epsilon  }{}^{\varepsilon  \mu  } 
H_{\varepsilon  }{}^{\zeta  \eta  } \nabla^{\alpha  }\Phi  
\nabla^{\beta  }\Phi  \nabla_{\eta  }H_{\beta  \mu  \zeta  } \nn\\&\!\!\!\!\!\!\!\!\!\!&
+   c_{329}  
    H_{\alpha  }{}^{\gamma  \delta  } H_{\beta  }{}^{\epsilon  
\varepsilon  } H_{\gamma  }{}^{\mu  \zeta  } H_{\mu  \zeta  
}{}^{\eta  } \nabla^{\alpha  }\Phi  \nabla^{\beta  }\Phi  
\nabla_{\eta  }H_{\delta  \epsilon  \varepsilon  } +   
   c_{330}  
    H_{\alpha  }{}^{\gamma  \delta  } H_{\beta  }{}^{\epsilon  
\varepsilon  } H_{\gamma  }{}^{\mu  \zeta  } H_{\epsilon  \mu  
}{}^{\eta  } \nabla^{\alpha  }\Phi  \nabla^{\beta  }\Phi  
\nabla_{\eta  }H_{\delta  \varepsilon  \zeta  }\nn\\&\!\!\!\!\!\!\!\!\!\!& +   
   c_{331}  
    H_{\alpha  }{}^{\gamma  \delta  } H_{\beta  }{}^{\epsilon  
\varepsilon  } H_{\gamma  \epsilon  }{}^{\mu  } H_{\mu  
}{}^{\zeta  \eta  } \nabla^{\alpha  }\Phi  \nabla^{\beta  
}\Phi  \nabla_{\eta  }H_{\delta  \varepsilon  \zeta  } +   
   c_{332}  
    H_{\alpha  }{}^{\gamma  \delta  } H_{\beta  }{}^{\epsilon  
\varepsilon  } H_{\gamma  }{}^{\mu  \zeta  } H_{\delta  \mu  
}{}^{\eta  } \nabla^{\alpha  }\Phi  \nabla^{\beta  }\Phi  
\nabla_{\eta  }H_{\epsilon  \varepsilon  \zeta  }\nn\\&\!\!\!\!\!\!\!\!\!\!& +   
   c_{333}  
    H_{\alpha  }{}^{\gamma  \delta  } H_{\beta  }{}^{\epsilon  
\varepsilon  } H_{\gamma  \delta  }{}^{\mu  } H_{\mu  
}{}^{\zeta  \eta  } \nabla^{\alpha  }\Phi  \nabla^{\beta  
}\Phi  \nabla_{\eta  }H_{\epsilon  \varepsilon  \zeta  } +   
   c_{334}  
    H_{\beta  \gamma  }{}^{\epsilon  } H^{\beta  \gamma  
\delta  } H_{\delta  }{}^{\varepsilon  \mu  } H_{\varepsilon  
}{}^{\zeta  \eta  } \nabla_{\alpha  }\Phi  \nabla^{\alpha  
}\Phi  \nabla_{\eta  }H_{\epsilon  \mu  \zeta  }\nn\\&\!\!\!\!\!\!\!\!\!\!& +   
   c_{335}  
    H_{\alpha  }{}^{\gamma  \delta  } H_{\beta  \gamma  
}{}^{\epsilon  } H_{\delta  }{}^{\varepsilon  \mu  } 
H_{\varepsilon  }{}^{\zeta  \eta  } \nabla^{\alpha  }\Phi  
\nabla^{\beta  }\Phi  \nabla_{\eta  }H_{\epsilon  \mu  \zeta  
} +   c_{336}  
    H_{\beta  }{}^{\epsilon  \varepsilon  } H^{\beta  \gamma  
\delta  } H_{\gamma  \epsilon  }{}^{\mu  } H_{\delta  
}{}^{\zeta  \eta  } \nabla_{\alpha  }\Phi  \nabla^{\alpha  
}\Phi  \nabla_{\eta  }H_{\varepsilon  \mu  \zeta  }\nn\\&\!\!\!\!\!\!\!\!\!\!& +   
   c_{337}  
    H_{\alpha  }{}^{\gamma  \delta  } H_{\beta  }{}^{\epsilon  
\varepsilon  } H_{\gamma  \epsilon  }{}^{\mu  } H_{\delta  }{}^{
\zeta  \eta  } \nabla^{\alpha  }\Phi  \nabla^{\beta  }\Phi  
\nabla_{\eta  }H_{\varepsilon  \mu  \zeta  } +   c_{338} 
     H_{\alpha  }{}^{\gamma  \delta  } H_{\beta  }{}^{\epsilon  
\varepsilon  } H_{\gamma  \delta  }{}^{\mu  } H_{\epsilon  }{}^{
\zeta  \eta  } \nabla^{\alpha  }\Phi  \nabla^{\beta  }\Phi  
\nabla_{\eta  }H_{\varepsilon  \mu  \zeta  }\nn\\&\!\!\!\!\!\!\!\!\!\!& +   c_{167} 
     H_{\alpha  }{}^{\gamma  \delta  } H_{\gamma  \delta  }{}^{
\epsilon  } H_{\epsilon  }{}^{\varepsilon  \mu  } H_{\varepsilon 
 }{}^{\zeta  \eta  } \nabla^{\alpha  }\Phi  \nabla^{\beta  
}\Phi  \nabla_{\mu  }H_{\beta  \zeta  \eta  } +   
   c_{185}  
    H_{\beta  \gamma  }{}^{\epsilon  } H^{\beta  \gamma  
\delta  } H_{\delta  }{}^{\varepsilon  \mu  } H_{\varepsilon  
}{}^{\zeta  \eta  } \nabla_{\alpha  }\Phi  \nabla^{\alpha  
}\Phi  \nabla_{\mu  }H_{\epsilon  \zeta  \eta  }\nn\\&\!\!\!\!\!\!\!\!\!\!& +   
   c_{186}  
    H_{\alpha  }{}^{\gamma  \delta  } H_{\beta  \gamma  
}{}^{\epsilon  } H_{\delta  }{}^{\varepsilon  \mu  } 
H_{\varepsilon  }{}^{\zeta  \eta  } \nabla^{\alpha  }\Phi  
\nabla^{\beta  }\Phi  \nabla_{\mu  }H_{\epsilon  \zeta  \eta  
} +   c_{189}  
    H_{\alpha  }{}^{\gamma  \delta  } H_{\beta  }{}^{\epsilon  
\varepsilon  } H_{\gamma  \epsilon  }{}^{\mu  } H_{\delta  }{}^{
\zeta  \eta  } \nabla^{\alpha  }\Phi  \nabla^{\beta  }\Phi  
\nabla_{\mu  }H_{\varepsilon  \zeta  \eta  } \nn\\&\!\!\!\!\!\!\!\!\!\!&+   c_{190} 
     H_{\alpha  }{}^{\gamma  \delta  } H_{\beta  }{}^{\epsilon  
\varepsilon  } H_{\gamma  \delta  }{}^{\mu  } H_{\epsilon  }{}^{
\zeta  \eta  } \nabla^{\alpha  }\Phi  \nabla^{\beta  }\Phi  
\nabla_{\mu  }H_{\varepsilon  \zeta  \eta  },\nn
\eeqa
\beqa
&&[H^4\nabla H\nabla^2\Phi]_{21}=\nn\\&&  c_{133}  
    H_{\alpha  }{}^{\gamma  \delta  } H_{\gamma  }{}^{\epsilon  
\varepsilon  } H_{\mu  \zeta  \eta  } H^{\mu  \zeta  \eta  } 
\nabla^{\beta  }\nabla^{\alpha  }\Phi  \nabla_{\varepsilon  
}H_{\beta  \delta  \epsilon  } +   c_{147}  
    H_{\alpha  }{}^{\gamma  \delta  } H_{\beta  }{}^{\epsilon  
\varepsilon  } H_{\gamma  }{}^{\mu  \zeta  } H_{\mu  \zeta  
}{}^{\eta  } \nabla^{\beta  }\nabla^{\alpha  }\Phi  
\nabla_{\varepsilon  }H_{\delta  \epsilon  \eta  }\nn\\&& +   
   c_{358}  
    H_{\alpha  }{}^{\gamma  \delta  } H_{\gamma  }{}^{\epsilon  
\varepsilon  } H_{\delta  }{}^{\mu  \zeta  } H_{\epsilon  
\varepsilon  }{}^{\eta  } \nabla^{\beta  }\nabla^{\alpha  
}\Phi  \nabla_{\zeta  }H_{\beta  \mu  \eta  } +   
   c_{359}  
    H_{\alpha  }{}^{\gamma  \delta  } H_{\gamma  }{}^{\epsilon  
\varepsilon  } H_{\epsilon  }{}^{\mu  \zeta  } H_{\mu  \zeta  
}{}^{\eta  } \nabla^{\beta  }\nabla^{\alpha  }\Phi  
\nabla_{\eta  }H_{\beta  \delta  \varepsilon  } \nn\\&&+   
   c_{360}  
    H_{\alpha  }{}^{\gamma  \delta  } H_{\gamma  }{}^{\epsilon  
\varepsilon  } H_{\epsilon  }{}^{\mu  \zeta  } H_{\varepsilon  
\mu  }{}^{\eta  } \nabla^{\beta  }\nabla^{\alpha  }\Phi  
\nabla_{\eta  }H_{\beta  \delta  \zeta  } +   c_{361}  
    H_{\alpha  }{}^{\gamma  \delta  } H_{\gamma  }{}^{\epsilon  
\varepsilon  } H_{\epsilon  \varepsilon  }{}^{\mu  } H_{\mu  
}{}^{\zeta  \eta  } \nabla^{\beta  }\nabla^{\alpha  }\Phi  
\nabla_{\eta  }H_{\beta  \delta  \zeta  } \nn\\&&+   c_{362}  
    H_{\alpha  }{}^{\gamma  \delta  } H_{\gamma  }{}^{\epsilon  
\varepsilon  } H_{\delta  }{}^{\mu  \zeta  } H_{\epsilon  
\varepsilon  }{}^{\eta  } \nabla^{\beta  }\nabla^{\alpha  
}\Phi  \nabla_{\eta  }H_{\beta  \mu  \zeta  } +   
   c_{363}  
    H_{\alpha  }{}^{\gamma  \delta  } H_{\gamma  }{}^{\epsilon  
\varepsilon  } H_{\delta  \epsilon  }{}^{\mu  } H_{\varepsilon  
}{}^{\zeta  \eta  } \nabla^{\beta  }\nabla^{\alpha  }\Phi  
\nabla_{\eta  }H_{\beta  \mu  \zeta  }\nn\\&& +   c_{364}  
    H_{\alpha  }{}^{\gamma  \delta  } H_{\gamma  \delta  
}{}^{\epsilon  } H_{\epsilon  }{}^{\varepsilon  \mu  } 
H_{\varepsilon  }{}^{\zeta  \eta  } \nabla^{\beta  
}\nabla^{\alpha  }\Phi  \nabla_{\eta  }H_{\beta  \mu  \zeta  
} +   c_{365}  
    H_{\alpha  }{}^{\gamma  \delta  } H_{\beta  }{}^{\epsilon  
\varepsilon  } H_{\gamma  }{}^{\mu  \zeta  } H_{\mu  \zeta  
}{}^{\eta  } \nabla^{\beta  }\nabla^{\alpha  }\Phi  
\nabla_{\eta  }H_{\delta  \epsilon  \varepsilon  } \nn\\&&+   
   c_{366}  
    H_{\alpha  }{}^{\gamma  \delta  } H_{\beta  }{}^{\epsilon  
\varepsilon  } H_{\gamma  }{}^{\mu  \zeta  } H_{\epsilon  \mu  
}{}^{\eta  } \nabla^{\beta  }\nabla^{\alpha  }\Phi  
\nabla_{\eta  }H_{\delta  \varepsilon  \zeta  } +   
   c_{367}  
    H_{\alpha  }{}^{\gamma  \delta  } H_{\beta  }{}^{\epsilon  
\varepsilon  } H_{\gamma  \epsilon  }{}^{\mu  } H_{\mu  
}{}^{\zeta  \eta  } \nabla^{\beta  }\nabla^{\alpha  }\Phi  
\nabla_{\eta  }H_{\delta  \varepsilon  \zeta  }\nn\\&& +   
   c_{368}  
    H_{\alpha  }{}^{\gamma  \delta  } H_{\beta  }{}^{\epsilon  
\varepsilon  } H_{\gamma  }{}^{\mu  \zeta  } H_{\delta  \mu  
}{}^{\eta  } \nabla^{\beta  }\nabla^{\alpha  }\Phi  
\nabla_{\eta  }H_{\epsilon  \varepsilon  \zeta  } +   
   c_{369}  
    H_{\alpha  }{}^{\gamma  \delta  } H_{\beta  }{}^{\epsilon  
\varepsilon  } H_{\gamma  \delta  }{}^{\mu  } H_{\mu  
}{}^{\zeta  \eta  } \nabla^{\beta  }\nabla^{\alpha  }\Phi  
\nabla_{\eta  }H_{\epsilon  \varepsilon  \zeta  }\nn\\&& +   
   c_{370}  
    H_{\alpha  }{}^{\gamma  \delta  } H_{\beta  \gamma  
}{}^{\epsilon  } H_{\delta  }{}^{\varepsilon  \mu  } 
H_{\varepsilon  }{}^{\zeta  \eta  } \nabla^{\beta  
}\nabla^{\alpha  }\Phi  \nabla_{\eta  }H_{\epsilon  \mu  
\zeta  } +   c_{371}  
    H_{\alpha  }{}^{\gamma  \delta  } H_{\beta  }{}^{\epsilon  
\varepsilon  } H_{\gamma  \epsilon  }{}^{\mu  } H_{\delta  }{}^{
\zeta  \eta  } \nabla^{\beta  }\nabla^{\alpha  }\Phi  
\nabla_{\eta  }H_{\varepsilon  \mu  \zeta  } \nn\\&&+   c_{372} 
     H_{\alpha  }{}^{\gamma  \delta  } H_{\beta  }{}^{\epsilon  
\varepsilon  } H_{\gamma  \delta  }{}^{\mu  } H_{\epsilon  }{}^{
\zeta  \eta  } \nabla^{\beta  }\nabla^{\alpha  }\Phi  
\nabla_{\eta  }H_{\varepsilon  \mu  \zeta  } +   c_{168} 
     H_{\alpha  }{}^{\gamma  \delta  } H_{\gamma  \delta  }{}^{
\epsilon  } H_{\epsilon  }{}^{\varepsilon  \mu  } H_{\varepsilon 
 }{}^{\zeta  \eta  } \nabla^{\beta  }\nabla^{\alpha  }\Phi  
\nabla_{\mu  }H_{\beta  \zeta  \eta  }\nn\\&& +   c_{187}  
    H_{\alpha  }{}^{\gamma  \delta  } H_{\beta  \gamma  
}{}^{\epsilon  } H_{\delta  }{}^{\varepsilon  \mu  } 
H_{\varepsilon  }{}^{\zeta  \eta  } \nabla^{\beta  
}\nabla^{\alpha  }\Phi  \nabla_{\mu  }H_{\epsilon  \zeta  
\eta  } +   c_{191}  
    H_{\alpha  }{}^{\gamma  \delta  } H_{\beta  }{}^{\epsilon  
\varepsilon  } H_{\gamma  \epsilon  }{}^{\mu  } H_{\delta  }{}^{
\zeta  \eta  } \nabla^{\beta  }\nabla^{\alpha  }\Phi  
\nabla_{\mu  }H_{\varepsilon  \zeta  \eta  } \nn\\&&+   c_{192} 
     H_{\alpha  }{}^{\gamma  \delta  } H_{\beta  }{}^{\epsilon  
\varepsilon  } H_{\gamma  \delta  }{}^{\mu  } H_{\epsilon  }{}^{
\zeta  \eta  } \nabla^{\beta  }\nabla^{\alpha  }\Phi  
\nabla_{\mu  }H_{\varepsilon  \zeta  \eta  },\nn
\eeqa
\beqa
&&[H^3(\nabla H)^2\nabla\Phi]_{34}=\nn\\&&  c_{279}  
    H_{\beta  }{}^{\epsilon  \varepsilon  } H^{\beta  \gamma  
\delta  } H_{\gamma  }{}^{\mu  \zeta  } \nabla^{\alpha  }\Phi  
\nabla_{\epsilon  }H_{\alpha  \delta  }{}^{\eta  } 
\nabla_{\eta  }H_{\varepsilon  \mu  \zeta  } +   c_{280} 
     H_{\beta  \gamma  }{}^{\epsilon  } H^{\beta  \gamma  
\delta  } H_{\delta  }{}^{\varepsilon  \mu  } \nabla^{\alpha  }
\Phi  \nabla_{\epsilon  }H_{\alpha  }{}^{\zeta  \eta  } 
\nabla_{\eta  }H_{\varepsilon  \mu  \zeta  }  \nn\\&&+   c_{281} 
     H_{\alpha  }{}^{\beta  \gamma  } H_{\delta  }{}^{\mu  
\zeta  } H^{\delta  \epsilon  \varepsilon  } \nabla^{\alpha  
}\Phi  \nabla_{\epsilon  }H_{\beta  \gamma  }{}^{\eta  } 
\nabla_{\eta  }H_{\varepsilon  \mu  \zeta  } +   c_{283} 
     H_{\beta  \gamma  }{}^{\epsilon  } H^{\beta  \gamma  
\delta  } H_{\delta  }{}^{\varepsilon  \mu  } \nabla^{\alpha  }
\Phi  \nabla_{\zeta  }H_{\epsilon  \mu  \eta  } \nabla^{\eta  
}H_{\alpha  \varepsilon  }{}^{\zeta  }  \nn\\&&+   c_{284}  
    H_{\alpha  }{}^{\beta  \gamma  } H_{\delta  }{}^{\mu  
\zeta  } H^{\delta  \epsilon  \varepsilon  } \nabla^{\alpha  
}\Phi  \nabla_{\zeta  }H_{\varepsilon  \mu  \eta  } 
\nabla^{\eta  }H_{\beta  \gamma  \epsilon  } +   c_{285} 
     H_{\alpha  }{}^{\beta  \gamma  } H_{\delta  }{}^{\mu  
\zeta  } H^{\delta  \epsilon  \varepsilon  } \nabla^{\alpha  
}\Phi  \nabla_{\eta  }H_{\varepsilon  \mu  \zeta  } 
\nabla^{\eta  }H_{\beta  \gamma  \epsilon  } \nn\\&& +   c_{287} 
     H_{\alpha  }{}^{\beta  \gamma  } H_{\delta  }{}^{\mu  
\zeta  } H^{\delta  \epsilon  \varepsilon  } \nabla^{\alpha  
}\Phi  \nabla_{\zeta  }H_{\gamma  \mu  \eta  } \nabla^{\eta  
}H_{\beta  \epsilon  \varepsilon  } +   c_{288}  
    H_{\alpha  }{}^{\beta  \gamma  } H_{\delta  }{}^{\mu  
\zeta  } H^{\delta  \epsilon  \varepsilon  } \nabla^{\alpha  
}\Phi  \nabla_{\zeta  }H_{\gamma  \varepsilon  \eta  } 
\nabla^{\eta  }H_{\beta  \epsilon  \mu  } \nn\\&& +   c_{290}  
    H_{\alpha  }{}^{\beta  \gamma  } H_{\beta  }{}^{\delta  
\epsilon  } H^{\varepsilon  \mu  \zeta  } \nabla^{\alpha  
}\Phi  \nabla_{\zeta  }H_{\epsilon  \mu  \eta  } \nabla^{\eta 
 }H_{\gamma  \delta  \varepsilon  } +   c_{291}  
    H_{\alpha  }{}^{\beta  \gamma  } H_{\beta  }{}^{\delta  
\epsilon  } H^{\varepsilon  \mu  \zeta  } \nabla^{\alpha  
}\Phi  \nabla_{\eta  }H_{\epsilon  \mu  \zeta  } \nabla^{\eta 
 }H_{\gamma  \delta  \varepsilon  }  \nn\\&&+   c_{292}  
    H_{\alpha  }{}^{\beta  \gamma  } H_{\beta  }{}^{\delta  
\epsilon  } H_{\delta  }{}^{\varepsilon  \mu  } \nabla^{\alpha  
}\Phi  \nabla_{\zeta  }H_{\varepsilon  \mu  \eta  } 
\nabla^{\eta  }H_{\gamma  \epsilon  }{}^{\zeta  } +   
   c_{293}  
    H_{\alpha  }{}^{\beta  \gamma  } H_{\beta  }{}^{\delta  
\epsilon  } H_{\delta  }{}^{\varepsilon  \mu  } \nabla^{\alpha  
}\Phi  \nabla_{\eta  }H_{\varepsilon  \mu  \zeta  } 
\nabla^{\eta  }H_{\gamma  \epsilon  }{}^{\zeta  }  \nn\\&&+   
   c_{297}  
    H_{\alpha  }{}^{\beta  \gamma  } H_{\beta  }{}^{\delta  
\epsilon  } H_{\delta  }{}^{\varepsilon  \mu  } \nabla^{\alpha  
}\Phi  \nabla_{\zeta  }H_{\epsilon  \mu  \eta  } \nabla^{\eta 
 }H_{\gamma  \varepsilon  }{}^{\zeta  } +   c_{298}  
    H_{\alpha  }{}^{\beta  \gamma  } H_{\beta  }{}^{\delta  
\epsilon  } H_{\delta  }{}^{\varepsilon  \mu  } \nabla^{\alpha  
}\Phi  \nabla_{\eta  }H_{\epsilon  \mu  \zeta  } \nabla^{\eta 
 }H_{\gamma  \varepsilon  }{}^{\zeta  }  \nn\\&&+   c_{294}  
    H_{\alpha  }{}^{\beta  \gamma  } H_{\beta  }{}^{\delta  
\epsilon  } H^{\varepsilon  \mu  \zeta  } \nabla^{\alpha  
}\Phi  \nabla_{\zeta  }H_{\delta  \epsilon  \eta  } 
\nabla^{\eta  }H_{\gamma  \varepsilon  \mu  } +   
   c_{295}  
    H_{\alpha  }{}^{\beta  \gamma  } H_{\beta  }{}^{\delta  
\epsilon  } H^{\varepsilon  \mu  \zeta  } \nabla^{\alpha  
}\Phi  \nabla_{\eta  }H_{\delta  \epsilon  \zeta  } 
\nabla^{\eta  }H_{\gamma  \varepsilon  \mu  } \nn\\&& +   
   c_{299}  
    H_{\alpha  }{}^{\beta  \gamma  } H_{\beta  }{}^{\delta  
\epsilon  } H_{\delta  \epsilon  }{}^{\varepsilon  } 
\nabla^{\alpha  }\Phi  \nabla_{\eta  }H_{\varepsilon  \mu  
\zeta  } \nabla^{\eta  }H_{\gamma  }{}^{\mu  \zeta  } +   
   c_{300}  
    H_{\beta  }{}^{\epsilon  \varepsilon  } H^{\beta  \gamma  
\delta  } H_{\gamma  }{}^{\mu  \zeta  } \nabla_{\alpha  
}H_{\mu  \zeta  \eta  } \nabla^{\alpha  }\Phi  \nabla^{\eta  
}H_{\delta  \epsilon  \varepsilon  } \nn\\&& +   c_{301}  
    H_{\alpha  }{}^{\beta  \gamma  } H_{\beta  }{}^{\delta  
\epsilon  } H^{\varepsilon  \mu  \zeta  } \nabla^{\alpha  
}\Phi  \nabla_{\gamma  }H_{\mu  \zeta  \eta  } \nabla^{\eta  
}H_{\delta  \epsilon  \varepsilon  } +   c_{302}  
    H_{\alpha  }{}^{\beta  \gamma  } H_{\beta  \gamma  
}{}^{\delta  } H^{\epsilon  \varepsilon  \mu  } \nabla^{\alpha  
}\Phi  \nabla_{\zeta  }H_{\varepsilon  \mu  \eta  } 
\nabla^{\eta  }H_{\delta  \epsilon  }{}^{\zeta  } \nn\\&& +   
   c_{303}  
    H_{\alpha  }{}^{\beta  \gamma  } H_{\beta  \gamma  
}{}^{\delta  } H^{\epsilon  \varepsilon  \mu  } \nabla^{\alpha  
}\Phi  \nabla_{\eta  }H_{\varepsilon  \mu  \zeta  } 
\nabla^{\eta  }H_{\delta  \epsilon  }{}^{\zeta  } +   
   c_{306}  
    H_{\beta  }{}^{\epsilon  \varepsilon  } H^{\beta  \gamma  
\delta  } H_{\gamma  \epsilon  }{}^{\mu  } \nabla_{\alpha  }H_{
\mu  \zeta  \eta  } \nabla^{\alpha  }\Phi  \nabla^{\eta  }H_{
\delta  \varepsilon  }{}^{\zeta  } \nn\\&& +   c_{304}  
    H_{\beta  \gamma  }{}^{\epsilon  } H^{\beta  \gamma  
\delta  } H^{\varepsilon  \mu  \zeta  } \nabla_{\alpha  
}H_{\epsilon  \zeta  \eta  } \nabla^{\alpha  }\Phi  
\nabla^{\eta  }H_{\delta  \varepsilon  \mu  } +   
   c_{305}  
    H_{\alpha  }{}^{\beta  \gamma  } H_{\beta  }{}^{\delta  
\epsilon  } H^{\varepsilon  \mu  \zeta  } \nabla^{\alpha  
}\Phi  \nabla_{\gamma  }H_{\epsilon  \zeta  \eta  } 
\nabla^{\eta  }H_{\delta  \varepsilon  \mu  } \nn\\&& +   
   c_{308}  
    H_{\beta  \gamma  }{}^{\epsilon  } H^{\beta  \gamma  
\delta  } H_{\delta  }{}^{\varepsilon  \mu  } \nabla_{\alpha  
}H_{\mu  \zeta  \eta  } \nabla^{\alpha  }\Phi  \nabla^{\eta  
}H_{\epsilon  \varepsilon  }{}^{\zeta  } +   c_{309}  
    H_{\alpha  }{}^{\beta  \gamma  } H_{\beta  }{}^{\delta  
\epsilon  } H_{\delta  }{}^{\varepsilon  \mu  } \nabla^{\alpha  
}\Phi  \nabla_{\gamma  }H_{\mu  \zeta  \eta  } \nabla^{\eta  
}H_{\epsilon  \varepsilon  }{}^{\zeta  } \nn\\&& +   c_{310}  
    H_{\alpha  }{}^{\beta  \gamma  } H_{\beta  \gamma  
}{}^{\delta  } H^{\epsilon  \varepsilon  \mu  } \nabla^{\alpha  
}\Phi  \nabla_{\delta  }H_{\mu  \zeta  \eta  } \nabla^{\eta  
}H_{\epsilon  \varepsilon  }{}^{\zeta  } +   c_{311}  
    H_{\alpha  }{}^{\beta  \gamma  } H_{\beta  }{}^{\delta  
\epsilon  } H_{\delta  \epsilon  }{}^{\varepsilon  } 
\nabla^{\alpha  }\Phi  \nabla_{\gamma  }H_{\mu  \zeta  \eta  
} \nabla^{\eta  }H_{\varepsilon  }{}^{\mu  \zeta  } \nn\\&& +   
   c_{278}  
    H_{\beta  }{}^{\epsilon  \varepsilon  } H^{\beta  \gamma  
\delta  } H^{\mu  \zeta  \eta  } \nabla^{\alpha  }\Phi  
\nabla_{\eta  }H_{\delta  \varepsilon  \zeta  } \nabla_{\mu  
}H_{\alpha  \gamma  \epsilon  } +   c_{289}  
    H_{\alpha  }{}^{\beta  \gamma  } H_{\delta  \epsilon  }{}^{
\mu  } H^{\delta  \epsilon  \varepsilon  } \nabla^{\alpha  
}\Phi  \nabla^{\eta  }H_{\beta  \varepsilon  }{}^{\zeta  } 
\nabla_{\mu  }H_{\gamma  \zeta  \eta  }  \nn\\&&+   c_{282}  
    H_{\beta  \gamma  }{}^{\epsilon  } H^{\beta  \gamma  
\delta  } H_{\delta  }{}^{\varepsilon  \mu  } \nabla^{\alpha  }
\Phi  \nabla^{\eta  }H_{\alpha  \varepsilon  }{}^{\zeta  } 
\nabla_{\mu  }H_{\epsilon  \zeta  \eta  } +   c_{296}  
    H_{\alpha  }{}^{\beta  \gamma  } H_{\beta  }{}^{\delta  
\epsilon  } H_{\delta  }{}^{\varepsilon  \mu  } \nabla^{\alpha  
}\Phi  \nabla^{\eta  }H_{\gamma  \varepsilon  }{}^{\zeta  } 
\nabla_{\mu  }H_{\epsilon  \zeta  \eta  } \nn\\&& +   c_{307}  
    H_{\alpha  }{}^{\beta  \gamma  } H_{\beta  }{}^{\delta  
\epsilon  } H_{\gamma  }{}^{\varepsilon  \mu  } \nabla^{\alpha  
}\Phi  \nabla^{\eta  }H_{\delta  \varepsilon  }{}^{\zeta  } 
\nabla_{\mu  }H_{\epsilon  \zeta  \eta  } +   c_{286}  
    H_{\alpha  }{}^{\beta  \gamma  } H_{\delta  \epsilon  }{}^{
\mu  } H^{\delta  \epsilon  \varepsilon  } \nabla^{\alpha  
}\Phi  \nabla^{\eta  }H_{\beta  \gamma  }{}^{\zeta  } \nabla_{
\mu  }H_{\varepsilon  \zeta  \eta  },\nn
\eeqa

\beqa
[H(\nabla H)^2(\nabla\Phi)^3]_3&\!\!\!\!\!=\!\!\!\!\!& c_{137}  
   \! H^{\delta  \epsilon  \varepsilon  } \nabla^{\alpha  }\Phi  
\nabla_{\beta  }H_{\alpha  \delta  }{}^{\mu  } \nabla^{\beta  
}\Phi  \nabla^{\gamma  }\Phi  \nabla_{\varepsilon  }H_{\gamma  
\epsilon  \mu  }\! + \!  c_{201}  
    H^{\delta  \epsilon  \varepsilon  } \nabla^{\alpha  }\Phi  
\nabla^{\beta  }\Phi  \nabla_{\gamma  }H_{\beta  \varepsilon  
\mu  } \nabla^{\gamma  }\Phi  \nabla^{\mu  }H_{\alpha  \delta 
 \epsilon  } \nn\\&\!\!\!\!\!\!\!\!\!&+   c_{210}  
    H^{\gamma  \delta  \epsilon  } \nabla_{\alpha  }\Phi  
\nabla^{\alpha  }\Phi  \nabla^{\beta  }\Phi  \nabla_{\mu  
}H_{\delta  \epsilon  \varepsilon  } \nabla^{\mu  }H_{\beta  
\gamma  }{}^{\varepsilon  },\nn 
\eeqa
\beqa
[H(\nabla H)^2R\nabla\Phi]_{20}&=&
   c_{258}  
    H^{\beta  \gamma  \delta  } R_{\delta  \zeta  
\varepsilon  \mu  } \nabla^{\alpha  }\Phi  \nabla_{\beta  }H_{
\alpha  }{}^{\epsilon  \varepsilon  } \nabla^{\zeta  }H_{\gamma 
 \epsilon  }{}^{\mu  } +   c_{260}  
    H^{\beta  \gamma  \delta  } R_{\delta  \zeta  
\epsilon  \mu  } \nabla^{\alpha  }\Phi  \nabla^{\varepsilon  
}H_{\alpha  \beta  }{}^{\epsilon  } \nabla^{\zeta  }H_{\gamma  
\varepsilon  }{}^{\mu  } \nn\\&&+   c_{259}  
    H^{\beta  \gamma  \delta  } R_{\delta  \mu  
\epsilon  \zeta  } \nabla^{\alpha  }\Phi  \nabla^{\varepsilon  
}H_{\alpha  \beta  }{}^{\epsilon  } \nabla^{\zeta  }H_{\gamma  
\varepsilon  }{}^{\mu  } +   c_{261}  
    H_{\alpha  }{}^{\beta  \gamma  } R_{\delta  \mu  
\epsilon  \zeta  } \nabla^{\alpha  }\Phi  \nabla^{\varepsilon  
}H_{\beta  }{}^{\delta  \epsilon  } \nabla^{\zeta  }H_{\gamma  
\varepsilon  }{}^{\mu  }\nn\\&& +   c_{262}  
    H^{\beta  \gamma  \delta  } R_{\alpha  \zeta  
\varepsilon  \mu  } \nabla^{\alpha  }\Phi  \nabla^{\varepsilon 
 }H_{\beta  \gamma  }{}^{\epsilon  } \nabla^{\zeta  }H_{\delta  
\epsilon  }{}^{\mu  } +   c_{264}  
    H_{\alpha  }{}^{\beta  \gamma  } R_{\gamma  \zeta  
\varepsilon  \mu  } \nabla^{\alpha  }\Phi  \nabla^{\varepsilon 
 }H_{\beta  }{}^{\delta  \epsilon  } \nabla^{\zeta  }H_{\delta  
\epsilon  }{}^{\mu  }\nn\\&& +   c_{263}  
    H_{\alpha  }{}^{\beta  \gamma  } R_{\gamma  \mu  
\varepsilon  \zeta  } \nabla^{\alpha  }\Phi  
\nabla^{\varepsilon  }H_{\beta  }{}^{\delta  \epsilon  } 
\nabla^{\zeta  }H_{\delta  \epsilon  }{}^{\mu  } +   
   c_{266}  
    H^{\beta  \gamma  \delta  } R_{\alpha  \zeta  
\epsilon  \mu  } \nabla^{\alpha  }\Phi  \nabla^{\varepsilon  
}H_{\beta  \gamma  }{}^{\epsilon  } \nabla^{\zeta  }H_{\delta  
\varepsilon  }{}^{\mu  } \nn\\&&+   c_{265}  
    H^{\beta  \gamma  \delta  } R_{\alpha  \mu  
\epsilon  \zeta  } \nabla^{\alpha  }\Phi  \nabla^{\varepsilon  
}H_{\beta  \gamma  }{}^{\epsilon  } \nabla^{\zeta  }H_{\delta  
\varepsilon  }{}^{\mu  } +   c_{268}  
    H_{\alpha  }{}^{\beta  \gamma  } R_{\gamma  \zeta  
\epsilon  \mu  } \nabla^{\alpha  }\Phi  \nabla^{\varepsilon  
}H_{\beta  }{}^{\delta  \epsilon  } \nabla^{\zeta  }H_{\delta  
\varepsilon  }{}^{\mu  }\nn\\&& +   c_{267}  
    H_{\alpha  }{}^{\beta  \gamma  } R_{\gamma  \mu  
\epsilon  \zeta  } \nabla^{\alpha  }\Phi  \nabla^{\varepsilon  
}H_{\beta  }{}^{\delta  \epsilon  } \nabla^{\zeta  }H_{\delta  
\varepsilon  }{}^{\mu  } +   c_{269}  
    H^{\beta  \gamma  \delta  } R_{\epsilon  \zeta  
\varepsilon  \mu  } \nabla^{\alpha  }\Phi  \nabla_{\gamma  
}H_{\alpha  \beta  }{}^{\epsilon  } \nabla^{\zeta  }H_{\delta  
}{}^{\varepsilon  \mu  } \nn\\&&+   c_{270}  
    H_{\alpha  }{}^{\beta  \gamma  } R_{\epsilon  
\zeta  \varepsilon  \mu  } \nabla^{\alpha  }\Phi  
\nabla^{\epsilon  }H_{\beta  \gamma  }{}^{\delta  } 
\nabla^{\zeta  }H_{\delta  }{}^{\varepsilon  \mu  } +   
   c_{272}  
    H^{\beta  \gamma  \delta  } R_{\gamma  \mu  
\delta  \zeta  } \nabla^{\alpha  }\Phi  \nabla_{\beta  
}H_{\alpha  }{}^{\epsilon  \varepsilon  } \nabla^{\zeta  
}H_{\epsilon  \varepsilon  }{}^{\mu  } \nn\\&&+   c_{273}  
    H^{\beta  \gamma  \delta  } R_{\gamma  \mu  
\delta  \zeta  } \nabla^{\alpha  }\Phi  \nabla^{\varepsilon  
}H_{\alpha  \beta  }{}^{\epsilon  } \nabla^{\zeta  }H_{\epsilon 
 \varepsilon  }{}^{\mu  } +   c_{275}  
    H^{\beta  \gamma  \delta  } R_{\alpha  \zeta  
\delta  \mu  } \nabla^{\alpha  }\Phi  \nabla^{\varepsilon  
}H_{\beta  \gamma  }{}^{\epsilon  } \nabla^{\zeta  }H_{\epsilon 
 \varepsilon  }{}^{\mu  }\nn\\&& +   c_{274}  
    H^{\beta  \gamma  \delta  } R_{\alpha  \mu  
\delta  \zeta  } \nabla^{\alpha  }\Phi  \nabla^{\varepsilon  
}H_{\beta  \gamma  }{}^{\epsilon  } \nabla^{\zeta  }H_{\epsilon 
 \varepsilon  }{}^{\mu  } +   c_{276}  
    H^{\beta  \gamma  \delta  } R_{\delta  \zeta  
\varepsilon  \mu  } \nabla^{\alpha  }\Phi  \nabla_{\gamma  
}H_{\alpha  \beta  }{}^{\epsilon  } \nabla^{\zeta  }H_{\epsilon 
 }{}^{\varepsilon  \mu  }\nn\\&& +   c_{277}  
    H_{\alpha  }{}^{\beta  \gamma  } R_{\delta  \zeta  
\varepsilon  \mu  } \nabla^{\alpha  }\Phi  \nabla^{\epsilon  
}H_{\beta  \gamma  }{}^{\delta  } \nabla^{\zeta  }H_{\epsilon  
}{}^{\varepsilon  \mu  } +   c_{271}  
    H^{\beta  \gamma  \delta  } R_{\alpha  \zeta  
\gamma  \delta  } \nabla^{\alpha  }\Phi  \nabla^{\zeta  
}H_{\epsilon  \varepsilon  \mu  } \nabla^{\mu  }H_{\beta  }{}^{
\epsilon  \varepsilon  },\nn
\eeqa
\beqa
[(\nabla H)^3(\nabla\Phi)^2]_3&=&  c_{212} 
     \nabla^{\alpha  }\Phi  \nabla^{\beta  }\Phi  
\nabla^{\epsilon  }H_{\alpha  }{}^{\gamma  \delta  } 
\nabla_{\mu  }H_{\delta  \epsilon  \varepsilon  } \nabla^{\mu  
}H_{\beta  \gamma  }{}^{\varepsilon  } +   c_{219}  
    \nabla^{\alpha  }\Phi  \nabla^{\beta  }\Phi  
\nabla^{\epsilon  }H_{\alpha  }{}^{\gamma  \delta  } 
\nabla_{\mu  }H_{\gamma  \delta  \varepsilon  } \nabla^{\mu  
}H_{\beta  \epsilon  }{}^{\varepsilon  }\nn\\&& +   c_{223}  
    \nabla^{\alpha  }\Phi  \nabla_{\beta  }H_{\delta  
\varepsilon  \mu  } \nabla^{\beta  }\Phi  \nabla^{\epsilon  
}H_{\alpha  }{}^{\gamma  \delta  } \nabla^{\mu  }H_{\gamma  
\epsilon  }{}^{\varepsilon  },\nn
\eeqa
\beqa
[(\nabla H)^3\nabla^2\Phi]_3&=& 
   c_{213}  
    \nabla^{\beta  }\nabla^{\alpha  }\Phi  \nabla^{\epsilon  
}H_{\alpha  }{}^{\gamma  \delta  } \nabla_{\mu  }H_{\delta  
\epsilon  \varepsilon  } \nabla^{\mu  }H_{\beta  \gamma  
}{}^{\varepsilon  } +   c_{220}  
    \nabla^{\beta  }\nabla^{\alpha  }\Phi  \nabla^{\epsilon  
}H_{\alpha  }{}^{\gamma  \delta  } \nabla_{\mu  }H_{\gamma  
\delta  \varepsilon  } \nabla^{\mu  }H_{\beta  \epsilon  
}{}^{\varepsilon  } \nn\\&&+   c_{224}  
    \nabla_{\beta  }H_{\delta  \varepsilon  \mu  } 
\nabla^{\beta  }\nabla^{\alpha  }\Phi  \nabla^{\epsilon  
}H_{\alpha  }{}^{\gamma  \delta  } \nabla^{\mu  }H_{\gamma  
\epsilon  }{}^{\varepsilon  },\nn 
\eeqa

\beqa
[H(\nabla H)^2\nabla\Phi\nabla^2\Phi]_{23}&\!\!\!\!\!=\!\!\!\!\!& 
   c_{114}  
    H^{\delta  \epsilon  \varepsilon  } \nabla^{\alpha  }\Phi  
\nabla_{\gamma  }H_{\beta  \varepsilon  \mu  } \nabla^{\gamma  
}\nabla^{\beta  }\Phi  \nabla_{\epsilon  }H_{\alpha  \delta  
}{}^{\mu  } +   c_{120}  
    H_{\beta  }{}^{\delta  \epsilon  } \nabla^{\alpha  }\Phi  
\nabla^{\gamma  }\nabla^{\beta  }\Phi  \nabla_{\delta  
}H_{\alpha  }{}^{\varepsilon  \mu  } \nabla_{\epsilon  
}H_{\gamma  \varepsilon  \mu  }\nn\\&\!\!\!\!\!\!\!\!\!\!& +   c_{138}  
    H^{\delta  \epsilon  \varepsilon  } \nabla^{\alpha  }\Phi  
\nabla_{\beta  }H_{\alpha  \delta  }{}^{\mu  } \nabla^{\gamma  
}\nabla^{\beta  }\Phi  \nabla_{\varepsilon  }H_{\gamma  
\epsilon  \mu  } +   c_{139}  
    H^{\delta  \epsilon  \varepsilon  } \nabla^{\alpha  }\Phi  
\nabla^{\gamma  }\nabla^{\beta  }\Phi  \nabla_{\delta  
}H_{\alpha  \beta  }{}^{\mu  } \nabla_{\varepsilon  }H_{\gamma  
\epsilon  \mu  } \nn\\&\!\!\!\!\!\!\!\!\!\!&+   c_{173}  
    H^{\delta  \epsilon  \varepsilon  } \nabla^{\alpha  }\Phi  
\nabla^{\gamma  }\nabla^{\beta  }\Phi  \nabla_{\delta  
}H_{\alpha  \beta  }{}^{\mu  } \nabla_{\mu  }H_{\gamma  
\epsilon  \varepsilon  } +   c_{178}  
    H_{\beta  }{}^{\delta  \epsilon  } \nabla^{\alpha  }\Phi  
\nabla_{\gamma  }H_{\alpha  }{}^{\varepsilon  \mu  } 
\nabla^{\gamma  }\nabla^{\beta  }\Phi  \nabla_{\mu  
}H_{\delta  \epsilon  \varepsilon  } \nn\\&\!\!\!\!\!\!\!\!\!\!&+   c_{179}  
    H_{\alpha  }{}^{\delta  \epsilon  } \nabla^{\alpha  }\Phi  
\nabla_{\gamma  }H_{\beta  }{}^{\varepsilon  \mu  } 
\nabla^{\gamma  }\nabla^{\beta  }\Phi  \nabla_{\mu  
}H_{\delta  \epsilon  \varepsilon  } +   c_{197}  
    H^{\delta  \epsilon  \varepsilon  } \nabla^{\alpha  }\Phi  
\nabla^{\gamma  }\nabla^{\beta  }\Phi  \nabla_{\varepsilon  
}H_{\gamma  \epsilon  \mu  } \nabla^{\mu  }H_{\alpha  \beta  
\delta  } \nn\\&\!\!\!\!\!\!\!\!\!\!&+   c_{198}  
    H^{\delta  \epsilon  \varepsilon  } \nabla^{\alpha  }\Phi  
\nabla^{\gamma  }\nabla^{\beta  }\Phi  \nabla_{\mu  
}H_{\gamma  \epsilon  \varepsilon  } \nabla^{\mu  }H_{\alpha  
\beta  \delta  } +   c_{199}  
    H_{\beta  }{}^{\delta  \epsilon  } \nabla^{\alpha  }\Phi  
\nabla^{\gamma  }\nabla^{\beta  }\Phi  \nabla_{\varepsilon  
}H_{\delta  \epsilon  \mu  } \nabla^{\mu  }H_{\alpha  \gamma  
}{}^{\varepsilon  } \nn\\&\!\!\!\!\!\!\!\!\!\!&+   c_{200}  
    H_{\beta  }{}^{\delta  \epsilon  } \nabla^{\alpha  }\Phi  
\nabla^{\gamma  }\nabla^{\beta  }\Phi  \nabla_{\mu  
}H_{\delta  \epsilon  \varepsilon  } \nabla^{\mu  }H_{\alpha  
\gamma  }{}^{\varepsilon  } +   c_{202}  
    H^{\delta  \epsilon  \varepsilon  } \nabla^{\alpha  }\Phi  
\nabla_{\gamma  }H_{\beta  \varepsilon  \mu  } \nabla^{\gamma  
}\nabla^{\beta  }\Phi  \nabla^{\mu  }H_{\alpha  \delta  
\epsilon  } \nn\\&\!\!\!\!\!\!\!\!\!\!&+   c_{203}  
    H_{\beta  }{}^{\delta  \epsilon  } \nabla^{\alpha  }\Phi  
\nabla^{\gamma  }\nabla^{\beta  }\Phi  \nabla_{\epsilon  
}H_{\gamma  \varepsilon  \mu  } \nabla^{\mu  }H_{\alpha  
\delta  }{}^{\varepsilon  } +   c_{204}  
    H_{\beta  }{}^{\delta  \epsilon  } \nabla^{\alpha  }\Phi  
\nabla^{\gamma  }\nabla^{\beta  }\Phi  \nabla_{\varepsilon  
}H_{\gamma  \epsilon  \mu  } \nabla^{\mu  }H_{\alpha  \delta  
}{}^{\varepsilon  } \nn\\&\!\!\!\!\!\!\!\!\!\!&+   c_{205}  
    H_{\beta  }{}^{\delta  \epsilon  } \nabla^{\alpha  }\Phi  
\nabla^{\gamma  }\nabla^{\beta  }\Phi  \nabla_{\mu  
}H_{\gamma  \epsilon  \varepsilon  } \nabla^{\mu  }H_{\alpha  
\delta  }{}^{\varepsilon  } +   c_{209}  
    H^{\gamma  \delta  \epsilon  } \nabla^{\alpha  }\Phi  
\nabla^{\beta  }\nabla_{\alpha  }\Phi  \nabla_{\varepsilon  
}H_{\delta  \epsilon  \mu  } \nabla^{\mu  }H_{\beta  \gamma  
}{}^{\varepsilon  } \nn\\&\!\!\!\!\!\!\!\!\!\!&+   c_{211}  
    H^{\gamma  \delta  \epsilon  } \nabla^{\alpha  }\Phi  
\nabla^{\beta  }\nabla_{\alpha  }\Phi  \nabla_{\mu  
}H_{\delta  \epsilon  \varepsilon  } \nabla^{\mu  }H_{\beta  
\gamma  }{}^{\varepsilon  } +   c_{214}  
    H^{\delta  \epsilon  \varepsilon  } \nabla_{\alpha  
}H_{\gamma  \varepsilon  \mu  } \nabla^{\alpha  }\Phi  
\nabla^{\gamma  }\nabla^{\beta  }\Phi  \nabla^{\mu  }H_{\beta 
 \delta  \epsilon  }\nn\\&\!\!\!\!\!\!\!\!\!\!& +   c_{218}  
    H_{\alpha  }{}^{\delta  \epsilon  } \nabla^{\alpha  }\Phi  
\nabla^{\gamma  }\nabla^{\beta  }\Phi  \nabla_{\epsilon  
}H_{\gamma  \varepsilon  \mu  } \nabla^{\mu  }H_{\beta  \delta 
 }{}^{\varepsilon  } +   c_{221}  
    H^{\gamma  \delta  \epsilon  } \nabla^{\alpha  }\Phi  
\nabla_{\beta  }H_{\epsilon  \varepsilon  \mu  } \nabla^{\beta 
 }\nabla_{\alpha  }\Phi  \nabla^{\mu  }H_{\gamma  \delta  
}{}^{\varepsilon  }\nn\\&\!\!\!\!\!\!\!\!\!\!& +   c_{222}  
    H_{\beta  }{}^{\delta  \epsilon  } \nabla_{\alpha  
}H_{\epsilon  \varepsilon  \mu  } \nabla^{\alpha  }\Phi  
\nabla^{\gamma  }\nabla^{\beta  }\Phi  \nabla^{\mu  
}H_{\gamma  \delta  }{}^{\varepsilon  } +   c_{228}  
    H_{\alpha  \beta  }{}^{\delta  } \nabla^{\alpha  }\Phi  
\nabla^{\gamma  }\nabla^{\beta  }\Phi  \nabla_{\mu  
}H_{\delta  \epsilon  \varepsilon  } \nabla^{\mu  }H_{\gamma  
}{}^{\epsilon  \varepsilon  } \nn\\&\!\!\!\!\!\!\!\!\!\!&+   c_{229}  
    H_{\alpha  \beta  }{}^{\delta  } \nabla^{\alpha  }\Phi  
\nabla_{\gamma  }H_{\epsilon  \varepsilon  \mu  } 
\nabla^{\gamma  }\nabla^{\beta  }\Phi  \nabla^{\mu  
}H_{\delta  }{}^{\epsilon  \varepsilon  }.\nn
\eeqa


\end{document}